\documentclass[hidelinks, 10pt, twocolumn]{IEEEtran}
\IEEEoverridecommandlockouts
\def\BibTeX{{\rm B\kern-.05em{\sc i\kern-.025em b}\kern-.08em
		T\kern-.1667em\lower.7ex\hbox{E}\kern-.125emX}}
\usepackage[normalem]{ulem}
\usepackage{textcomp}
\usepackage{bm}
\usepackage{xurl}
\usepackage{subfigure}
\usepackage{epsfig}
\usepackage[table,xcdraw]{xcolor}
\usepackage[utf8x]{inputenc}
\usepackage{rotating}
\usepackage{graphicx}
\usepackage{tabularx}
\usepackage{array}
\usepackage{color,soul}

\usepackage{xtab, booktabs}
\usepackage{moreverb}
\usepackage{epsfig}
\usepackage{amsmath,amssymb,amsthm,mathrsfs,amsfonts,dsfont}
\usepackage{adjustbox,lipsum}
\usepackage{multirow}
\usepackage{caption}
\usepackage{anyfontsize}
\usepackage{epstopdf}
\usepackage{fancyhdr}
\usepackage[noblocks]{authblk}
\usepackage[subfigure]{tocloft}
\usepackage{tcolorbox}
\usepackage{algorithmic}
\usepackage[linesnumbered,ruled,vlined]{algorithm2e}
\SetKwInput{KwInput}{Input}
\SetKwInput{KwOutput}{Output}
\definecolor{azure(colorwheel)}{rgb}{0.0, 0.5, 1.0}
\usepackage[colorlinks=true,allcolors=azure(colorwheel)]{hyperref}
\usepackage{fancyhdr}
\usepackage{tablefootnote}
\usepackage{cite}
\allowdisplaybreaks
\usepackage{lettrine}
\usepackage{pifont}
\newcommand{\cmark}{\ding{51}}%
\newcommand{\xmark}{\ding{55}}%

\makeatletter
\def\endthebibliography{%
	\def\@noitemerr{\@latex@warning{Empty `thebibliography' environment}}%
	\endlist
}
\makeatother

\newcommand{\nosemic}{\renewcommand{\@endalgocfline}{\relax}}
\newcommand{\dosemic}{\renewcommand{\@endalgocfline}{\algocf@endline}}
\let\oldnl\nl
\newcommand{\nonl}{\renewcommand{\nl}{\let\nl\oldnl}}
\makeatother

\makeatletter
\def\endthebibliography{%
	\def\@noitemerr{\@latex@warning{Empty `thebibliography' environment}}%
	\endlist
}
\makeatother

\renewcommand{\thetable}{\arabic{table}}

\begin{document}
	\title{A {C}omprehensive {S}urvey of {S}pectrum {S}haring {S}chemes from a Standardization and Implementation Perspective}
		\author{Mohammad Parvini, \textit{Student Member, IEEE},~Amir Hossein Zarif, \textit{Student Member, IEEE},~Ali Nouruzi, \textit{Student Member, IEEE},
		~Nader Mokari,~\textit{Senior Member, IEEE}, ~Mohammad Reza Javan,~\textit{Senior Member, IEEE}, Bijan Abbasi,~\textit{Senior Member, IEEE}, Amir Ghasemi~\textit{Senior Member, IEEE}, and Halim Yanikomeroglu,~\textit{Fellow, IEEE}
		\thanks{M. Parvini,~ A.H. Zarif,~ A. Nouruzi, N. Mokari, and B. Abbasi are with the Department of Electrical and Computer Engineering,~Tarbiat Modares University,~Tehran,~Iran, (e-mails: \{m.parvini, a.zarif, ali\_nouruzi nader.mokari, abbasi\}@modares.ac.ir). M.~R.~Javan is with the Department of Electrical and Computer Engineering, Shahrood University of Technology, Iran, (e-mail: javan@shahroodut.ac.ir). A. Ghasemi is with the Commutations Research Centre, Innovation,  Science  and  Economic  Development  Canada,  Ottawa,  Ontario, K2H 8S2 Canada (e-mail: amir.ghasemi@canada.ca). H. Yanikomeroglu is with the Department of Systems and Computer Engineering, Carleton University, Ottawa, Canada (e-mail: halim@sce.carleton.ca).}}
	\maketitle
	\begin{abstract}
		As the services and requirements of next-generation wireless networks become increasingly diversified, it is estimated that the current frequency bands of mobile network operators (MNOs) will be unable to cope with the immensity of anticipated demands. Due to spectrum scarcity, there has been a growing trend among stakeholders toward identifying practical solutions to make the most productive use of the exclusively allocated bands on a shared basis through spectrum sharing mechanisms. However, due to the technical complexities of these mechanisms, their design presents challenges, as it requires coordination among multiple entities. To address this challenge, in this paper, we begin with a detailed review of the recent literature on spectrum sharing methods, classifying them on the basis of their operational frequency regime—that is, whether they are implemented to operate in licensed bands (e.g., licensed shared access (LSA), spectrum access system (SAS), and dynamic spectrum sharing (DSS)) or unlicensed bands (e.g., LTE-unlicensed (LTE-U), licensed assisted access (LAA), MulteFire, and new radio-unlicensed (NR-U)). Then, in order to narrow the gap between the standardization and vendor-specific implementations, we provide a detailed review of the potential implementation scenarios and necessary amendments to legacy cellular networks from the perspective of telecom vendors and regulatory bodies. Next, we analyze applications of artificial intelligence (AI) and machine learning (ML) techniques for facilitating spectrum sharing mechanisms and leveraging the full potential of autonomous sharing scenarios. Finally, we conclude the paper by presenting open research challenges, which aim to provide insights into prospective research endeavors.
	\end{abstract}
	\section{Introduction}
	\lettrine[lraise=0, findent=0.1em, nindent=0.2em, slope=-.5em, lines=2]{S}{ince}  the advent of the first-generation (1G) mobile communication system, there has been a tremendous growth in the number of wireless broadband and multimedia devices. In addition, with the introduction of the fifth-generation (5G) wireless networks, new services and use cases have been foreseen, namely, enhanced mobile broadband (eMBB), which focuses on enhancements to user data rates, ultra-reliable low latency communications (URLLC), which aims to provide low-latency and ultra-high reliability for mission-critical and time-sensitive applications, and massive machine type communications (mMTC), which enables communications between a great number of devices \cite{agiwal2016next}. Furthermore, the current industry trend of ubiquitous Internet connections for everything has lead to the evolution of the Internet of things (IoT) and its subset, industrial IoT (IIoT), which paves the way toward efficient and sustainable production \cite{sisinni2018industrial}. 
	5G and the forthcoming sixth-generation (6G) communication systems will benefit numerous industry areas and accelerate IoT implementation. By 2030, it is estimated that over 50 billion devices will be connected as part of the IoT. In order to cope with the explosive growth of new applications and mobile data traffic and to  address their immense demands for bandwidth, new solutions must be sought as the available frequency resources are limited. Therefore, it is essential to explore supplementary frequency bands in higher spectrums or revise the utilization of currently available bands. The so-called millimeter-wave (mmW) bands, ranging from 30 to 300 GHz, can provide greater bandwidths compared to the current cellular allocations constrained to the sub-3 GHz band. Due to the smaller wavelengths of mmW signals, greater numbers of antennas can be established in the same dimensions, which can lead to very high gains. However, despite these encouraging developments, challenges in the use of mmW bands still exist that limit their large-scale implementation. These challenges can be summarized as follows: 1) due to the directional behavior of mmWs, specific changes have to be applied to current cellular systems; 2) mmW bands are susceptible to shadowing and can be readily affected by multiple objects; and 3) due to the small coherence time of mmW bands, channels sustain rapid fluctuations, leading to intermittent connectivity and necessitating adaptable communication \cite{rangan2014millimeter}. 
	
	\begin{figure*}[!t] 
		\centering
		\includegraphics[width=.9\textwidth]{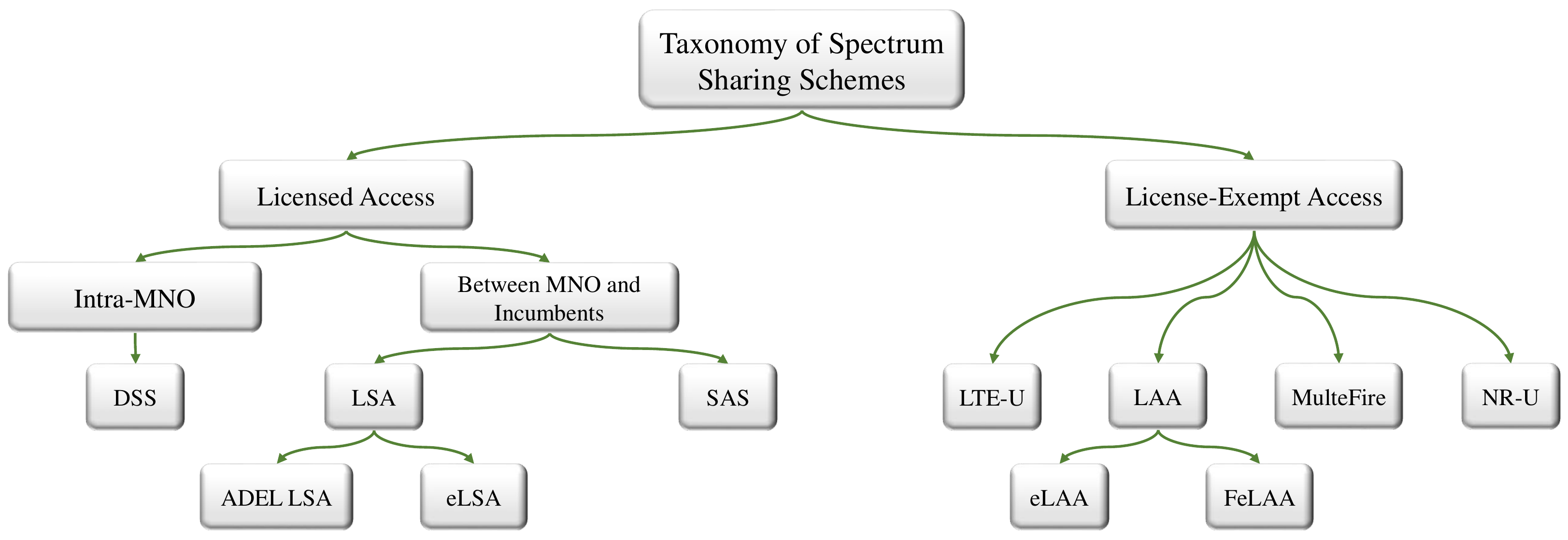}
		\caption{Taxonomy of spectrum sharing schemes.} 
		\label{Taxonomy}
	\end{figure*}
	These challenges call for an investigation into alternative solutions. One solution would involve improving the utilization of available radio frequency bands in the conventional microwave bands  (i.e., $ < $ 6 GHz), which would address the issue of spectrum scarcity.  However, most sub-6 GHz bands, due to their desired propagation properties, have already been fragmented and allocated to various non-mobile users on an exclusive basis (e.g., radars, TV broadcasting, Federal aeronautical users, military services, medical, non-federal fixed-satellite services, and event production). Despite this rigid allocation of spectrum, there are many reasons for mobile network operators (MNOs) to be optimistic about the future. Several spectrum measurements in frequencies up to 6 GHz indicate that the utilization rate of users who currently maintain an exclusive license to use a spectrum band is on the scale of 10–20 percent, which is relatively inefficient \cite{yang2016advanced}. Thus, much prime spectrum remains unused in the spatial and temporal domains, and there is growing pressure on commercial mobile bands, especially in dense urban areas. This is the primary motivation for the development of novel sharing methods, which permits the shared use of the spectrum. This notion, which is called spectrum sharing, allows multiple classes of users to share the same frequency bands in a reliable manner. 

	\vspace*{-1em}
	\subsection{Motivation}
	Compared to the fixed allocation of spectrum, where primary users (known as incumbents) can access the frequency bands on an exclusive basis, spectrum sharing facilitates the shared use of the spectrum.  This lets MNOs access the incumbents' dedicated frequency chunks under specific rules and conditions. Although it may seem trivial, the spectrum sharing process is inherently complicated, involving various organizations and parties. Spectrum sharing can limit the activities of incumbents on the band due to MNOs operating simultaneously; therefore, proper incentives are needed to motivate incumbents to participate in spectrum sharing schemes and to view participation in such schemes as a net positive outcome for their organization. However, if conditions are too prohibitive or the sharing agreements with the users too onerous, MNOs may become discouraged from taking part in the sharing process. Such a situation would impair the economic benefits for both sides.
	Some of the potential benefits and risks that spectrum sharing can offer to incumbents and MNOs are described here: \cite{gsmalsa}:
	\subsubsection{Potential Advantages}
	By participating in sharing agreements, incumbents can profit from the additional funding through spectrum licensing auctions and fees. Spectrum sharing can depreciate capital expenditures for both sides as incumbents and MNOs utilize the spectrum on a shared basis. On the other side, MNOs can enrich their users' quality of services with the additional spectrum they have access to on a shared basis.
	\subsubsection{Potential Risks}
	Under a sharing contract, incumbents have a smaller geographic spectrum footprint, less flexibility regarding how to utilize the spectrum, and are more confined in their time of use. Furthermore, incumbents' services and activities may deteriorate sharply if MNOs do not adhere to the negotiated performance levels. The shared spectrum, which is naturally less reliable, limits the MNOs from applying it to meet their diverse marketplace services.
	
	\begin{table*}[!t]
		\centering
		\label{surveyed_docs}
		\captionsetup{font=scriptsize} 
		\setlength\tabcolsep{0pt} 
		\setlength\lightrulewidth{0.051pt}
		\caption{CLASSIFICATION OF SURVEY WORKS IN THE FIELD OF SPECTRUM SHARING AND COGNITIVE RADIO.}
		{\renewcommand{\arraystretch}{1.2}
			\begin{tabular}{p{3.4cm} p{13.9cm} p{1cm}}
				\toprule[1pt]
				\textbf{Main domain} & \textbf{Paper title} & \textbf{Ref.}  \\
				\midrule[0.5pt]
				& Advances on spectrum sensing for cognitive radio networks: Theory and applications & \cite{ali2016advances}\\
				\textbf{Spectrum Sensing}& Application of compressive sensing in cognitive radio communications: A survey & \cite{sharma2016application}\\
				& Energy-efficient cooperative spectrum sensing: A survey & \cite{cichon2016energy}\\\midrule[0.1pt]
				& Resource allocation for underlay cognitive radio networks: A survey & \cite{el2016resource}\\
				\textbf{Resource Allocation}&A survey on radio resource allocation in cognitive radio sensor networks&\cite{ahmad2015survey}\\
				& Radio resource allocation techniques for efficient spectrum access in cognitive radio networks & \cite{tsiropoulos2014radio}\\\midrule
				\textbf{Spectrum Access}& Cooperative overlay spectrum access in cognitive radio networks &\cite{liang2017cooperative}\\\midrule
				\textbf{Security Aspects} & Advances on security threats and countermeasures for cognitive radio networks: A survey & \cite{sharma2014advances}\\\midrule
				\multirow{2}{*}{\textbf{Artificial Intelligence}}&A survey on machine-learning techniques in cognitive radios&\cite{bkassiny2012survey}\\
				& Learning and reasoning in cognitive radio networks & \cite{gavrilovska2013learning}\\\midrule
				\multirow{8}{=}{\textbf{Coexistence Analysis in}\newline \textbf{Licensed and Unlicensed bands}}& Licensed spectrum sharing schemes for mobile operators: A survey and outlook & \cite{tehrani2016licensed}\\
				&Dynamic spectrum sharing in 5G wireless networks with full-duplex technology: Recent advances and research challenges&\cite{sharma2017dynamic}\\
				&Survey of spectrum sharing for inter-technology coexistence& \cite{voicu2018survey}\\
				&A spectrum sharing framework for intelligent next generation wireless networks&\cite{de2018spectrum}\\
				&Full spectrum sharing in cognitive radio networks toward 5G: A survey&\cite{hu2018full}\\
				& Coexistence of LTE-LAA and Wi-Fi on 5 GHz with corresponding deployment scenarios: A survey & \cite{chen2016coexistence}\\
				&Coexistence of wireless technologies in the 5 GHz bands: A survey of existing solutions and a roadmap for future research&\cite{naik2018coexistence}\\
				&New radio beam-based access to unlicensed spectrum: Design challenges and solutions&\cite{lagen2019new}\\
				\bottomrule	
		\end{tabular}}
	\end{table*}
	
	A taxonomy of spectrum sharing methods is presented in Fig. \ref{Taxonomy}. In general, spectrum sharing can be classified into two different types: (i) sharing methods in licensed bands that enable the MNOs to access the incumbents' licensed bands; and (ii) sharing in unlicensed bands that enable the MNOs to utilize the unlicensed spectrum \cite{labib2017extending}. Type (i) can be further divided into Dynamic Spectrum Sharing (DSS), Licensed Shared Access (LSA), and Spectrum Access System (SAS). DSS has been envisioned as a seamless global transition towards universal 5G deployment, and it is based on high-level scheduling algorithms between 4G and 5G \cite{xin2021dynamic}. LSA was promoted by the Electronic Communications Committee (ECC) of CEPT, aiming at 2.3–2.4 GHz band. Two principal components of LSA are the LSA Repository (LR), which is a centralized database that contains the incumbent protection information, and the LSA Controller (LC), which obtains the LSA spectrum resource availability information (LSRAI) from the LR. ETSI is currently working on the enhanced version of LSA, named eLSA, which seeks to extend LSA functionality to support spectrum access to local high-quality wireless networks operated by vertical sector operators. Furthermore, an innovative system architecture for the LSA scheme developed out of the EU project ADEL with the purpose of offering more dynamic behavior into the LSA model by introducing various sensing networks and optimization algorithms for better spectrum utilization. Somewhat similar to the LSA, the Federal Communications Commission (FCC) introduced the SAS to enable the shared use of the Citizens Broadband Radio Service (CBRS) in the 3.55–3.7 GHz frequency band, which was initially used by the US Federal Government for Navy  radar systems and commercial satellite operators.

	Type (ii) spectrum sharing methods are split into LTE-Unlicensed (LTE-U), LTE-Licensed Assisted Access (LTE-LAA), MulteFire, and NR-U \cite{qualcommunlicensed, hirzallah20205g}. LTE-U and LTE-LAA rely on Carrier Aggregation (CA) implementation, as they leverage the unlicensed 5 GHz band, traditionally used by Wi-Fi technologies, anchored to the licensed carrier serving as the Primary Cell (PCell) to provide users access to both licensed and unlicensed spectrum and enhance their performance. New features and improvements have been added to LAA in the context of Enhanced LAA (eLAA) and Further Enhanced LAA (feLAA) in the following work items. One feature that distinguishes LAA from LTE-U is the support of Listen Before Talk (LBT) to check the channel's availability before transmitting, which facilitates the fair coexistence of LTE with Wi-Fi. LBT is a contention-based protocol based on the Energy Detection (ED) procedure that determines the activity of the other users in the medium by measuring their propagated energy. Whenever this energy is beyond a certain ED threshold, the medium is considered busy, and the operation suspends. However, LTE-U operates on the basis of a duty-cycling mechanism. Whenever the duty cycle pattern turns from OFF to ON state, LTE-U will access the channel despite the presence of other devices in the band, which is one of the principal drawbacks of this technology. By contrast, MulteFire operates solely in the unlicensed band without a licensed anchor channel requirement, and it aims to promote the standalone operation of LTE in the unlicensed frequency bands \cite{multefire_alliance}. 
	Recently, the 3rd Generation Partnership Project  (3GPP) has introduced a new radio access technology called 5G NR Unlicensed (5G NR-U), extending 5G NR to unlicensed bands. Similar to LAA and MulteFire, NR-U employs the LBT mechanism to promote fair coexistence in the unlicensed band; however, due to the beam-based transmissions in NR, the legacy LBT procedure has to be modified. 
	
	Although there has been a concerted effort from the regulators and standards organizations in defining and developing protocols for each of these sharing methods; there has always been the question of how the MNOs and telecom vendors would take these features into account. Regulatory rules and standards typically do not specify the exact technical requirements, and, in most cases, the telecom vendors might take different approaches to better align with the needs of the practice. In order to bridge the gap between standardization and vendor-specific implementation, it is required that we study the spectrum sharing frameworks both from the standardization and implementation perspective to meaningfully relate theory and practice.

	Besides the aforementioned spectrum sharing methods, another promising technology is Cognitive Radio (CR). Introduced by Mitola in 1999 \cite{mitola1999cognitive}, CR is an intelligent wireless communication system that permits secondary users (SUs), also called cognitive users, to sense the spectrum usage of surrounding primary users (PUs) and whether any unused spectrum (spectral holes) exist. SUs can then reconfigure their software and hardware parameters to utilize the idle spectrum in a way that does not interfere with PUs.  The core idea behind CR is to facilitate dynamic and flexible access to the idle spectrum of PUs \cite{haykin2005cognitive}. One of the critical enablers towards this implication is the spectrum sensing capabilities of SUs, which empower them to continuously monitor the frequency bands used by PUs and flexibly detect spectral holes and use them.
	Consequently, next generation wireless communications not only have to employ the aforementioned spectrum sharing methods but also have to accommodate intelligently collaborative radios to maximize spectral efficiency and avoid any interference with other incumbents to reap the full potential of licensed and unlicensed frequency bands. It is also expected that ML and AI will be used to augment spectrum sharing methods to provide more collaborative and enhanced policies for more complex environments, in which with today's often centralized approaches is not applicable \cite{bkassiny2012survey}. 
	\newpage
	{
		\small
		\renewcommand{\arraystretch}{1}
		\xentrystretch{-0.15}
		\begin{xtabular}{p{1.5cm}p{7cm}}
			\toprule[1.1pt]
			\textbf{Acronym} & \textbf{Definition}  \\
			\midrule
			AAA			 & Authentication, Authorization, and Accounting\\
			AI			 & Artificial Intelligence\\
			B-IFDMA		 & Block-Interleaved Frequency Division Multiple Access\\
			CA           & Carrier Aggregation\\
			CAPEX	     & Capital Expenditures\\
			CBRS		 & Citizens Broadband Radio Service\\
			CBSD         & Citizens Broadband Radio Service Device \\ 
			CCA		 	 & Clear Channel Assessment\\
			CEPT		 & Conference of European Postal \& Telecommunications\\
			CNN          & Convolutional Neural Network            \\ 
			CQI			 & Channel Quality Indicator\\
			CRN			 & Cognitive Radio Network\\
			CRS			 & Cell-specific Reference Signal\\
			CSAT		 & Carrier Sense Adaptive Transmission\\
			CST			 & Concurrent Sensing and Transmission\\ 
			CSI          & Channel State Information\\
			CW			 & Contention Window\\
			DAG-SVM		 & Directed Acyclic Graph Support Vector Machine\\
			DC			 & Dual Connectivity\\
			DMTC		 & DRS Measurement Timing Configuration\\
			DNN          & Deep Neural Network                     \\ 
			DRS			 & Discovery Reference Signal\\
			DSRC		 & Dedicated Short Range Communication\\
			DSS          & Dynamic Spectrum Sharing                \\ 
			ECC			 & Electronic Communications Committee\\
			ED			 & Energy Detection\\
			ELM			 & Extreme Learning Machine\\
			eMBB		 & Enhanced Mobile Broadband\\
			eMBMS		 & Evolved Multimedia Broadcast Multicast Services\\
			ENN          & Encoder Neural Network                  \\
			ESC			 & Environmental Sensing Capability\\
			ETSI         & European Telecommunications Standards Institute\\ 
			EUD		     & End User Devices\\
			FCC			 & Federal Communications Commission\\
			FD			 & Full Duplex\\
			FF-DNN		 & Feed-Forward Deep Neural	Network\\
			FDD			 & Frequency Division Duplex\\
			GAA          & General Authorized Access               \\ 
			GBR          & Guaranteed  Bit  \\ 
			HARQ		 & Hybrid Automatic Repeat Request\\
			IMT			 & International Mobile Telecommunication\\
			IoT			 & Internet of Things\\
			IIoT		 & Industrial IoT\\
			LAA          & Licensed-Assisted Access                \\ 
			LBT          & Listen Before Talk                      \\ 
			LC           & LSA Controller                      \\ 
			LR			 & LSA Repository\\
			LSA          & Licensed-Assisted Access                \\ 
			LSRAI        & LSA Spectrum Resource Availability Information\\
			LSTM         & Long Short Term Memory                  \\ 
			LTE			 & Long Term Evolution\\
			LTE-A		 & LTE-Advanced\\
			LTE-U		 & LTE-Unlicensed\\
			MAC			 & Medium Access Control\\
			MBSFN		 & Multi-Broadcast Single-Frequency Network\\
			MCOT		 & Maximum Channel Occupancy Time\\
			MFCN		 & Mobile/Fixed Communications Networks\\
			MIB			 & Master Information Block\\
			MIMO         & Multi Input  and Multi Output           \\ 
			ML			 & Machine Learning\\
			MME			 & Mobility Management Entity\\
			mMTC		 & Massive Machine Type Communications\\
			mmWave		 & millimeter Wave\\
			MNO          & Mobile Network Operator Output           \\ 
			MF-PBCH		 & MulteFire Physical Broadcast Channel\\
			NBC			 & Naive Bayes Classifier\\
			NHN			 & Neutral Host Network\\
			NRA			 & National Regulatory Authority\\
			NR-U		 & New Radio Unlicensed\\
			OAM 		 & Operation Administration and Management\\
			OFDM		 & Orthogonal Frequency Division Multiplexing\\
			OPEX		 & Operational Expenditure\\
			PAL          & Priority Access Licensees               \\ 
			PBCH         & Physical Broadcast Channel\\
			PCC			 & Primary Component Carrier\\
			PCell		 & Primary Cell\\
			PDCCH		 & Physical Downlink Control Channel\\
			PDSCH  		 & Physical Download Shared Channel\\
			PLMN		 & Public Land Mobile Network\\
			PMSE		 & Program Making and Special Events\\
			PPDR         & Public Protection Disaster Relief\\
			PSS			 & Primary Synchronization Signal\\
			PU			 & Primary User\\
			PUSCH  		 & Physical Uplink Shared Channel\\
			QoS			 & Quality of Service\\
			RAN			 & Radio Access Network\\
			RAT			 & Radio Access Technology\\
			RB			 & Resource Block\\
			RE			 & Resource Element\\
			RL			 & Reinforcement Learning\\
			RNN          & Recurrent Neural Network                \\ 
			RRC			 & Radio Resource Control \\
			RRM			 & Radio Resource management\\
			SAS          & Spectrum Access System                  \\ 
			SCC			 & Secondary Component Carrier\\
			SCell		 & Secondary Cell\\
			SCS			 & Sub-carrier Spacing\\
			SDL		     & Supplemental Downlink\\
			SI			 & Self Interference\\
			SINR         & Signal to Interference-plus-Noise Ratio\\
			SSB			 & Synchronization Signal Blocks\\
			SSS			 & Secondary Synchronization Signal\\
			SU			 & Secondary Users\\
			SVM			 & Support Vector Machine\\
			TDD			 & Time Division Duplex\\ 
			TR           & Technical Report         \\ 
			TS           & Technical Specification         \\ 
			UAV          & Unmanned Aerial Vehicle                \\ 
			UE			 & User Equipment\\
			URLLC		 & Ultra-Reliable Low Latency Communications\\
			VSP          & Vertical Sector Players\\
			WISP         & Wireless Internet Service Providers\\
			1G			 & First Generation\\
			5G			 & Fifth Generation\\
			6G			 & Sixth Generation\\
			3GPP         &3rd Generation Partnership Project\\ 
			\bottomrule[1pt]	
		\end{xtabular}
	}
	\vspace*{+1em}
	
	We should emphasize that spectrum sharing is inherently complex and might not work for all combinations of incumbents and secondary users. As an alternative, incumbents may benefit from newer technologies, which would allow them to deliver the same service by using less spectrum. Furthermore, regulators and stakeholders might consider other options (e.g., relocating incumbents to other bands or confining them to smaller channels within the same band under a new band plan). This procedure can leave extra bandwidth to be auctioned for mobile services. 

	\subsection{Review of Related Survey Articles}
	Spectrum sharing has been broadly investigated in the literature. In this subsection, we provide a concise overview of survey works in domains covered by this paper. In \cite{ali2016advances}, the authors introduced a classification of spectrum sensing methods and discussed the technical challenges of conventional spectrum sensing techniques, while also presenting insights into more novel spectrum sensing systems. The authors of \cite{sharma2016application} analyzed  compressive sensing theory and its use cases in CRNs. In \cite{cichon2016energy},  spectrum sensing was examined on the basis of energy efficiency. \cite{el2016resource} provided an overview of  resource allocation approaches for CRNs and compared various resource allocation algorithms on the basis of strategies, criteria, standard techniques, and network architecture. The authors of \cite{ahmad2015survey} reviewed state-of-the-art resource allocation schemes designed for CR Sensor Networks (CRSNs) and classified these schemes on the basis of their performance optimization criteria. 
	
	In \cite{tsiropoulos2014radio}, the authors surveyed various optimization resource allocation designs in CRNs, such as Signal to Interference-plus-Noise Ratio (SINR)-based, transmission power-based, centralized, and distributed decision-making methods. They reviewed spectrum assignment challenges while focusing on dynamic spectrum allocation and spectrum aggregation.
	\cite{liang2017cooperative} investigated the overlay spectrum access scheme in a cooperative cognitive radio (CCR) network setting. In so doing, they concentrated on two preliminary designs, namely the frequency-division-based channel as well as the time-division-based channel. The authors of \cite{liang2017cooperative} reviewed a  game-based model of the overlay-based CR network. They considered  families of non-cooperative, cooperative, and matching games. In \cite{sharma2014advances}, the authors studied novel approaches to  security perils and countermeasures in CRNs. In particular, the authors focused on the physical layer by classifying challenges and countermeasures by type to protect unlicensed SUs and licensed PUs. \cite{bkassiny2012survey} and \cite{gavrilovska2013learning} discussed the role of ML and AI in CRs and emphasized the importance of autonomous learning for better CRN performance.
	
	The authors in \cite{tehrani2016licensed} provided an in-depth survey of various authorization regimes and their characteristics. Their work investigated existing spectrum sharing scenarios with different network topologies alongside different coordination protocols applied to licensed sharing scenarios. 
	\cite{sharma2017dynamic} presented a comprehensive overview of full-duplex (FD)-enabled dynamic spectrum sharing and recent advances in this domain. The authors of this work also proposed an innovative communication framework to facilitate the concurrent sensing and transmission (CST) in DSS systems by applying a power control-based self-interference (SI) mitigation design. 
	\cite{voicu2018survey} investigated spectrum sharing mechanisms for wireless inter-technology coexistence by taking into account both technical and non-technical parameters. The authors of \cite{de2018spectrum} emphasized the importance of employing more all-embracing intelligence and collaboration to circumvent  interference while enhancing spectrum utilization in the network. They also presented an open-source software define radio (SDR)-based structure that could enhance the inefficient use of radio spectrum. A comprehensive review of full-spectrum sharing in CR 5G networks was presented in \cite{hu2018full}. The authors outlined the prospective enhancements from the perspective of enhancing spectrum and energy efficiency in CRNs. 
	The authors of \cite{chen2016coexistence} provided a comprehensive survey of LTE-LAA and Wi-Fi coexistence in 5 GHz and analyzed various deployment scenarios in detail. \cite{naik2018coexistence} offered a thorough summary of the different coexistence scenarios in the 5 GHz bands. The authors delineated concerns about coexistence among several wireless technologies (i.e., LTE and Wi-Fi, radar and Wi-Fi, dedicated short range communication (DSRC) and Wi-Fi), and coexistence among various 802.11 protocols operating in the 5 GHz bands. 
	NR beam-based access to the Unlicensed spectrum (NR-U) along with its technical specifications were investigated in \cite{lagen2019new}. The authors discussed various NR-U scenarios and LBT procedures while describing their critical challenges by considering the regulatory necessities and the narrow-beam transmission influence.	
	
	\begin{figure}[!t]
		\centering
		\includegraphics[width=.498\textwidth]{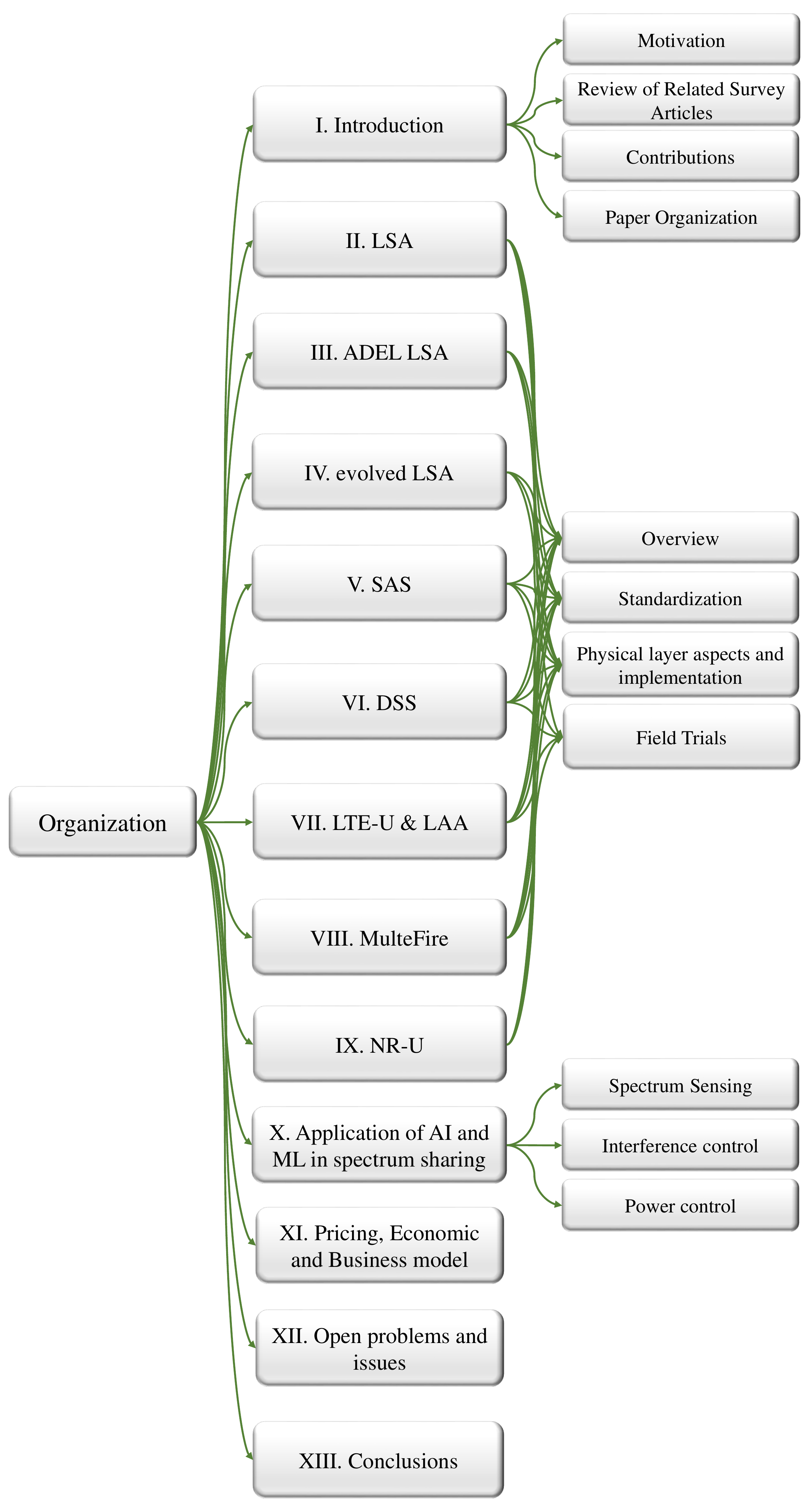}
		\caption{Structure of the paper.} 
		\label{structure_paper}
	\end{figure}
	
	\begin{figure*}[!t] 
		\centering
		\includegraphics[width=1\textwidth]{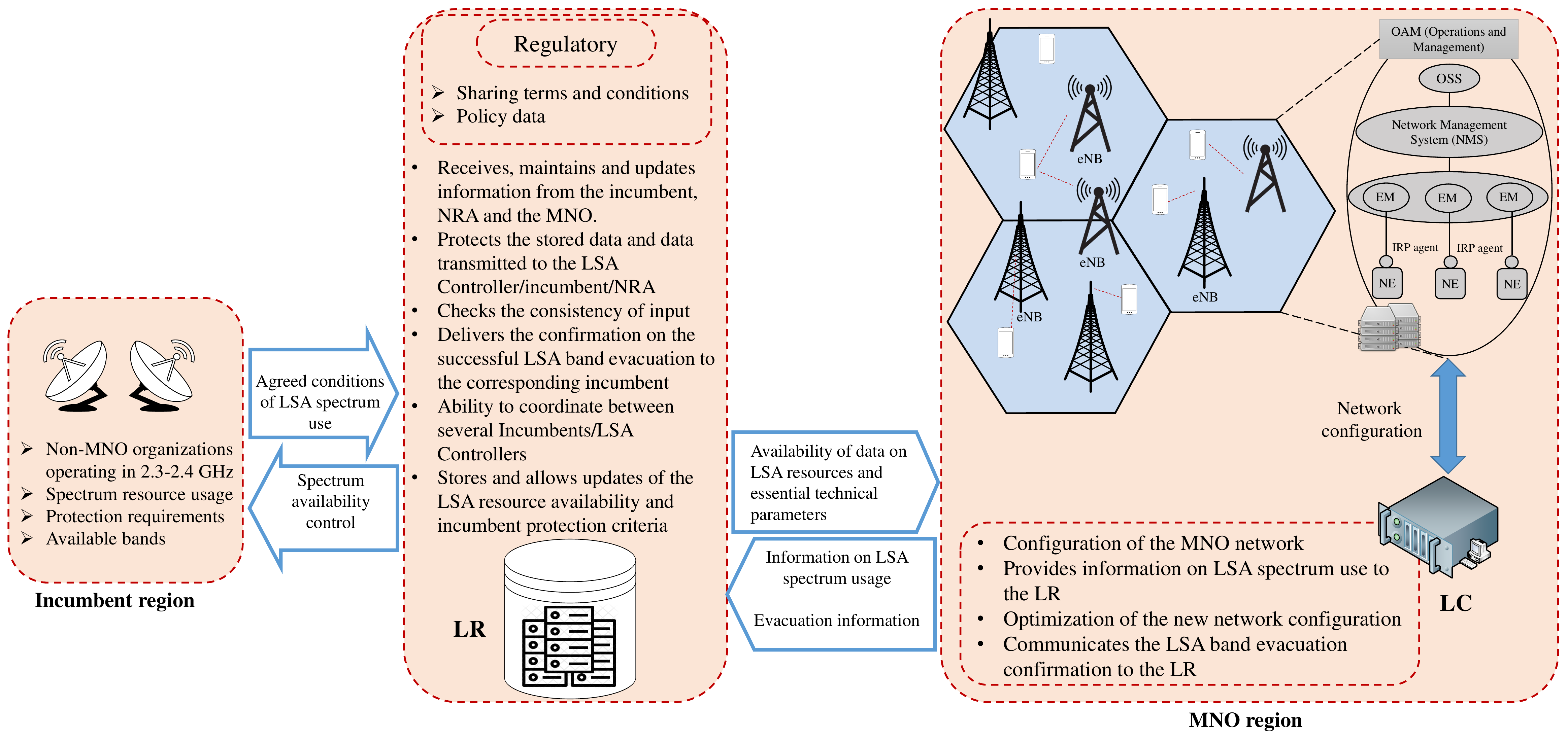}
		\caption{Technical overview of the LSA architecture\cite{ETSI_TS_103_235}.} 
		\label{LSA_archtecture}
	\end{figure*}
	\subsection{Contributions}
	Although there is an extensive body of literature on the applications of CRNs and spectrum sharing methods in wireless communications, few have focused on advances in sharing methods (e.g., DSS, and MulteFire). Except for DSS and NR-U, most sharing schemes reviewed in this paper are 4G-related; however, since these form the basis of the prospective sharing methods that are envisioned for next-generation wireless networks (i.e., DSS and NR-U), their thorough investigation is warranted. Furthermore, the current literature does not provide a clear picture of the implementation perspectives of the various spectrum sharing methods nor the necessary modifications to be made in the radio protocols of legacy cellular networks to better coexist and operate with the aforementioned sharing schemes. In addition, in this survey paper, we investigate the role of of AI and ML algorithms in different stages of a sharing process and how they can help towards an autonomous sharing paradigm. The primary goal of this survey is to provide forward-looking insight into practically viable spectrum sharing plans both in licensed and unlicensed bands that will offer MNOs better QoS and lower Capital (CAPEX) and Operational Expenditures (OPEX). 
	
	To summarize, this article has the following aims:
	\begin{itemize}
		\item To provide a comprehensive taxonomy of the current spectrum sharing methods both in licensed and unlicensed bands as well as their specifications and requirements.
		\item To offer a comprehensive review of recent survey articles and a thorough survey of existing specifications that various standards organizations and regulatory bodies have provided for each sharing scheme.
		\item To investigate AI and ML capabilities for leveraging the full potential of autonomous sharing scenarios. In essence, recent state-of-the-art literature concerning the AI impact on the performance of different spectrum sharing methods and spectrum sensing process in CR has been surveyed, each of which contributes significantly to a cooperative sharing procedure.
		\item To identify potential implementation scenarios and required amendments to the legacy cellular networks, both in radio protocols and the architecture, aiming at narrowing the gap between the theory and practice. These data are essentially drawn from the primary regulatory and standardization bodies documents (i.e., FCC, ETSI, 3GPP, MulteFire Alliance, etc.) and telecommunication bodies' white papers (Qualcomm, Nokia, Ericsson, etc.).
		\item To study the business and regulatory aspects of spectrum sharing, including pricing, economic, and business models.
		\item To identify existing challenges in deploying the aforementioned sharing schemes in licensed and unlicensed bands, and to recommend several research directions for additional improvements.
	\end{itemize}
	\begin{table*}[!t]
		\centering
		\caption{Standardization works in the field of LSA.}
		\begin{tabular}{p{0.75cm} p{2cm} p{14cm}}
			\toprule
			\textbf{Ref.} & \textbf{Standard} & \textbf{Subject}  \\
			\midrule
			\cite{RSPG_LSA_1}      & RSPG 11-392               &  Report on Collective Use of Spectrum (CUS) and other spectrum sharing approaches \\\midrule
			\cite{RSPG_LSA_2}      & RSPG 12-424               &  Request for Opinion on Licensed Shared Access (LSA) \\\midrule
			\cite{RSPG_LSA_3}      & RSPG 13-538               &  RSPG Opinion on Licensed Shared Access \\\midrule
			\cite{CEPT_LSA_203}    & ECC 203                   &  Least Restrictive Technical Conditions suitable for Mobile/Fixed Communication Networks (MFCN),
			including IMT, in the frequency bands 3400--3600 MHz and 3600--3800 MHz \\\midrule
			\cite{CEPT_LSA}        & ECC 205                   &  Licensed Shared Access (LSA) \\\midrule
			\cite{CEPT_LSA_1402}   & ECC 14 (02)               &  Harmonised technical and regulatory conditions for the
			use of the band 2300--2400 MHz for Mobile/Fixed
			Communications Networks (MFCN) \\\midrule
			\cite{CEPT_LSA_1404}   & ECC 14 (04)               &  Cross-border coordination for mobile/fixed
			communications networks (MFCN) and between MFCN
			and other systems in the frequency band 2300--2400 MHz \\\midrule
			\cite{ITU_R_1} & ITU-R M 2330  & Cognitive radio systems in the land mobile service \\\midrule
			\cite{ITU_R_2} & ITU-R M 2320  & Future technology trends of terrestrial IMT systems \\\midrule
			\cite{ITU_R_3} & ITU-R SM 2404 & Regulatory tools to support enhanced shared use of the spectrum \\\midrule
			\cite{ETSI_TR_103_113} & ETSI TR 103 113   & Electromagnetic compatibility and Radio spectrum Matters (ERM);
			System Reference document (SRdoc); Mobile broadband services in
			the 2300 MHz--2400 MHz frequency band under Licensed Shared Access regime  \\\midrule
			\cite{ETSI_TS_103_154} & ETSI TR 103 154   & Reconfigurable Radio Systems (RRS); System requirements for operation of Mobile Broadband
			Systems in the 2300 MHz - 2400 MHz band under Licensed Shared Access (LSA)  \\\midrule 
			\cite{ETSI_TS_103_235} & ETSI TR 103 235    & Reconfigurable Radio Systems (RRS); System architecture and high level procedures
			for operation of Licensed Shared Access (LSA) in the 2300 MHz--2400 MHz band  \\\midrule
			\cite{ETSI_TS_103_379} & ETSI TR 103 379   & Reconfigurable Radio Systems (RRS); Information elements and protocols for the interface
			between LSA Controller (LC) and LSA Repository (LR) for operation of Licensed Shared Access (LSA)
			in the 2300 MHz--2400 MHz band  \\\midrule
			\cite{3gpp_32_855}     & 3GPP TR 32.855   & Study on OAM support for Licensed Shared Access (LSA) \\\midrule
			\cite{3gpp_28_301}     & 3GPP TS 28.301   & Licensed Shared Access (LSA) Controller (LC) Integration Reference Point (IRP);
			Requirements  \\\midrule
			\cite{3gpp_28_302}     & 3GPP TS 28.302    & Licensed Shared Access (LSA) Controller (LC) Integration Reference Point (IRP);
			Information Service (IS)  \\\midrule
			\cite{3gpp_28_303}     & 3GPP TS 28.303   & Licensed Shared Access (LSA) Controller (LC) Integration Reference Point (IRP);
			Solution Set (SS) definitions  \\
			\bottomrule
		\end{tabular}
		\label{LSA_table}
	\end{table*}
	\subsection{Paper Organization}
	The remainder of this paper is structured as follows:
	Sections \ref{LSA} to \ref{nr_u} describe existing spectrum sharing methods, from LSA, SAS, and DSS in the licensed bands to LTE-U, LAA, MulteFire, and NR-U in the unlicensed bands, respectively. For each spectrum sharing method, we first provide a general overview of the ecosystem as well as the key players and their functionalities. Then, we provide a comprehensive review of the standardization and regulation activities as well as the implementation perspectives and physical layer aspects for each sharing scheme. Section \ref{app_AI} highlights the key enabling features that AI and ML can contribute to the various stages of a sharing process, including sensing, pricing, and interference control. Next, Section \ref{price} highlights pricing considerations as well as the economic and business model that must be taken into account during the sharing process. Finally, Section \ref{problem} addresses open research issues, and Section \ref{Conclusion} concludes the paper. An overview of the structure of the paper is provided in Fig. \ref{structure_paper}, and the definitions of acronyms are given on page 4.	
	\section{Licensed Shared Access}\label{LSA}
	\subsection{Overview}
	LSA is considered a controlled spectrum sharing procedure in which 
	incumbent users, maintaining a dedicated right of access to their spectrum, permit a limited number of LSA Licensees (MNOs) to share their spectrum resources based on predefined conditions. According to \cite{RSPG_LSA_3,CEPT_LSA}, the LSA framework is designed to meet the needs of the following stakeholders:
	\begin{itemize}
		\item Incumbent user(s): these are the primary users of the spectrum and have exclusive rights of access for their band; under certain conditions they can make their spectrum available to LSA licensee(s).
		\item LSA licensee(s): these are users who are allowed to use frequency bands licensed to the original incumbents under pre-arranged contracts.
		\item National regulation administration (NRA): this stakeholder manages negotiations and spectrum sharing activities between incumbents and mobile operators.
	\end{itemize}

	The technical overview of this sharing method is shown in Fig. \ref{LSA_archtecture}. The main functionality of the LSA system is accomplished through two central units on the existing mobile network, the LR and the LC.
	The LR is a database that stores and updates the information required for spectrum sharing. It contains the information about the spectrum usage of the incumbent user and protection requirements to balance operations between the incumbent user and the LSA licensee. It can notify the approved  LCs about the availability of the spectrum granted by the incumbent users. Furthermore, the LR provides an interface for the NRA to monitor the spectrum sharing process. The LR can be operated by a regulatory body or a third party organization \cite{ETSI_TS_103_154}.
	
	The LC resides within the LSA licensee's domain. It obtains the LSRAI from the LR and then conveys this information to the associated LSA licensee. The principal function of the LC is to ensure optimal and interference-free performance for both the MNO and the incumbent user based on the available information in the LR. This information is transmitted to the LC through a secure and reliable transmission path \cite{ETSI_TS_103_154}.
	
\begin{figure*}[!t] 
		\centering
		\includegraphics[width=1\textwidth]{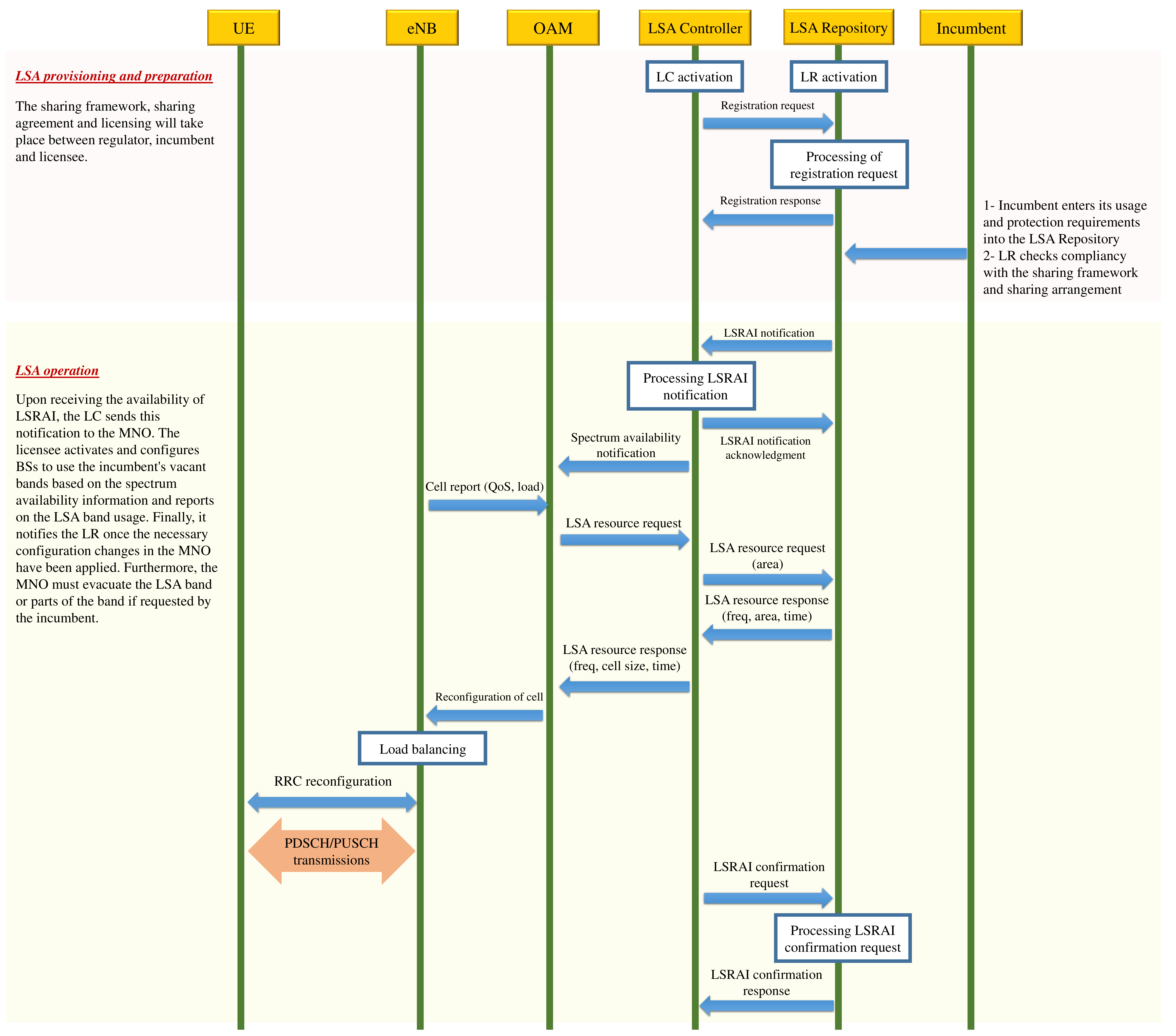}
		\caption{LSA message flow.} 
		\label{LSA_flow}
	\end{figure*}

	As we can see in Fig. \ref{LSA_archtecture}, incumbents make their spectrum consumption information available to the LR. The LC  communicates with the LR to obtain the vacant bands and then transmits this information to the MNO's operations, administration, and maintenance (OAM). Next, the MNO configures its base stations (BSs) on the released bands\cite{Cellular_LSA}. An illustration of the message flow and interactions between different parts of the LSA framework is shown in Fig \ref{LSA_flow}. 
	\subsection{Standardization}
	The regulation and standardization path for the LSA scheme is depicted in Fig. \ref{standardtimeline}. The relevant standards and their corresponding subject of discussion are listed in Table \ref{LSA_table}. In the following, we will elaborate on the technical details of these standards.

	The merits of the LSA method were first identified by the Radio Spectrum Policy Group (RSPG). In \cite{RSPG_LSA_1}, the authors began by analyzing the Collective Use of Spectrum approaches to investigate their potential in establishing a regulatory framework for the shared use of spectrum. The report then shifted its focus and examined the challenges of innovative sharing arrangements, especially those related to cognitive technologies. In this way forward, the RSPG proposed a new notion called “Licensed Shared Access.” Finally, a set of suggestions were offered to the European Commission to promote shared access to spectrum in Europe. On September 3, 2012, the European Commission expressed motivation to support  LSA as one form of sharing and stimulated further development on this front\cite{RSPG_LSA_2}. In \cite{RSPG_LSA_3}, RSPG responded to the European Commission’s request for an opinion on spectrum regulatory and economic perspectives of LSA and agreed on the following amended description for LSA:
	
	\textit{\textbf{``A regulatory approach aiming to facilitate the introduction of radiocommunication systems operated by a limited number of licensees under an individual licensing regime in a frequency band already assigned or expected to be assigned to one or more incumbent users. Under the Licensed Shared Access (LSA) approach, the additional users are authorized to use the spectrum (or part of the spectrum) in accordance with sharing rules included in their rights of use of spectrum, thereby allowing all the authorized users, including incumbents, to provide a certain Quality of Service (QoS)."}}
	
	Following the RSPG's definition for LSA \cite{RSPG_LSA_3}, Working Group Frequency Management (WGFM) entrusted Project Team FM53 \cite{CEPT_LSA} with developing nonexclusive guidelines to CEPT administrations in accordance with the implementation of LSA, and Project Team FM52 \cite{CEPT_LSA_1402,CEPT_LSA_1404} to extend ECC decision on the harmonized band plan for Mobile/Fixed Communication Networks (MFCN) in 2300--2400 MHz and also organized the regulatory prerequisites for LSA implementation. The following conditions were defined by these documents.
	\begin{itemize}
		\item Frequency arrangement for MFCN using LSA in the 2300–2400 MHz band is based on 20 blocks of 5 MHz in time-division duplex (TDD) mode.
		\item The least restrictive technical condition (LRTC) is defined in the form of a block-edge mask (BEM), which is related to the interference restriction between the users of a spectrum, and applies to MFCN in the 2300--2400 MHz band \cite{CEPT_LSA_203}.
		\item For cross-border coordination between MFCN and other types of applications, the regulators of the countries should agree on solutions to coordinate the operation of those systems.
	\end{itemize}
	\begin{figure}[!b]
		\centering
		\includegraphics[width=0.5\linewidth]{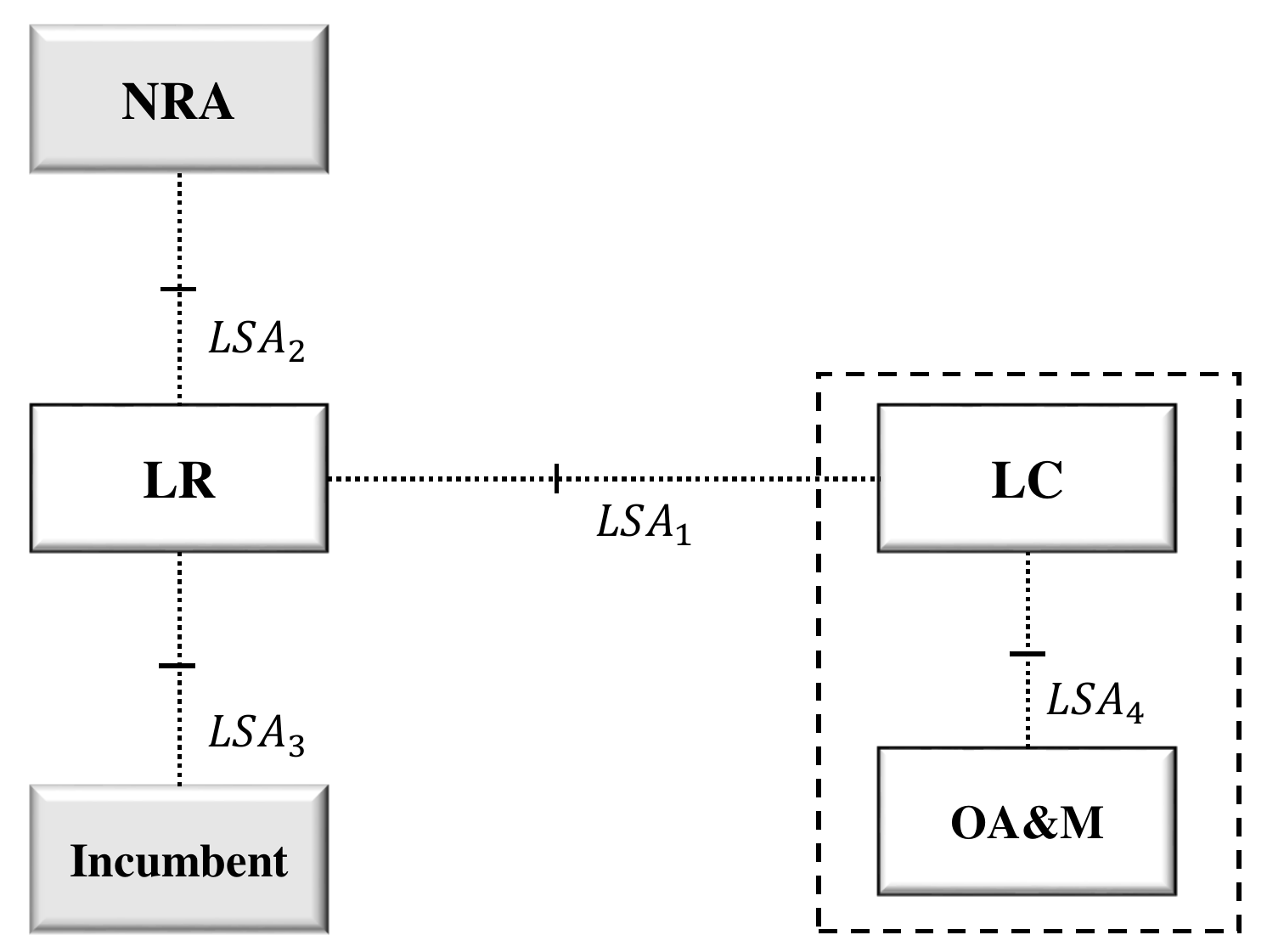}
		\caption{LSA architecture reference model \cite{ETSI_TS_103_235}.}
		\label{lsareferencemodel}
	\end{figure}

	Although LSA is a national issue, global spectrum harmonization is the essential preliminary step for any radio innovation to proliferate and succeed. In this regard, 
	the International Telecommunication Union Radiocommunication Sector (ITU-R) plays a  crucial role in defining the technical parameters to facilitate widespread LSA implementation. The ITU-R considers LSA to be a possible cognitive radio solution for vertical sharing and addresses possible regulatory solutions for the shared use of spectrum in \cite{ITU_R_1,ITU_R_2,ITU_R_3}.

	Meanwhile, the European Telecommunications Standards Institute (ETSI) in cooperation with the ECC section of CEPT, and the standardization Mandate from EC \cite{EC_mandate}, drafted the reports \cite{ETSI_TR_103_113,ETSI_TS_103_154,ETSI_TS_103_235,ETSI_TS_103_379}, which were intended to provide system reference information on the operating features of MFCN services under the LSA regime. ETSI's initial report on this was \cite{ETSI_TR_103_113}. This report presented a summary of the LSA concept and introduced its operational features, performance requirements, high-level architecture, and related functions which empower the LSA concept.

	\begin{figure}[!b] 
		\centering
		\includegraphics[width=.35\textwidth]{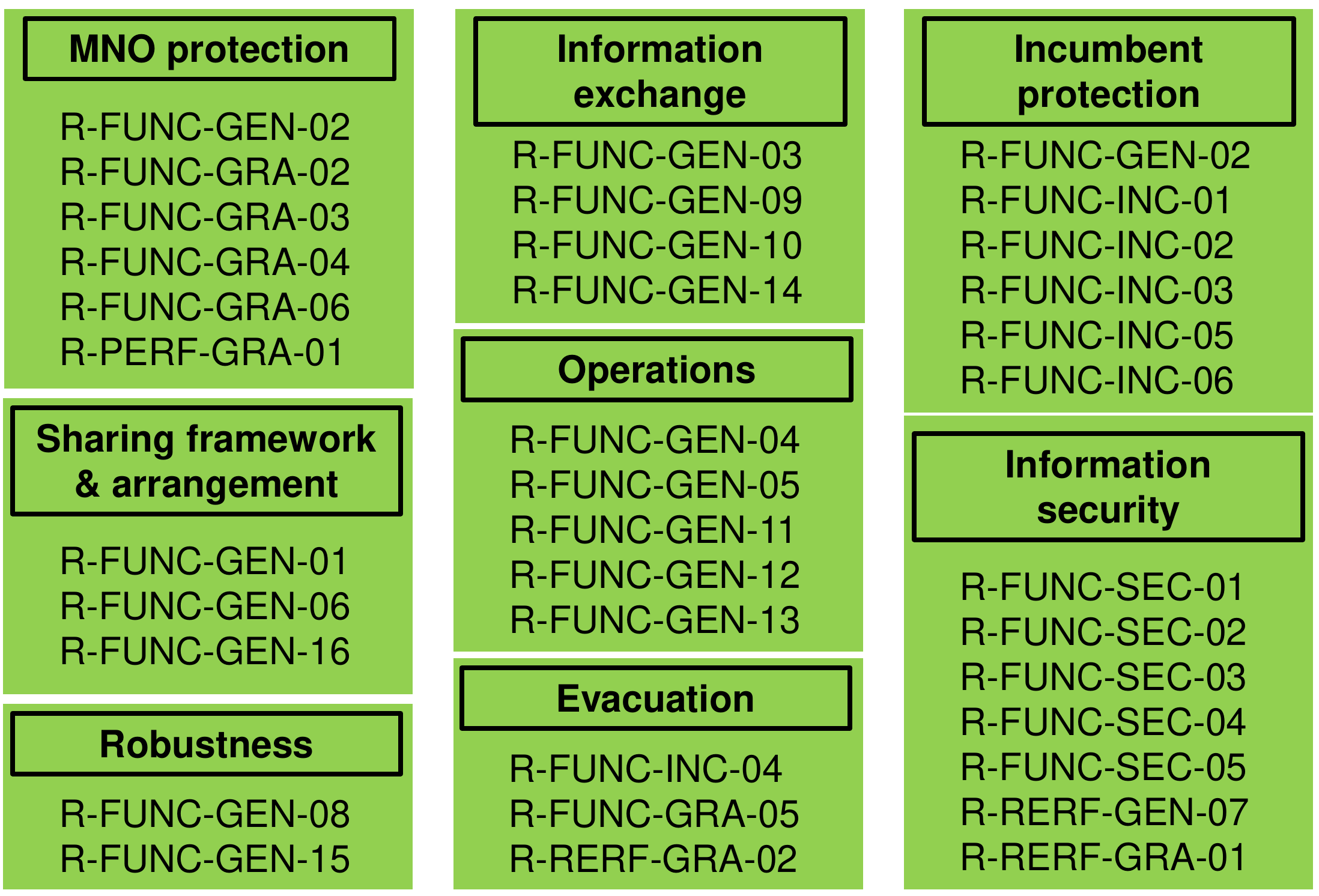}
		\caption{Grouping of requirements for the LSA system from ETSI RRS to support implementation, \cite{mustonen2015analysis}.}
		\label{requirements}
	\end{figure}
	
	The second report of ETSI RRS \cite{ETSI_TS_103_154} introduced a total of 36 system requirements needed for mobile broadband access to the 2300--2400 MHz band under  LSA. These requirements were further divided into general system operation requirements (GEN), incumbent protection requirements (INC), resource grant requirement (GRA), security requirements (SEC), and performance requirements (PERF). As  a complement to the work in \cite{ETSI_TS_103_154}, the authors in \cite{mustonen2015analysis} translated the so-called system requirements into the LSA system architecture building blocks. An illustration of these requirements is given in Fig. \ref{requirements}. Then, in \cite{ETSI_TS_103_235}, ETSI defined the system architecture for the operation of MFCN services under LSA. Moreover, this document included a thorough description of the logical elements, reference points, functions supported by the architecture, high-level procedures, and various deployment architecture examples. Besides, in \cite{ETSI_TS_103_379}, drawing on the system requirements, system architecture, and high-level procedures defined in preceding works, ETSI further discussed the application protocol for the LSA1 interface. This protocol is defined between the LC and LR (LSA1 protocol) and conveys the LSRAI. Although ETSI's standardization was precise; nevertheless, the organizations supporting LSA standardization were somehow optimistic that the LSA standard would also be a facilitator for the deployment of LSA in the 2300–2400 MHz band using LTE technology. However, even though the ETSI considers the LC within the MNO domain, no genuine standardization of the interface between the LC and LSA licensee was provided; this had to be done by the 3GPP \cite{3gpp_32_855,3gpp_28_301,3gpp_28_302,3gpp_28_303}.
	
	\subsection{Physical Layer Aspects and Implementation}
	\begin{table*}[!t]
		\centering
		\caption{LSA field trials and projects.}
		\scriptsize
		\begin{tabular}{p{0.6cm} p{2cm} p{1cm} p{2.5cm} p{3cm} p{2cm} p{4cm}}
			\toprule
			\multirow{2}{*}{\textbf{Ref.}} & \multirow{2}{=}{\textbf{Project name}} & \multirow{2}{=}{\textbf{Country}} & \textbf{Industry companies \newline \& vendors} & \multirow{2}{=}{\textbf{Governmental organizations}} & \textbf{Participating incumbents} & \textbf{Project outcomes} \\
			\midrule
			\multirow{8}{*}{\cite{coreplus}} & \multirow{8}{*}{CORE+}  & \multirow{8}{*}{Finland} & \multirow{8}{=}{1. Nokia Siemens Networks \newline 2. EXFO \newline 3. Elektrobit \newline 4.PehuTec \newline 5.Rugged Tooling} & \multirow{8}{=}{1. The Finnish Defense Forces \newline
				2. Finnish Communications Regulatory Authority (FICORA)
			}	& \multirow{8}{=}{PMSE (cordless cameras)} & 
			
			1. The new LSA concept can be implemented with existing network elements and a minimum amount of new components. \newline
			2. LSA band evacuation times indicate that the LSA band can be evacuated and released in good time and the incumbents’ rights can be protected.
			\\\midrule
			\multirow{3}{*}{\cite{comora}} & \multirow{3}{*}{CoMoRa} & \multirow{3}{*}{Germany} & \multirow{3}{=}{1. Nokia \newline 2.Fraunhofer heinrich hertz institute} & \multirow{3}{=}{BNetzA} & \multirow{3}{=}{Commercial systems} &
			
			1. The LSA concept can be combined with the carrier aggregation \newline
			2. Increased data rates up to 450 Mbps
			\\\midrule
			\multirow{5}{*}{\cite{anacom}}& \multirow{5}{=}{Study on the LSA spectrum sharing model}&\multirow{5}{*}{Portugal}& \multirow{5}{=}{1.Nokia\newline 2.Huawei\newline 3.Fairspectrum} &\multirow{5}{=}{Autoridade Nacional de Comunicações (AACOM)}&\multirow{5}{=}{TV operators:\newline 1. RTP\newline 2.SIC \newline 3.TVI}& 
			1. Compatibility test between PMSE and LTE signals was approved \newline
			2. Efficient monitoring of the related spectrum by associating the incumbent users to the LSA warner system
			\\\midrule
			\multirow{4}{=}{\cite{francelsa}}&  \multirow{4}{=}{Towards more \newline dynamic spectrum \newline sharing: LSA} & \multirow{4}{=}{France} & 1.RED technologies \newline 2. Ericsson \newline 3.Qualcomm \newline 4. Alcatel-Lucent & \multirow{4}{=}{L'Agence nationale des \newline fréquences (ANFR)}&\multirow{4}{=}{Ministry of Defence for remote aeronautical measurements}&
			
			1. More dense spectrum usage and increased capacity to mobile high speed is envisaged through LSA\newline
			2. Increase of MNO data rate by 10\%
			\\\midrule
			\multirow{6}{=}{\cite{italylsa}} & \multirow{6}{=}{LSA pilot} & \multirow{6}{=}{Italy} & \multirow{6}{=}{1. Nokia \newline 2. Cumocore \newline 3. Athonet \newline 4. RED technologies  \newline 5.PosteMobile} & \multirow{6}{=}{Fondazione Ugo Bordoni \newline(FUB)} & \multirow{6}{=}{1. Fixed service \newline 2. PMSE} & 1. Coexistence of LTE systems operating under LSA with has been demonstrated as feasible \newline 2. Based on the results, the evacuation time is always below 40 seconds. \newline 3. Increase of MNO data rates up to 30\% \\\bottomrule
		\end{tabular}
		\label{LSA_field_trial_table}
	\end{table*}
	
	The LSA's architecture reference model is depicted in Fig. \ref{lsareferencemodel}, \cite{ETSI_TS_103_235}. There are four major reference points:
	\begin{enumerate}
		\item $ LSA_1 $:  the reference point between the LC and LR; it  supports the communication mechanisms to exchange the LSA resource availability information and the corresponding acknowledgment reports.
		\item $ LSA_2 $:  the reference point between the administration/NRA and the LR. 
		\item $ LSA_3 $:  the reference point that enables the incumbent to interact with an LR.
		\item $ LSA_4 $:  the reference point between the operator's Network Management System (NMS) and the LC.
	\end{enumerate}
	\subsubsection{$ LSA_1 $ Interface}	
	In \cite{ETSI_TS_103_235}, multiple high-level procedures were introduced, which define the interactions between the LC and LR and must be supported by this interface. It is anticipated that these procedures and their implementation along with the required latency and response times will comply with the sharing framework and arrangement. Registration and deregistration, LSRAI request, notification, and confirmation, connectivity check requests, and notification are some of the high-level procedures described in \cite{ETSI_TS_103_235}.
	\begin{table*}[!t]
		\centering
		\caption{ADEL standardization works in the field of LSA.}
		\begin{tabular}{p{1cm} p{2cm} p{10cm}}
			\toprule
			\textbf{Ref.} & \textbf{Standard} & \textbf{Subject}  \\
			\midrule
			\cite{ADEL4_2}      & ADEL D4.2                &  Resource Allocation Schemes and Optimized Algorithms \\\midrule
			\cite{ADEL4_3}      & ADEL D4.3                &  Spectrum Sharing Policy Reinforcement \\\midrule
			\cite{ADEL5_3}      & ADEL D5.3                &  Cooperative Communication Techniques for LSA \\\midrule
			\cite{ADEL5_4}      & ADEL D5.4                &  Medium Access and Control Plane Protocols \\\midrule
			\cite{ADEL6_2}      & ADEL D6.2                &  System-Level Simulator: Overall Functionalities and Consolidated Results \\\midrule
			\cite{ADEL6_3}      & ADEL D6.3                &  Testbed Trials \\\midrule
			\cite{ADEL1_5_2}    & ADEL D1.5.2              &  PROJECT FINAL REPORT, Public Summary\\
			\bottomrule
		\end{tabular}
		\label{eLSA_table}
	\end{table*}
	\begin{figure}[!b]
		\centering
		\includegraphics[width=0.5\linewidth]{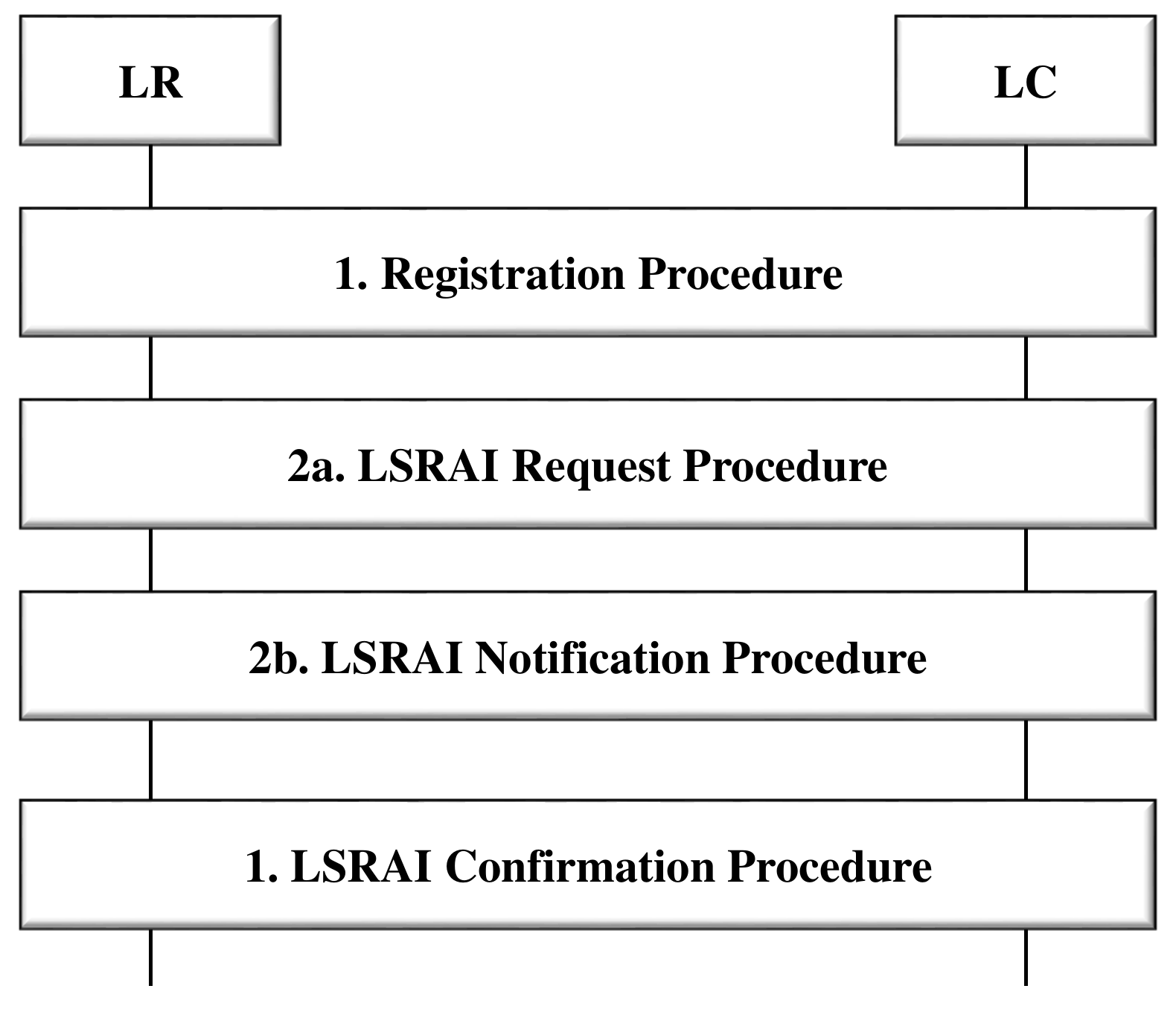}
		\caption{LC operation start-up procedure flow \cite{ETSI_TS_103_235}.}
		\label{lcoperation}
	\end{figure}
	ETSI has defined various procedure flows, each of which may be composed of several procedures, defined earlier. For instance, the LC operation start-up procedure is shown in Fig. \ref{lcoperation}. This message flow is to be employed during the LC start-up and incorporates the Registration procedure, depending on the LR decision it can be followed by LSRAI request or notification procedure. On completion of steps 2a or 2b, the LC has received the necessary LSRAI. Upon successful configuration of the LSA spectrum resources in the MFCN, the LC proceeds with the LSRAI confirmation procedure. A complete list of these procedure flows can be found in \cite{ETSI_TS_103_235}.
	\subsubsection{$ LSA_2 $ Interface}	
	This interface has to support the potential changes in the sharing rules and arrangement (clause 6.3.4 of \cite{ETSI_TS_103_154}), protection of the licensee's operations (6.3.3 of \cite{ETSI_TS_103_154}), and the security aspects and requirements (section 6.4 of \cite{ETSI_TS_103_154}). 
	
	\subsubsection{$ LSA_3 $ Interface}	
	Through this interface, the incumbent can enter the spectrum usage and protection requirements. This also allows the LSA spectrum availability information to be modified as needed. (6.2.1 -- 6.2.6 of \cite{ETSI_TS_103_154}). Whenever the incumbent wants to enter the LSRAI or protection information, they simply specify the exact time at which the change will occur, the authentication indication, the required protection aspects, and the tolerable time (clauses 6.3.6 and 7.2.1 of \cite{ETSI_TS_103_154}). This interface has to also support the interactions of multiple incumbents with the LR (clause 6.1.5 of \cite{ETSI_TS_103_154}), changes in the sharing arrangement (clauses 6.1.16 and 6.3.4 of \cite{ETSI_TS_103_154}), secure information exchange with the incumbent (clauses 6.4.1 -- 6.4.5 of \cite{ETSI_TS_103_154}), and the failure indication to the incumbent. 
	
	\subsubsection{$ LSA_4 $ Interface}	
	This interface is between the NM and LC and is a Type 7 interface \cite{3gpp_28_301}.
	Since the LC is implemented within the operator's domain, it is also called an MNO internal interface. The information provided by the LC to the OA\&M has to accommodate the eNB controlling information whenever a new spectrum band is reported to be idle. This interface  must also include the MNO's resource evacuation messages and other MNO-related network information. In \cite{3gpp_28_301}, two different deployment scenarios were proposed. In scenario 1, the LC acted as a relay for the LSRAI and only forwarded it to the NM. In scenario 2, the LSRAI is not directly transmitted to the NM. Instead, the LC calculated the radio configuration constraints on the basis of the radio planning parameters received from the NM and then forwarded the constraints to the NM.
	
	\subsection{LSA Field Trials}
	LSA has been studied in several research and trial projects. A summary of these projects has been gathered in Table \ref{LSA_field_trial_table}.
	\color{black}
	\section{ADEL Improved LSA}\label{ADEL_LSA}
	
	\subsection{Overview}
	\begin{figure}[!b] 
		\centering
		\includegraphics[width=.5\textwidth]{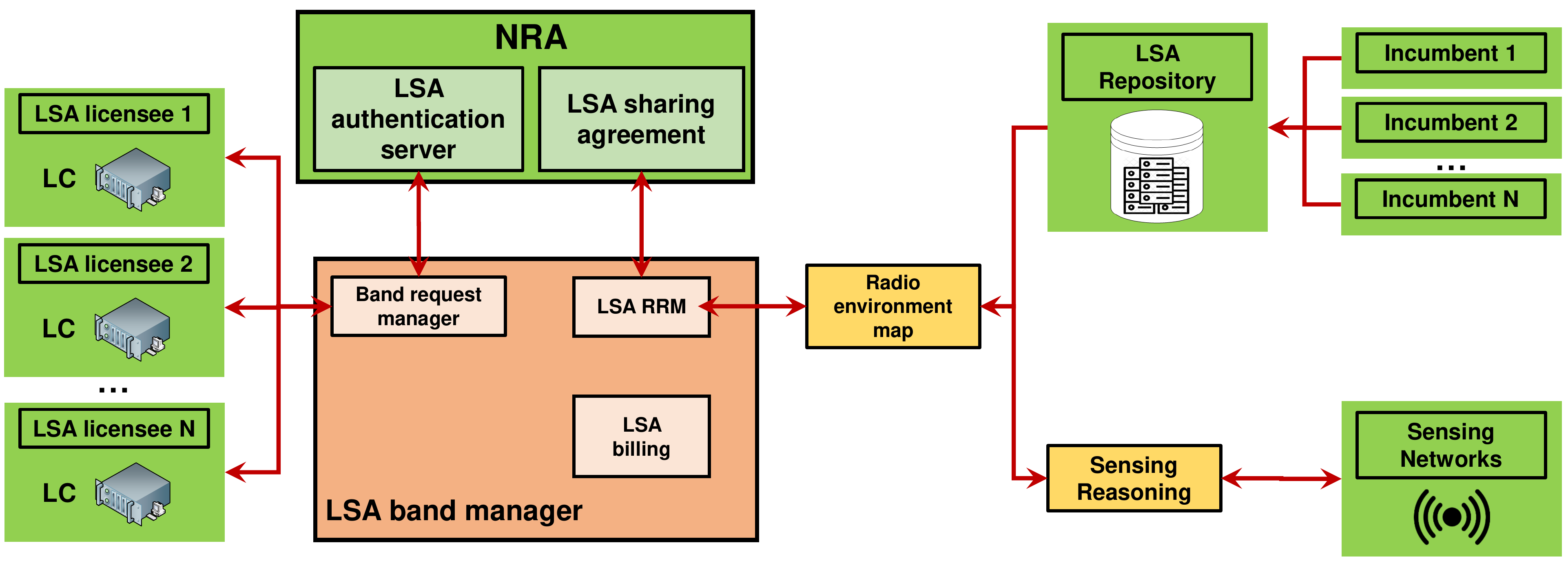}
		\caption{Proposed LSA architecture by ADEL\cite{ADEL5_4}.} 
		\label{ADEL_LSA_fram}
	\end{figure}
	Although the proposed LSA framework is able to facilitate spectrum sharing between the MNOs and incumbents in the licensed bands  successfully, it nevertheless lacks certain qualifications in terms of spectral efficiency. This rigid behavior is due to interference protection for stakeholders in LSA being handled statically on the basis of predefined situations and methods. 
	This confinement to such an inflexible method diminishes the performance of spectrum sharing. 
	
\begin{figure}[!b] 
		\centering
		\includegraphics[width=0.5\textwidth]{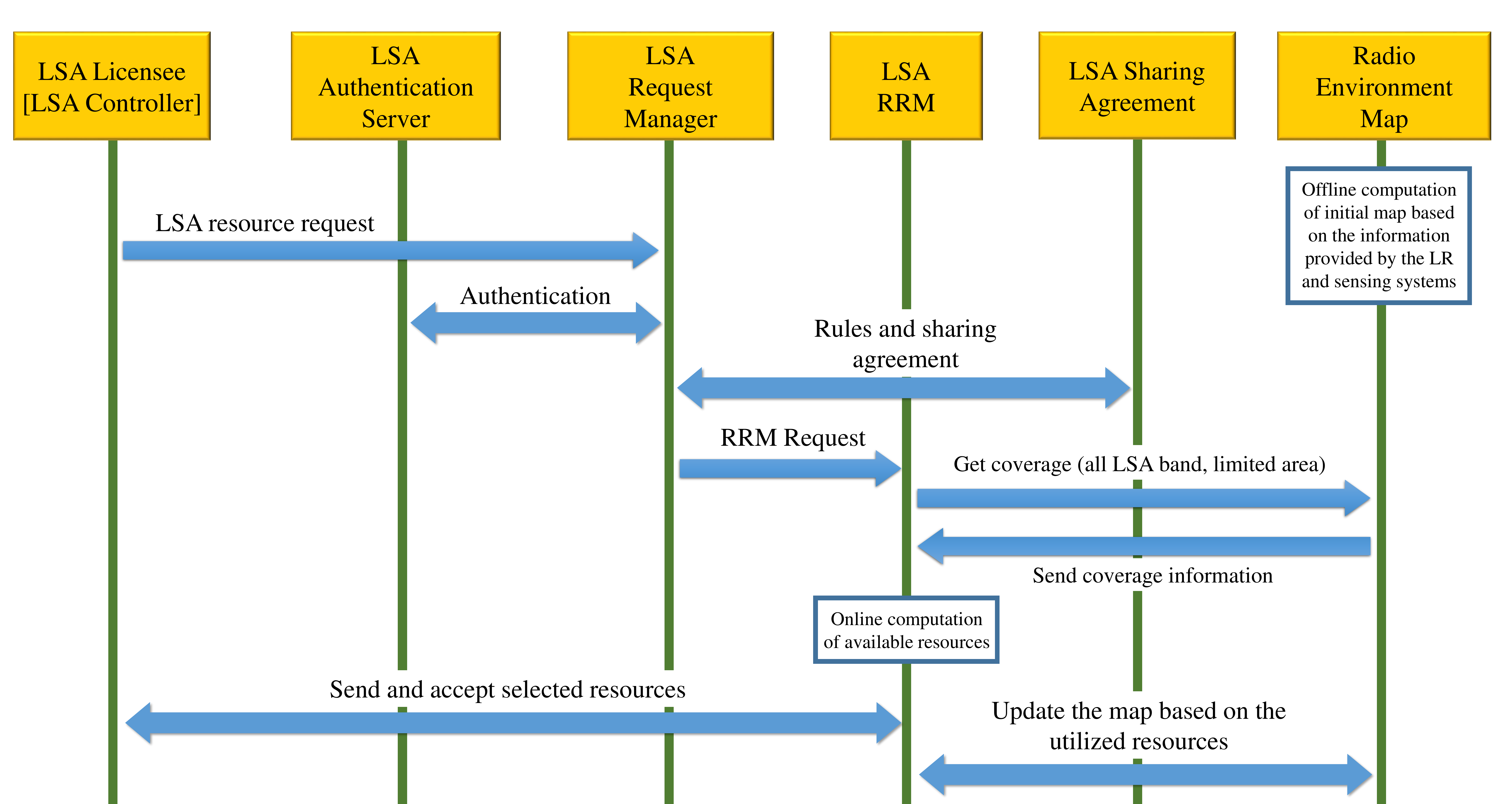}
		\caption{ADEL message flow.} 
		\label{adel_flow}
	\end{figure}

	\begin{table*}[!t]
		\centering
		\caption{ETSI standardization works in the field of eLSA.}
		\begin{tabular}{p{0.75cm} p{2cm} p{14cm}}
			\toprule[1pt]
			\textbf{Ref.} & \textbf{Standard} & \textbf{Subject}  \\
			\midrule
			\cite{elsa_1}      & ETSI TR 103 588               &  Reconfigurable Radio Systems (RRS); Feasibility study on temporary spectrum access for local high-quality wireless networks\\\midrule
			\cite{elsa_2}      & ETSI TS 103 652-1               &  Reconfigurable Radio Systems (RRS); evolved Licensed Shared Access (eLSA); Part 1: System requirements\\\midrule
			\cite{elsa_3}      & ETSI TS 103 652-2               & Reconfigurable Radio Systems (RRS); evolved Licensed Shared Access (eLSA); Part 2: System architecture and high-level procedures \\\midrule
			\cite{elsa_4}      & ETSI TS 103 652-3               & Reconfigurable Radio Systems (RRS); evolved Licensed Shared Access (eLSA); Part 3: Information elements and protocols for the interface between eLSA Controller (eLC) and eLSA Repository (eLR) \\
			\bottomrule
		\end{tabular}
		\label{eLSA_table_etsi}
	\end{table*}
	To cope with these problems, the European Research Project ADEL \cite{ADEL1_5_2}, investigated more dynamic sharing situations for LSA in which the number of active incumbents, licensees, and their respective frequencies would not affect the foreseen QoS levels envisioned by the regulation. Moreover, the project aimed to define  supplementary business cases beyond those defined in LSA, particularly scenarios where both the incumbents and LSA could be MNOs. The ADEL project’s primary goal was to develop collaborative sensing techniques, resource allocation, and dynamic spectrum access protocols in the context of LSA, which would pave the way for better spectral and energy efficiency in 5G mobile broadband networks. For more information regarding the message flow of the steps and interactions between different parts of the ADEL framework, refer to Fig \ref{adel_flow}.
	\subsection{Standardization}

	In conjunction with the improvements ADEL introduced to enhance the LSA scheme, ADEL also established the required algorithms to support these enhancements. In \cite{ADEL4_2}, ADEL describes the various centralized and decentralized resource allocation schemes that outperform the conventional protocols in terms of fairness, efficiency, convergence time while considering ADEL's description framework. In \cite{ADEL4_3}, the issue of how to translate the LSA policies into ADEL architecture was investigated. In this regard, the problem of discerning the misbehaving wireless nodes which violate the LSA policies was studied. Furthermore, in \cite{ADEL5_3}, collaborative communication and cooperative sensing in terms of both centralized and decentralized procedures were studied in greater detail. To maintain the data interchange among different elements of wireless nodes, controllers and databases, medium access and control plane protocols were developed in \cite{ADEL5_4}. ADEL in  \cite{ADEL6_2}, elaborated on the system-level simulator (SLS) to evaluate its proposed framework for LSA in terms of its overall performance and efficiency. Finally, in \cite{ADEL6_3}, the authors provided the results of the experimentation in the ADEL LSA setting. The executive summary of the project can be found in \cite{ADEL1_5_2}.

	\subsection{Physical Layer Aspects and Implementation}
	The system architecture that ADEL proposed for LSA, is depicted in Fig. \ref{ADEL_LSA_fram}. The first building block is the LSA band manager, which coordinates multiple licensees' access to the LSA band and performs the resource allocation procedures. This module consists of three submodules: the  LSA request manager, the LSA RRM module, and the LSA billing modules.
	
	Depending on the sharing agreement, the LSA request manager performs  priority management and inquires about the authentication of LSA licensees from the authentication server.
	The LSA RRM calculates the available resources to be assigned for LSA licensees based on the information accessible in the radio environment map. Furthermore, it implements admission control of the LSA licensee spectrum requests to assign appropriate resources for them. The financial accounting tasks are handled on the LSA authentication server. Meanwhile, the radio environment maps collect  data from the sensing networks and LR to update itself periodically  about possible changes in the system. The LR is a database that  collects the information about the incumbent's operating frequencies and bandwidth, locations; and it has the same functionality as defined in LSA.
	The framework proposed by ADEL also supports sensing capabilities, and this is initiated by the spectrum sensing reasoning block that dictates the prerequisites for each sensing network. This facilitates the quick detection of any activities or changes in the environment caused by other incumbents or licensees.
	\begin{figure*}[!t] 
		\centering
		\includegraphics[width=1\textwidth]{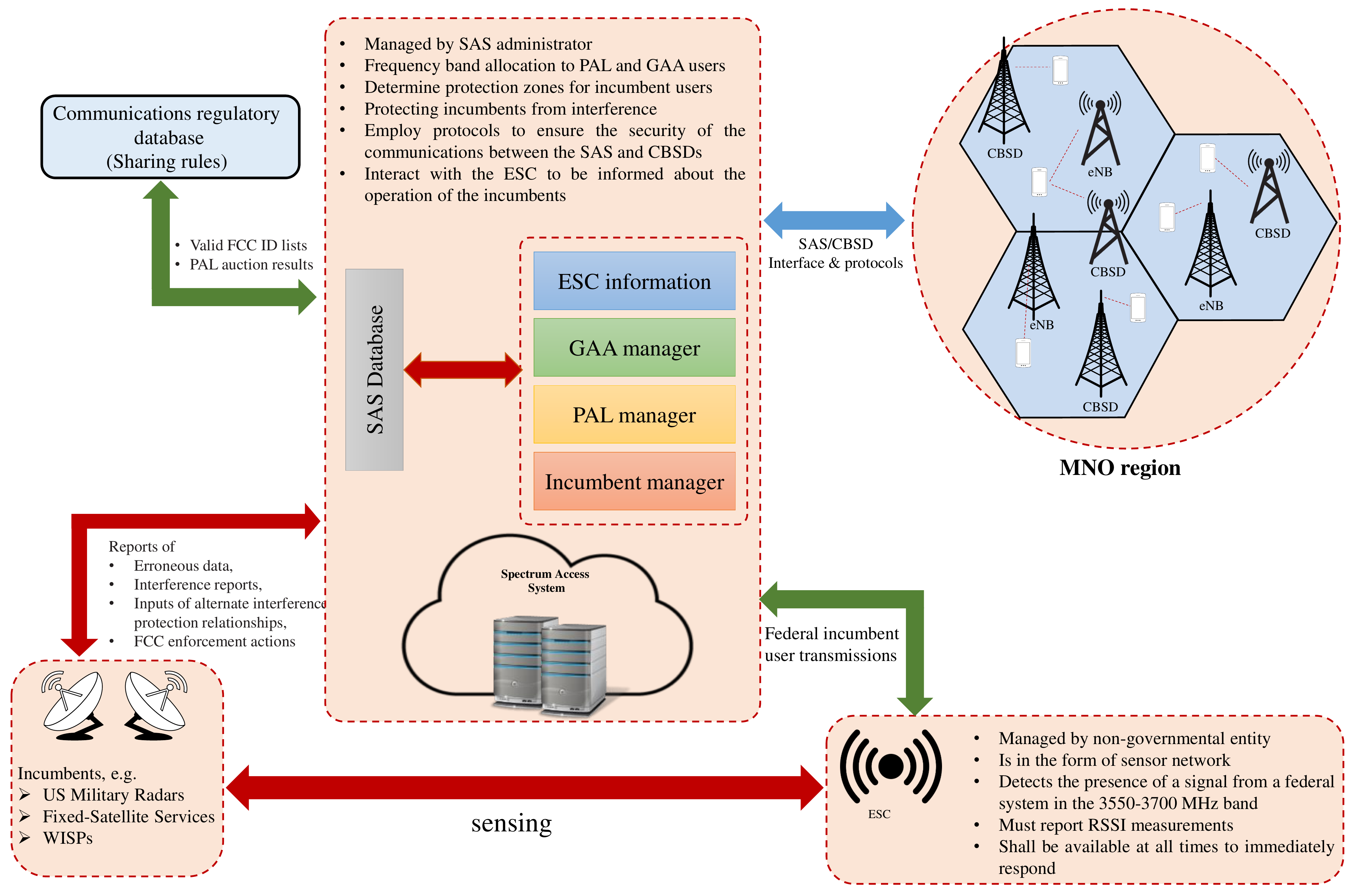}
		\caption{Technical overview of the SAS architecture\cite{FCC_15_47}.} 
		\label{SAS_architecture}
	\end{figure*}
	\section{Evolved LSA}\label{eLSA}
	\subsection{Overview}
	In order to extend LSA functionality to incorporate the applications of vertical sectors, ETSI introduced the concept of evolved LSA (eLSA). The objective of eLSA system is to support spectrum access to local high-quality wireless networks (e.g., industrial automation, programme making and special events (PMSE), culture and creative industry, public protection disaster relief (PPDR), e-Health, etc.). These are operated by vertical sector operators (often referred to as MFCN operators) and address vertical sector-specific local connectivity requirements.
	
	The support of additional MFCN operator types, which demand  predictable QoS levels, calls for both shorter and longer times for spectrum resource deployments. Furthermore, some LSA functionalities have to be redefined or enhanced to facilitate the deployment of  eLSA sharing scheme.
	\subsection{Standardization}
	Clause 4 of \cite{elsa_1} introduced various high-level use cases of local high-quality wireless networks. These use cases pertained to local area networks designed to serve applications that require predictable levels of QoS, such as PMSE, audio-visual content production, Culture and Creative Industry, PPDR, and e-health vertical sectors. Furthermore, ETSI in \cite{elsa_1} specified different spectrum access strategies to meet the requirements of such verticals. Spectrum sharing methods in licensed bands were identified as the most likely way of providing local area services with high QoS. Three possible spectrum access schemes were proposed, which are summarized as follows:
	\begin{itemize}
		\item MNOs can offer dedicated local area services in their licensed frequencies (spectrum as a service).
		\item MNOs can sublease part of their spectrum locally to local area service providers (leasing or subleasing).
		\item Spectrum can be licensed to local high-quality wireless networks (local licensing).
	\end{itemize}
	ETSI in \cite{elsa_1} also elaborated on potential enhancements that should be applied to LSA architecture to support newly introduced services. \cite{elsa_2} specified system requirements to support  spectrum access to the local high-quality wireless networks in licensed bands operated by vertical sector operators through the eLSA framework. As in LSA, these requirements are further divided into four categories, namely general system operation requirements, incumbent protection requirements, resource grant requirements, and security requirements. 
	
	In conjunction with the specified prerequisites in \cite{elsa_2}, \cite{elsa_3} defined high-level functions, which have to be performed by the eLSA System. The eLSA's high-level functions are quite similar to those defined for LSA architecture; however, the eLSA has to further incorporate the vertical sector operator's interfaces and interactions with the eLSA system. Furthermore, \cite{elsa_3} specified operational use cases such as possible deployments of vertical sector players (VSPs) with nomadic and temporary operations. According to the system requirements and high-level procedures defined in \cite{elsa_2} and \cite{elsa_3}, ETSI TS 103 652-3 \cite{elsa_4} outlined the application protocol for the eLSA1 interface, which connects the eLSA controller (eLC) and eLSA repository (eLR). \cite{elsa_4} also elaborated on the eLSA1 protocol principles and procedures, describing the ELSRAI information elements that convey the spectrum availability information.

	\subsection{Physical Layer Aspects and Implementation}
	The eLSA reference model is similar to the LSA framework and includes the same interfaces. The difference is that the so-called interfaces have to incorporate the information exchange regarding the VSPs either (See \cite{elsa_2,elsa_3} for more information).
	
	\section{Spectrum Access System}\label{SAS}
	\subsection{Overview}
\begin{figure}[!t] 
		\centering
		\includegraphics[width=.45\textwidth]{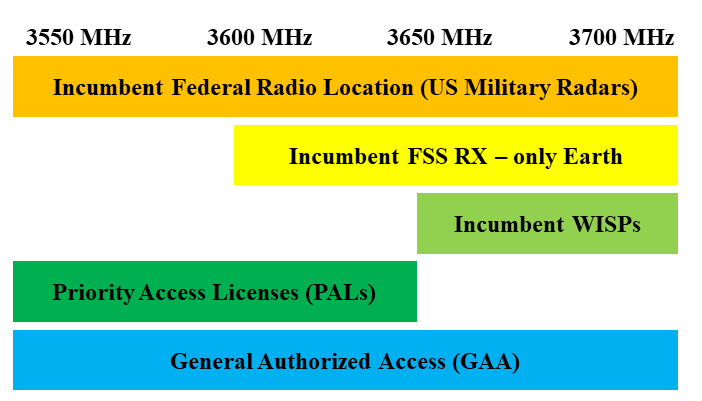}
		\caption{CBRS three-tier shared	spectrum licensing structure \cite{CBRS_SAS}.} 
		\label{SAS_users}
	\end{figure}

	Following a Notice of Inquiry (NOI) and Notice of Proposed Rulemaking (NPRM), the FCC formally issued a Report and Order (R\&O) to the CBRS 3.5 GHz band in April 2015\cite{FCC_15_47}. CBRS is referred to a group of radio frequency bands from 3.55 GHz to 3.7 GHz that the FCC has designated for sharing between three categories of users:
	\begin{itemize}
		\item Incumbent user(s): Refers to the primary users of the spectrum. These users always have priority to access the spectrum, and the interference will be managed for them so that they do not face any difficulties when they need to access the spectrum. Federal aeronautical users, non-federal fixed-satellite services, and wireless internet service providers (WISPs) are examples of such incumbents (Fig.\ref{SAS_users}).
		\item Priority access licensees (PALs): Like LSA licensees, PALs gain access to a part of the frequency band and are guaranteed protection from  interference caused by  third tier consumers (GAA).  PALs will be assigned in up to 70 MHz of the 3.55–3.65 GHz portion of the band\cite{FCC_15_47}. Critical access users, like hospitals, governmental users, and noncritical users (e.g.,  MNOs), are examples of PAL users.
		\item General authorized access (GAA): These users can only access the spectrum by obtaining a license from the regulator for the equipment they use. Of course, they are not guaranteed protection from interference by higher tier users. Interestingly, MNOs can also access the spectrum for free with this type of access in cases where a guaranteed QoS level is not needed. Access to at least 80 MHz of CBRS bandwidth is guaranteed to this group. A comparative summary of PAL and GAA users is shown in Table \ref{gaa_pal_comparisn}.
	\end{itemize}
	
	The technical overview of SAS is shown in Fig. \ref{SAS_architecture}. At the center of the CBRS ecosystem lies the SAS, which is a functionality provided by 3rd party administrators (such as Google, Federated Wireless, etc.) \cite{sasadministerlist}, which authorizes and regulates the spectrum sharing process between the previously mentioned tiers. PAL and GAA users who want to access the CBRS band must submit a request to the SAS system to reserve the unused bands in a specific geographic area. These users interact with SAS through the citizens broadband radio service devices (CBSDs), which are fixed stations or networks of such stations running on a PAL or GAA basis. If there exists a vacant channel, the SAS will provide it to the user. SAS adjusts the maximum permissible transmission power and available bandwidth for users. The information needed to perform such operations is provided either by a database center that communicates with the SAS or by sensors that have environmental sensing capabilities (ESC). The ESC capability coordinates the transmissions inside the Protection zones (The geographical areas within which the incumbent receivers will not be subject to any destructive interference). It informs the SAS system about the presence of an incumbent's signals Another alternative in the CBRS is ``incumbent informing capability" (IIC), where instead of sensing via ESC, the incumbents directly inform the SAS of their spectrum needs or any interference within their geographical area (similar to LSA).	Based on the available information in the database or the information transmitted from the ESC sensors, the SAS  can allocate the spectrum to PAL and GAA users effectively.

	The primary functions of SAS include protecting incumbent users from interference caused by PAL and GAA users, and also protecting PALs from interference caused by GAA users. In this regard, the FCC has set up a set of rules in which CBSD devices must first register with the SAS and provide it with location information and other related details. The SAS then assigns the channels for PAL and GAA use. The SAS can also limit the maximum transmit power of CBSDs to reduce  interference between incumbents, PALs and GAA users. End-user devices wait for access permission from the respective CBSD before operating in the spectrum. The SAS effectively resembles resembles the LC in terms of functionality. But according to ETSI standards, the LC is installed within the mobile network territory. At the same time, the SAS is expected to be located outside the mobile network so as to perform impeccably in restricting the potential interference between different users\cite{FCC_15_47}. 
	\begin{table}[!t]
		\scriptsize
		\caption{Comparison of GAA and PAL.}
		\begin{tabular}{p{2cm} p{2.7cm} p{2.8cm}}
			\toprule[1.1pt]
			&	\textbf{PAL} & \textbf{GAA} \\\midrule
			\textbf{Operating bands} &	3550--3650 MHz (will be assigned 
			up to 70 MHz of the band) & 3550--3700 MHz (will be guaranteed at least 80 MHz of the band) \\
			\textbf{Spectrum access \newline scheme} & through an auction & free access to the 
			spectrum  \\
			\textbf{Protection criteria} & protected from other PAL and GAA users & no guaranteed interference protection \\
			\multirow{9}{*}{\textbf{Use cases}} & 
			\begin{enumerate}
				\item QoS-managed enterprise networks
				\item sensitive use \newline cases (e.g., \newline hospitals, government users, MNO) 
			\end{enumerate} & 
			\begin{enumerate}
				\item Personal hot \newline spots
				\item Small business \newline hot spots
				\item Campus hot spots
				\item WISPs
			\end{enumerate} \\
			\bottomrule[1.1pt]
		\end{tabular}
		\label{gaa_pal_comparisn}
	\end{table}
	\subsection{Standardization}
	\begin{table*}[!t]
		\centering
		\caption{FCC standardization works in the field of SAS.}
		\begin{tabular}{p{1cm} p{2cm} p{14.25cm}}
			\toprule
			\textbf{Ref.} & \textbf{Standard} & \textbf{Subject}  \\
			\midrule
			\cite{FCC_18_149}      & FCC 18-149               &  Report and Order\\\midrule
			\cite{FCC_17_134}      & FCC 17-134               &  Notice of Proposed Rulemaking and Order Terminating Petitions\\\midrule
			\cite{FCC_16_55}       & FCC 16-55                &  Order on Reconsideration and Second Report and Order\\\midrule
			\cite{FCC_15_47}       & FCC 15-47                &  Report and Order and Second Further Notice of Proposed Rulemaking\\\midrule
			\cite{FCC_14_49}       & FCC 14-49                &  Further Notice of Proposed Rulemaking\\\midrule
			\cite{FCC_12_148}      & FCC 12-148               &  Notice of Proposed Rulemaking and order\\
			\bottomrule
		\end{tabular}
		\label{SAS_FCC_table}
	\end{table*}

	\begin{table*}[!t]
		\centering
		\caption{WInnForum standardization works in the field of SAS (Release1).}
		\begin{tabular}{p{0.75cm} p{2cm} p{14cm}}
			\toprule
			\textbf{Ref.} & \textbf{Standard} & \textbf{Subject}  \\
			\midrule
			\cite{WINNF_0016}      & WINNF-TS-0016                &  Signaling Protocols and Procedures for Citizens Broadband Radio Service (CBRS):
			Spectrum Access System (SAS) - Citizens Broadband Radio Service Device (CBSD)
			Interface Technical Specification \\\midrule
			\cite{WINNF_0022}      & WINNF-TS-0022                &  WInnForum CBRS Certificate Policy	Specification \\\midrule
			\cite{WINNF_0061}      & WINNF-TS-0061                &  Test and Certification for Citizens Broadband Radio Service (CBRS);
			Conformance and Performance Test Technical Specification; SAS as Unit Under Test (UUT) \\\midrule
			\cite{WINNF_0065}      & WINNF-TS-0065                &  CBRS Communications Security Technical Specification \\\midrule
			\cite{WINNF_0071}      & WINNF-TS-0071                &  CBRS Operational Security \\\midrule
			\cite{WINNF_0096}      & WINNF-TS-0096                &  Signaling Protocols and Procedures for Citizens Broadband Radio Service (CBRS): Spectrum
			Access System (SAS) - SAS Interface Technical Specification \\\midrule
			\cite{WINNF_0112}      & WINNF-TS-0112                &  Requirements for Commercial Operation in the U.S. 3550-3700 MHz Citizens Broadband
			Radio Service Band \\\midrule
			\cite{WINNF_0122}      & WINNF-TS-0122                &  Test and Certification for Citizens Broadband Radio Service (CBRS);
			Conformance and Performance Test Technical Specification; CBSD/DP as Unit
			Under Test (UUT) \\\midrule
			\cite{WINNF_0245}      & WINNF-TS-0245                &  Operations for Citizens Broadband Radio Service (CBRS):
			Priority Access License (PAL) Database Technical Specification \\\midrule
			\cite{WINNF_0247}      & WINNF-TS-0247                &  CBRS Certified Professional Installer Accreditation Technical Specification \\
			\bottomrule
		\end{tabular}
		\label{SAS_WINNF1_table}
	\end{table*}
	\begin{table*}[!t]
		\centering
		\caption{WInnForum standardization works in the field of SAS (Release2).}
		\begin{tabular}{p{0.75cm} p{2cm} p{14cm}}
			\toprule
			\textbf{Ref.} & \textbf{Standard} & \textbf{Subject}  \\
			\midrule
			\cite{WINNF_1001}      & WINNF-TS-1001               &  CBRS Operational and Functional	Requirements\\\midrule
			\cite{WINNF_3002}      & WINNF-TS-3002               &  Signaling Protocols and Procedures for Citizens Broadband Radio Service (CBRS):
			Extensions to Spectrum Access System (SAS)- Citizens Broadband Radio Service Device (CBSD) Interface Technical Specification\\\midrule
			\cite{WINNF_3003}      & WINNF-TS-3003               &  Signaling Protocols and Procedures for Citizens	Broadband Radio Service (CBRS):
			Extensions to Spectrum Access System (SAS) - SAS Interface Technical Specification\\\midrule
			\cite{WINNF_4004}      & WINNF-TS-4004               &  Test and Certification for Citizens	Broadband Radio Service (CBRS);
			Conformance and Performance Test Technical Specification; CBSD/DP as Unit
			Under Test (UUT)\\\midrule
			\cite{WINNF_4005}      & WINNF-TS-4005               &  CBRS Release 2 Self-Testing Policy\\\midrule
			\cite{WINNF_5006}      & WINNF-TS-5006               &  CBSD Antenna Pattern Database Technical	Specification\\
			\bottomrule
		\end{tabular}
		\label{SAS_WINNF2_table}
	\end{table*}
	
	The FCC in a report from 2012  \cite{FCC_12_148} proposed the creation of a new citizens broadband service in the 3550-3650 MHz  band, which had previously be used by the military and satellite operations. Then, in 2014, in another report \cite{FCC_14_49}, the FCC extended this band to 3700 MHz. They also specified up to 70 MHz of the CBRS 150 MHz band for PAL use and adopted rules for commercial use of this band, stipulating conditions for the license terms of PALs in \cite{FCC_15_47}. However, these provisions elicited controversy in the mobile cellular sector and among WISPs, which led to sweeping changes in FCC's proposed original rules. 
	Consequently, the FCC reestablished the investigation of the requirements for the CBRS \cite{FCC_16_55,FCC_17_134}, releasing a decision on new conditions.
	In 2018, the FCC stipulated that the terrestrial area of a PAL could cover the area of a country and operate for about ten years with permissions being renewable. New rules allow for the development of up to seven PALs in each license area, while also permitting for partitioning and disaggregation of PALs, promoting transmission over wider channels \cite{FCC_18_149}.
	
	To address the requirements stated in \href{https://www.ecfr.gov/cgi-bin/retrieveECFR?gp=&SID=0076fe7586178336d9db4c5146da8797&mc=true&n=pt47.5.96&r=PART&ty=HTML}{47 CFR Part 96}, the Wireless Innovation Forum (WInnForum) created the CBRS Baseline  Standards, which are summarized in Tables \ref{SAS_WINNF1_table} and \ref{SAS_WINNF2_table}. These Standards aim to develop an ecosystem of cooperative SAS and CBRS devices technologies. In technical specifications \cite{WINNF_0016} and \cite{WINNF_0096}, the signaling protocol and procedures for SAS-CBSD and SAS-SAS interfaces were respectively specified. In particular, \cite{WINNF_0016} introduced various procedures and corresponding message parameters that have to be supported by the SAS-CBSD interface. Similarly, \cite{WINNF_0096} addressed the SAS-SAS interface procedures, synchronization, and respective message parameters. \cite{WINNF_0022} examined the policies for Public Key Infrastructure (PKI), which is crucial in regulating and providing communications and authorization procedures within the CBRS and SAS ecosystem. Communications and operational security aspects in the CBRS ecosystem have been discussed in \cite{WINNF_0065} and  \cite{WINNF_0071}, respectively. More specifically, \cite{WINNF_0065} covered the security policies that govern the SAS and CBSD communication interfaces along with the authorization and authentication of messages interchanged within the CBRS ecosystem.
	In addition, the requirements for the ESC development and security of incumbents for their operation in the 3550--3700 MHz CBRS band were presented in \cite{WINNF_0071}. In \cite{WINNF_0112}, the requirements were specified for commercial operations in the 3550--3700 MHz band. These prerequisites are based on FCC rules \cite{FCC_15_47,FCC_16_55,FCC_18_149}. The document also investigated the requirements for SAS, CBSDs, end user devices (EUDs), PAL, and GAA to further facilitate a properly functioning sharing ecosystem. 
	\cite{WINNF_0061} and \cite{WINNF_0122} include test procedures to ensure the conformity of the three-tiered SAS architecture with the specifications and necessities dictated by the FCC and WInnForum in \cite{WINNF_0016,WINNF_0065,WINNF_0096,WINNF_0112,WINNF_0122}. 
	The FCC has agreed to the test procedures in \cite{WINNF_0061} and \cite{WINNF_0122} to certify the CBSDs and SASs in the CBRS band. The PAL database schema is blueprinted in \cite{WINNF_0245}. This database is maintained through collaboration between SAS administrators, and it outlines the results of PAL auctions. It also contains the PAL-ID, PAL licensee identity from the auction results, PAL initiation, and termination date.
	Finally, \cite{WINNF_0247} includes information about the preparation of a certified professional installer (CPI) training program to guarantee the professional installation process of CBSDs, including categories A and B.
	
	On January 23, 2020, the WInnForum established a new Release 2 specification that aimed to improve the baseline CBRS operational and functional specifications.  
	Additional characteristics and functionalities were defined in Release 2 that can be incorporated anytime. No entity is obligated to support anything beyond Release 1; thus, supporting the Release 2 configurations is not mandatory \cite{WINNF_1001}. As a supplementary work for \cite{WINNF_0016} and \cite{WINNF_0112} in Release 1, and \cite{WINNF_1001} in Release 2, \cite{WINNF_3002} specified the extension to the SAS-CBSD interface. Furthermore, \cite{WINNF_3003} extended the predefined SAS-SAS interface in \cite{WINNF_0096} to incorporate Release 2 procedures. However, some parameters and features defined in \cite{WINNF_0096} were deprecated in Release 2 SAS-SAS interface, including the time-range request support, by-ID request support, push support, SAS Administrator object, SAS Implementation object, and terminated parameter. \cite{WINNF_4004} specified the test cases for CBSD and domain proxy components of the CBRS to guarantee the conformity of the spectrum sharing architecture with the Release 2 features and requirements. In \cite{WINNF_4005}, testing policies were defined for companies providing SAS and CBSD/DP products using Release 2 test specifications. As an improvement for Release 1 SAS operation, \cite{WINNF_5006} defined the CBSD antenna pattern database, which allows the SAS to calculate CBSD's antenna gains and use their patterns in interference calculation and estimation. SAS is one of the most complex spectrum sharing schemes that has been commercially deployed so far. Nonetheless, the stakeholder engagement, inclusive standardization process via a third-party body (i.e., WInnForum), and availability of sufficient spectrum to motivate operators are the primary reasons behind its success.
	\begin{figure}[!b] 
		\centering
		\includegraphics[width=.4\textwidth]{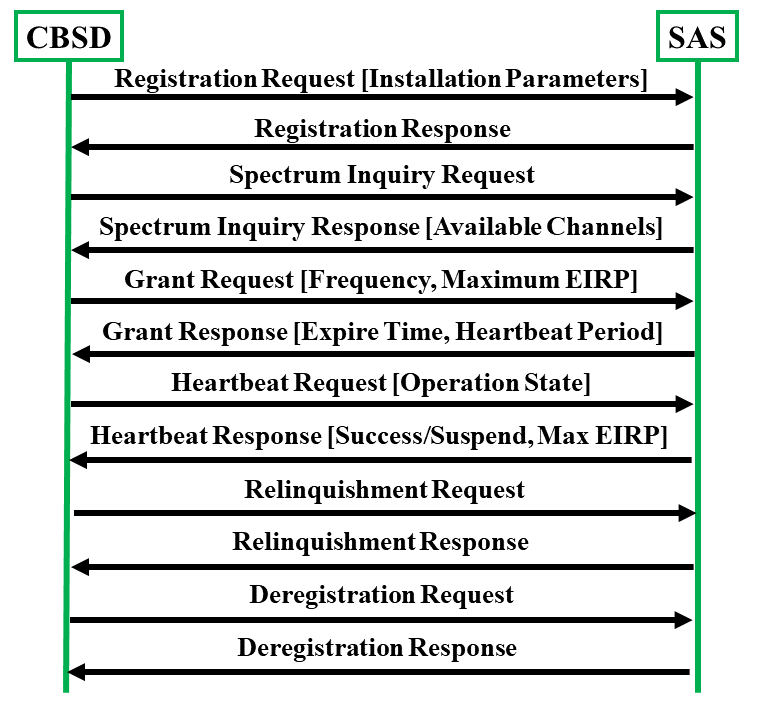}
		\caption{SAS message flow \cite{website_sas}.} 
		\label{SAS_Message}
	\end{figure}
	\subsection{Physical Layer Aspects and Implementation}
	At its core, an SAS is a cloud-based service that handles the wireless communications of devices transmitting in the CBRS band to protect higher priority users from harmful interference. 
	The main functions of an SAS are the following \cite{WINNF_0016}:
	\begin{itemize}
		\item Support and execute the policies and procedures developed by the SAS administrators.
		\item Determine and provide the authorized channels or frequencies to the respective CBSDs.
		\item Determine the maximum allowable power to CBSDs.
		\item Register and verify the identification information and the operation areas of CBSD devices.
		\item Communicate with the ESC to obtain information about the ongoing federal incumbent users' transmissions and notify the respective CBSDs to switch to another operating frequency range or stop sending.
		\item Protect the primary incumbents from the interference caused by PALs and GAA users.
		\item Protect PAL users from the interference caused by the other PALs and GAA users.
		\item Ensure reliable and secure communication between SAS and CBSD devices.
	\end{itemize}
	
\begin{figure}[!t] 
		\centering
		\includegraphics[width=.4\textwidth]{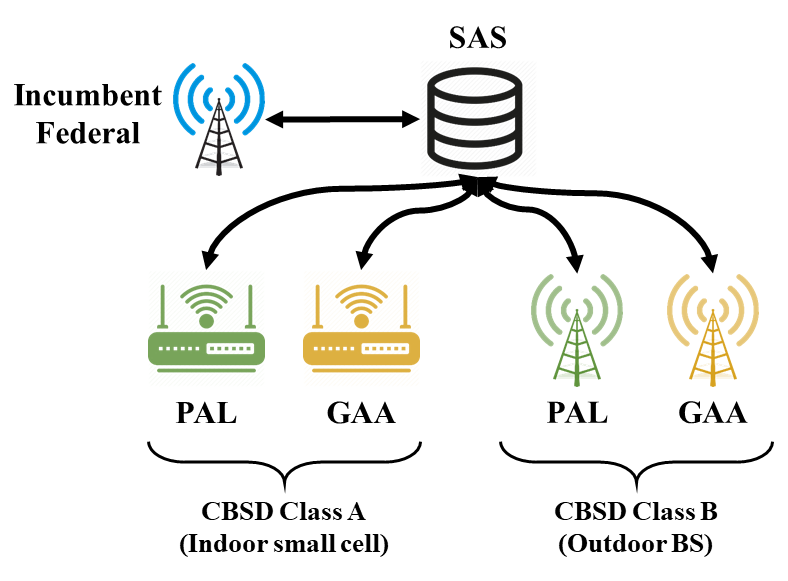}
		\caption{Different categories of CBSDs in an SAS.\cite{CBSD_SAS_BS}.} 
		\label{SAS_CBSD}
	\end{figure}

	CBSDs in the operator's network are set up to securely link to the SAS over the Internet.
	The FCC has envisaged two general classifications for CBSDs (Fig. \ref{SAS_CBSD}),
	\begin{itemize}
		\item \textit{CBSD type A}:
		\begin{enumerate}
			\item Should not be installed outdoors or have antennas with an average height of 6 meters above the ground.
			\item  Is permitted a maximum EIRP of 30 dBm (dBm/10 MHz) or 1 Watt.
		\end{enumerate}
		\item \textit{CBSD type B}:
		\begin{enumerate}
			\item Its deployment and operation is limited to outdoor, and antenna height is anticipated to be more than 6 meters above the terrain.
			\item  Is permitted a maximum EIRP of 47 dBm (dBm/10 MHz) or 50 Watt.
		\end{enumerate}
	\end{itemize}
	
	The SAS has a database of all CBSDs that includes their status, geolocation, and other related information to coordinate channel allocation and manage potential interference. 
	
	The FCC has developed a high-level architecture for SASs, which need to interface with other SASs, the ESC, and the FCC database. Besides, there may be some incumbent users who wish to report their bandwidth usage either directly to the SAS or through the database.
	There may also be certain types of SASs that are designed to meet the requirements of a particular service provider but need to interact with other SASs to protect the primary incumbent users and control the interference\cite{FCC_15_47}.

	The ESC's primary goal is to facilitate the coexistence of CBRS users with current federal incumbent users through signal measurement techniques and sensing capabilities and then informing the acquired information to one or more approved SAS while maintaining the security of the detected and transmitted signal information. The ESC is operated by a non-governmental organization. ESC equipment may be located adjacent to restricted and protected areas so that it can accurately detect the federal incumbents' operations.
	
	Since early 2020, the FCC has approved five SAS administrators to begin its initial commercial deployment and also has approved three ESC sensor registrations\cite{SAS_Administerator}.
	
	The SAS-CBSD protocol is based on the HTTPS protocol. The SAS-CBSD API is used by CBSDs to communicate with an SAS as follows \cite{WINNF_0016}:

	\subsubsection{Registration}
	A new CBSD must first register with an SAS before it can transmit. The CBSD sends the SAS a registration request containing information about the installation's parameters, such as location and transmission characteristics, and a registration response is returned by the SAS. The SAS will respond with a CBSD ID if the registration request is authorized. If the SAS refuses the registration request, an error message will be shown. The CBSD must correct the mistake and resend the registration request. All Category A and Category B CBSDs that it is not possible to determine their position automatically or using an antenna Height Above Average Terrain (HAAT) of more than 6 meters when performing outdoors need CPI (Certified Professional Installer) validation. The CPI sends the installation parameters to the respective CBSD, and are signed with their own CPI certificate.
	\cite{WINNF_0247}.
	
	\subsubsection{Spectrum inquiry}
	A registered CBSD will send a spectrum inquiry request to the SAS to determine the availability of spectrum, and the CBSD will receive the details of the available frequencies and information for efficient transmission. On the other hand, the SAS will refuse the spectrum inquiry request if it is malformed. An invalid frequency range or an invalid CBSD ID are examples of malformed requests. It should be noted that all SASs perform a procedure called Coordinated Periodic Activities among SASs (CPAS) to ensure they use the most up-to-date data.
	
	\subsubsection{Grant requests}
	Once the CBSD has determined which frequency band it is allowed to operate on, it sends a grant request to the SAS. To protect incumbents, the SAS accepts or refuses the request and sends the result in a grant response. If the request is accepted, the SAS responds with the grant ID and provides a grant to the CBSD. This indicates that a limited portion of the available spectrum is set aside by the SAS for exclusive use. The CBSD will not receive the grant if the grant request is denied; however, it can execute another spectrum inquiry. Typically, a request is denied if the parameters are incorrect or if the spectrum is unavailable. A CBSD may request a grant of any size between 5 and 150 MHz, but it has to be a multiple of 5 MHz.
	\subsubsection{Heartbeat requests}
	The CBSD will send a heartbeat request for every grant. The heartbeat request inquires whether the CBSD is capable of transmitting with the parameters specified in the grant request. The CBSD is allowed to transmit with the predetermined parameters until the Transmit Expire Time (TET), which is usually 240 seconds after the SAS heartbeat request authorization. The SAS can respond to a heartbeat in one of three ways: it can authorize the grant; it can terminate the grant (to protect incumbents); or it can suspend the grant (temporary disablement).
	\subsubsection{Relinquish a grant}
	The CBSD relinquishes a grant when it no longer wants to use it.
	\subsubsection{Deregistration}
	If a CBSD is relocated or decides to deregister from the SAS, it will make a deregistration request to the SAS and will cease transmission associated with any Grants. In order for the CBSD to be able to transmit again, it needs to send a registration request with new parameters later.
	
	\subsection{SAS Field Trials}
	Unlike LSA, there have been few field trials involving the SAS framework, particularly CORNET and CORE++. CORNET is a collection of 48 software-defined radio nodes deployed within Virginia Tech's main campus. A demonstration of SAS implementation using the CORNET testbed and its results were published in \cite{sasdemo1}. CORE++ is the extension of the Finnish CORE+ project to SAS. The corresponding results have been demonstrated on multiple occasions in \cite{sascore}.
	
	\section{Dynamic Spectrum Sharing}\label{DSS}
	
	\begin{figure}[!b] 
		\centering
		\includegraphics[width=0.48\textwidth]{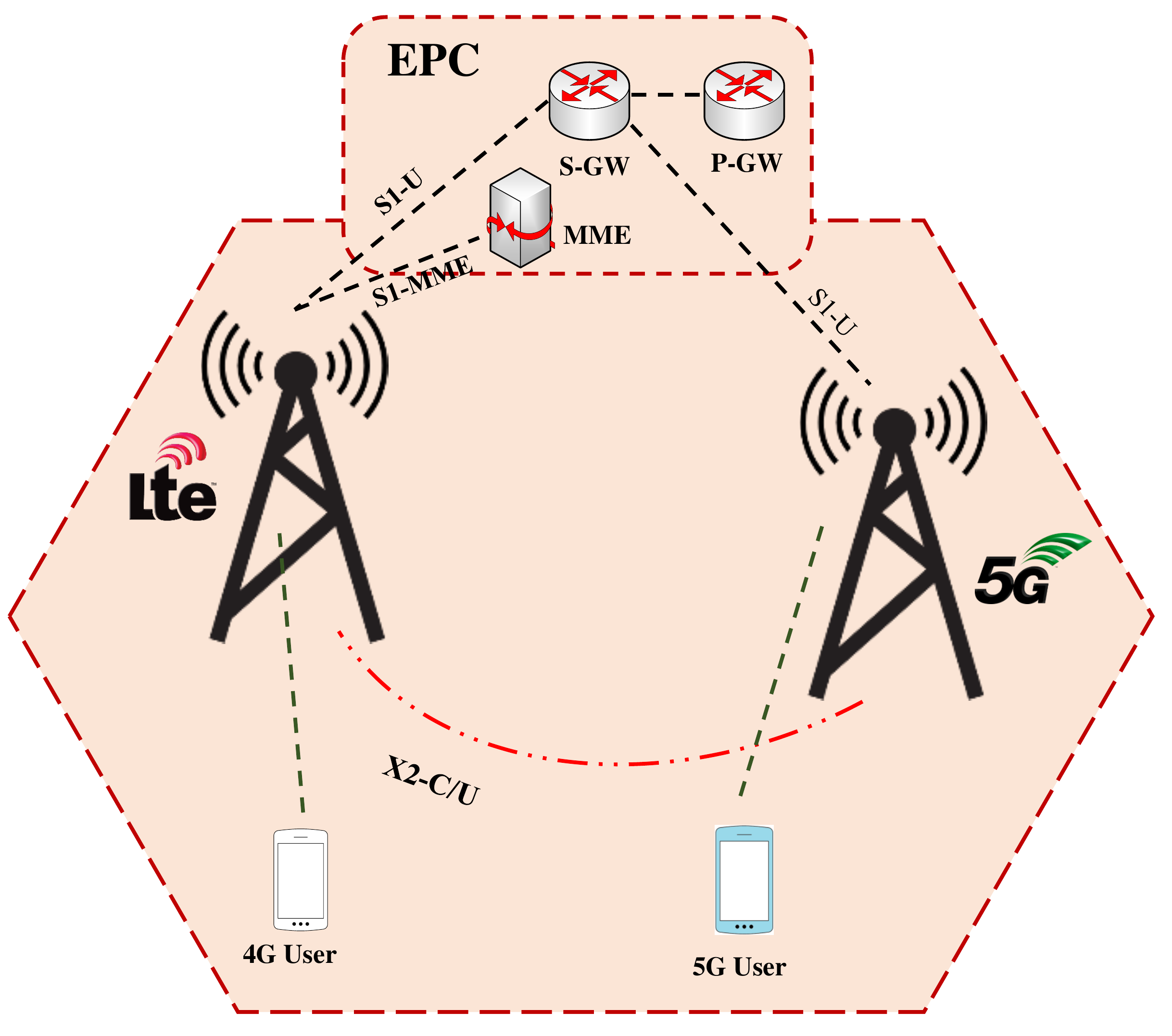}
		\caption{Technical overview of the DSS architecture.} 
		\label{DSS_architecture}
	\end{figure}
	\subsection{Overview}
	\begin{table*}[!t]
		\centering
		\caption{3GPP standardization works in the field of DSS.}
		\begin{tabular}{p{0.75cm} p{5cm} p{10cm}}
			\toprule[1.1 pt]
			\textbf{Ref.} & \textbf{Standard} & \textbf{Subject}  \\
			\midrule
			\cite{3gpp_38_912}      & 3GPP TS 38.912               & Study on New Radio (NR) access technology \\\midrule
			\cite{3gpp_38_101_1} & 3GPP TS 38.101-1 & NR; User Equipment (UE) radio transmission and reception; Part 1: Range 1 Standalone \\\midrule
			\cite{3gpp_38_212} & 3GPP TS 38.212 &NR; Multiplexing and channel coding  \\\midrule
			\cite{3gpp_38_331}& 3GPP TS 38.331&NR; Radio Resource Control (RRC) protocol specification\\\midrule
			\cite{3gpp_38_211} & 3GPP TS 38.211 & NR; Physical channels and modulation \\\midrule 
			\cite{3gpp_211345}	&RP-211345; 3GPP TSG RAN Meeting &New WID on NR Dynamic spectrum sharing (DSS)  \\\midrule
			\cite{3gpp_201314}	& RP-201314; 3GPP TSG RAN Meeting&  LTE/NR spectrum sharing in Band 38/n38 (TDD mode in 2600 MHz)\\\midrule
			\cite{3gpp_202084}	& RP-202084; 3GPP TSG RAN Meeting&
			LTE/NR spectrum sharing in Band 40/n40 (TDD mode in 2400 MHz)\\\midrule
			\cite{3gpp_2100694}&R1-2100694; 3GPP TSG RAN WG1& Discussion on cross carrier scheduling for NR DSS\\\midrule
			\cite{3gpp_2101362}&R1-2101362; 3GPP TSG RAN WG1&	Views on Rel-17 DSS SCell scheduling PCell\\
			\bottomrule[1.1 pt]
		\end{tabular}
		\label{DSS_table}
	\end{table*}
	One of the key challenges preventing MNOs from migrating to globally deployed 5G cellular systems is spectrum scarcity. On the one hand, the traditional way of solving this problem (i.e., by buying new spectrum) means heavy financial costs on behalf of MNOs and vendors. On the other hand, spectrum re-farming would be a short-term solution that relies on the inefficient division of spectrum between various generations of the cellular system. One promising alternative to address this problem is to increase the utilization of frequency bands by employing novel spectrum sharing methods that facilitate flexible interworking between various technology generations within an MNO's network \cite{ahokangas2013simple}. 
	

	DSS is considered a smooth migration from LTE to NR that dynamically allocates resources between 4G and 5G users according to the traffic requirements of LTE and NR networks (see Fig. \ref{DSS_architecture}). In DSS, the gNB and its corresponding users can use the available spectrum in LTE frequency bands. Mid-band is the main frequency band that can be used in this method. The DSS   greatly depends on the flexible interworking of LTE and 5G NR networks; in other words, backward compatibility is indispensable. The principal idea in DSS is to map the resource element (RE) of the NR onto the LTE resource grid. One condition for this is that there must be no overlapping or interference between the NR and LTE signals. In DSS, the 5G NR plays a crucial role, as it offers a flexible physical layer design with diverse numerologies and sub-carrier spacing (SCS). This means that it can flexibly adapt its transmissions based on the LTE physical layer design. In the following, we will elaborate more on the physical layer aspects and implementation scenarios of DSS \cite{xin2021dynamic}.

	\subsection{Standardization}
	From the perspective of standardization, DSS has been covered in 3GPP Rel-15 and was further improved in Rel-16. To guarantee efficient coexistence between LTE and NR, the following requirements have been specified: 
	\begin{enumerate}
		\item DSS has to support the coexistence of LTE UL and NR UL as well as the coexistence of LTE DL and NR DL within the bandwidth of an LTE component carrier and has to stipulate at least one NR or LTE-NR band combination for the operation \cite{3gpp_38_912}. 
		\item In DSS, both LTE and NR must be mutually compatible; that is, the LTE operation should not violate the operation of the NR, and the NR should not interfere with LTE performance \cite{3gpp_38_912}. 
		\item There should not be any requirement for the UEs to support the simultaneous connection of NR and LTE \cite{3gpp_38_912}.
		\item In UL of NR, an optional 7.5 kHz channel raster shift has to be supported to ensure that the NR UL grid is in compliance with the LTE UL gird in an orthogonal manner since they both employ different multiple access schemes (LTE uses SC-FDMA and NR uses both OFDMA and SC-FDMA) \cite{3gpp_38_101_1}.
		\item Rate matching, which is a baseband processing method, has to be adopted to enable NR physical download shared channel (PDSCH) transmission in time-frequency resources where LTE cell-specific reference signals (CRSs) is not located \cite{3gpp_38_212, 3gpp_38_331}. 
		\item In the case of an additional demodulation reference signal (DMRS) configuration for NR PDSCH, DMRS has to be shifted from the 12\textsuperscript{th} symbol to the 13\textsuperscript{th} symbol to avoid any potential collision with the LTE CRS in the subframe \cite{3gpp_38_211}.
		\item It is mandatory for both the NR and LTE PDCCH to be incorporated within the first three OFDM symbols of a subframe. The level of allocation greatly depends on the traffic load related to each network.
	\end{enumerate}
	
	\begin{figure*}[!t]
		\centering
		\includegraphics[width=0.95\linewidth]{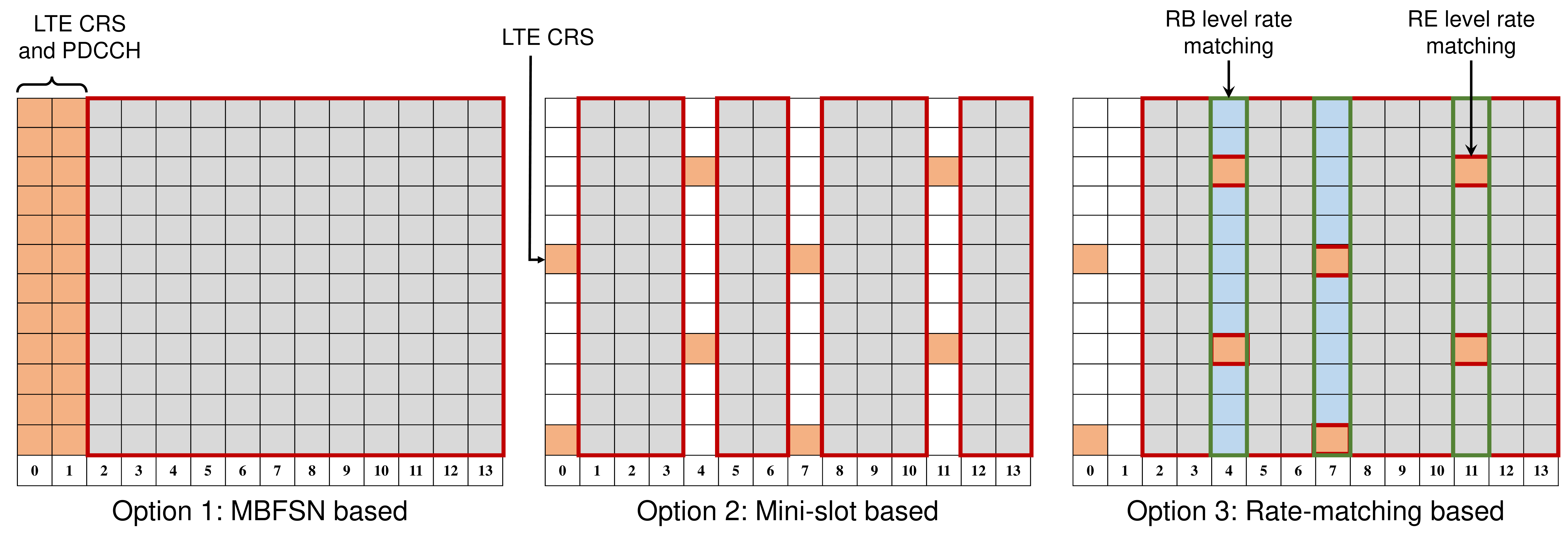}
		\caption{DSS different deployment scenarios \cite{mediatekdss}.}
		\label{mediadss}
	\end{figure*}
	
	In subsequent releases (i.e., Rel-16 and Rel-17 \cite{3gpp_211345,3gpp_201314,3gpp_202084,3gpp_2100694,3gpp_2101362}), further enhancements were introduced to boost the DSS performance. 
	In Rel-15, only the cases of 2, 4, and 7 symbol duration were identified for PDSCH, which resulted in inefficient utilization of resources in cases where the length of PDCCH exceeded 3 OFDM symbols. In response, in Rel-16, a new 10 symbol duration for PDSCH transmission was proposed to address the aforementioned problem.
	Furthermore, in order to enhance the scheduling rate of NR UEs in the shared carrier, a new concept named cross-carrier scheduling was introduced in REL-17.
	\subsection{Physical Layer Aspects and Implementation}
	Three different deployment options have been foreseen to permit the interworking between LTE and NR:
	\begin{itemize}
		\item \textbf{MBSFN based DSS}: The multi-broadcast single-frequency network (MBSFN) is used for point-to-multipoint transmission, such as evolved multimedia broadcast multicast services (eMBMS), and it is also used to provide broadcast services by MNOs. Depending on the MNO policy, the MBSFN resources that  occupy the last 12 OFDM symbols of the LTE resource grid can be allocated for NR users' PDSCH transmissions.
		\item \textbf{Mini-slot-based DSS}:  When dealing with non-MBSFN subframes that include LTE CRSs, this option can be used. The general idea is to schedule only the free OFDM symbols for NR users. Care must be taken not to interfere with symbols that carry the LTE CRS. This option is more suitable for URLLC applications that demand extremely low latency during the operation. 
		\item \textbf{Rate-matching-based DSS}: Similar to the preceding option, this option is also suitable for non-MBSFN subframes. This method works by puncturing REs that are used by LTE CRS. This can help the NR scheduler to discern which REs are not available for transmission. The deployment of this option can be either resource block (RB)-level or RE-level. In the first one, the whole RB that includes LTE CRS is taken out of NR scheduling, and in the latter case, only the REs that contain the CRS are ignored.
	\end{itemize}

	The details of these scenarios are depicted in Fig. \ref{mediadss}.
	The authors in \cite{samsung_whitepapaer} have evaluated the performance of DSS in comparison with dual connectivity (DC) and CA. Accordingly, when there is low demand on a cellular network, DSS can help enhance network performance; conversely, when network traffic is high, DSS can deteriorate network capacity and performance. As a result, using DSS for dense cellular network conditions can decrease the network capacity.
	\begin{figure*}
		\centering
		\includegraphics[width=0.99\linewidth]{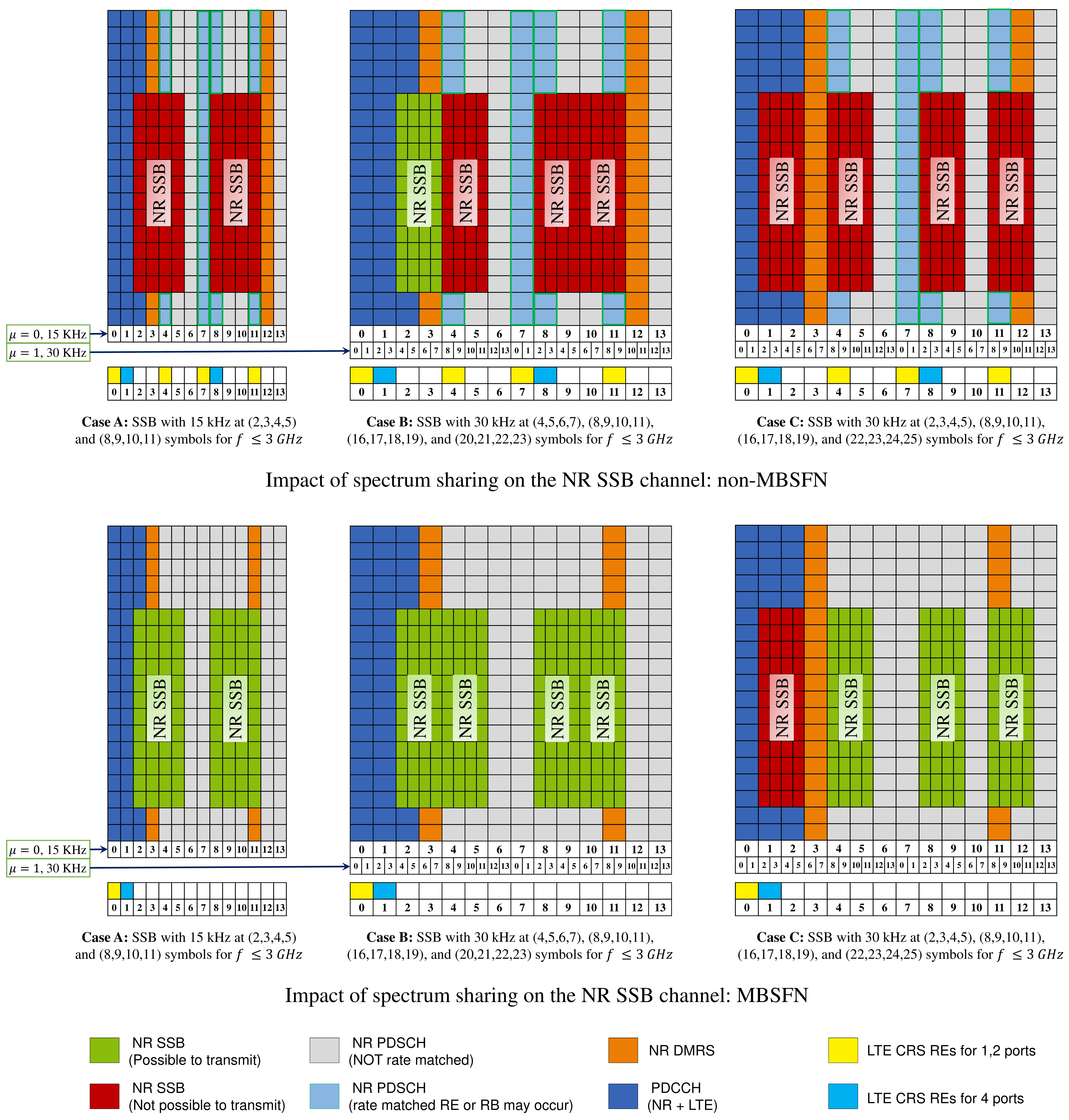}
		\caption{Comparison of NR-SSB transmission for non-MBSFN and MBSFN-based DSS \cite{mediatekdss}.}
		\label{SSB_Block}
	\end{figure*}

	An important consideration when designing for NR-LTE coexistence is that the DSS should support NR common signal transmission and the initial access procedure while taking into account the LTE frame structure and signal transmission properties. When designing the DSS with the purpose of balancing performance between NR and LTE, the NR signals and channels should be established in a way that does not overlap with LTE CRS in the frequency-time resources. By doing so, LTE operation will not be affected when NR operations need to share LTE resources. Previously, rate matching was introduced as a promising option to avoid CRS. However, during the initial access phase, an NR device does not know whether or not it has accessed a clean NR cell or an LTE-NR shared cell. Therefore, the NR device cannot simply assume any CRS rate matching to avoid LTE CRS.
	
	The NR synchronization signal blocks (SSBs), the essential broadcast signal that allows NR devices to detect a cell, is a one-shot transmission with a default 20 ms cycle when assuming 15 kHz SCS is used for spectrum sharing. To guarantee the necessary detection performance, SSB needs to be transmitted without any collision with legacy LTE signals. The SSB occupies 20 RBs and 4 OFDM symbols, which unfortunately cannot be transmitted in a normal subframe due to the conflict with LTE CRS transmission. To avoid such collision, SSB can be located in an LTE MBSFN subframe, which only contains LTE CRS in the LTE control region, referred to as LTE PDCCH. MBSFN is a legacy LTE functionality that helps regular LTE devices skip (or not expect CRS) certain subframe patterns. DSS can leverage this functionality to use the whole subframe (except the LTE PDCCH region) for NR dedicated use.
	
	Fig. \ref{SSB_Block} illustrates the NR SSB transmission alongside an LTE network with four CRSs configuration. The figure is shown for three different cases in the FR1 region (frequencies below 3 GHz), namely, case A where the SCS is 15 kHz, Case B, and C where the SCS is 30 kHz. SSB starting symbol indices are different for these cases. As we can see, for the non-MBSFN case, the SSB beams cannot be transmitted in either Case A or Case C, as they are always colliding with the LTE CRS. In Case C, only one out of four SSB beams can be successfully transmitted without any collisions from the gNB. For the MBSFN case, the results are thoroughly distinctive. For all the cases, the SSB beams can be transmitted except for Case C, where three out of four SSB beams are available. It is important that we mention that not all subframes can be allocated as MBSFN; therefore, subframe alignment with NR SSB slots is required.
	
	In an EN-DC non-standalone (NSA) system, where LTE serves as the control anchor, SSB and random access response (RAR) transmissions of NR are required. In an NR SA system, by contrast,  system, additional messages, such as system information block 1 (SIB1), other system information, and paging channel are transmitted. Since the LTE CRS rate matching for these signals is not defined for SSB, they have to be transmitted through the LTE MBSFN subframe.

	Based on the LTE network configuration and antenna ports, the number of REs that are used for transmitting the LTE CRS is different. For this reason, the total percentage of overhead will be different depending on the number of antenna ports. These details are summarized in Table \ref{DSS_overhead}.	
	\begin{table}[!b]
		\centering
		\caption{The total overhead RB that used in different configuration of LTE \cite{samsung_whitepapaer}.}
		
		\begin{tabular}{p{4cm} p{4cm}}
			\toprule
			\textbf{Configuration} & \textbf{Average overhead in a RB}    \\
			\midrule
			2 CRS, 1 LTE PDCCH   & 14\%
			\\\midrule
			2 CRS, 2 LTE PDCCH&  21\%
			\\\midrule
			4 CRS&  23\% 
			\\
			\bottomrule
		\end{tabular}
		\label{DSS_overhead}
	\end{table}

	\subsection{DSS Field Trials}
	Apart from the literature surveyed above,  some leading technology companies have also contributed various products in the field of DSS. For example, Qualcomm introduced the Snapdragon X55 modem \cite{qualcommdssssss} to provide coverage for different generations from 2G to 5G. This modem also has the ability to provide DSS technology to facilitate coexistence between 4G and 5G. In addition, Ericsson supports DSS technology with its own solution called Ericsson Spectrum Sharing (ESS) \cite{ericess}. ESS software can run on any of the five million 5G-ready radios Ericsson has delivered since 2015. Swisscom and Telstra plan to upgrade their networks with ESS. Furthermore, Telstra has launched its commercial 5G network and has now rolled out 5G coverage in different cities with the help of Ericsson. Ooredoo and Play are also interested in benefiting from Ericsson ESS. Nokia is another pioneer in the field of telecommunications solutions that announced a roadmap for DSS technology. The company also presented a document examining how different generations coexist and how radio resources are timed \cite{nokiadssyes}.

	\section{LTE-Unlicensed and Licensed Assisted Access}\label{LTE_LAA}
	\subsection{Overview}
	\begin{figure}[!t] 
		\centering
		\includegraphics[width=.48\textwidth]{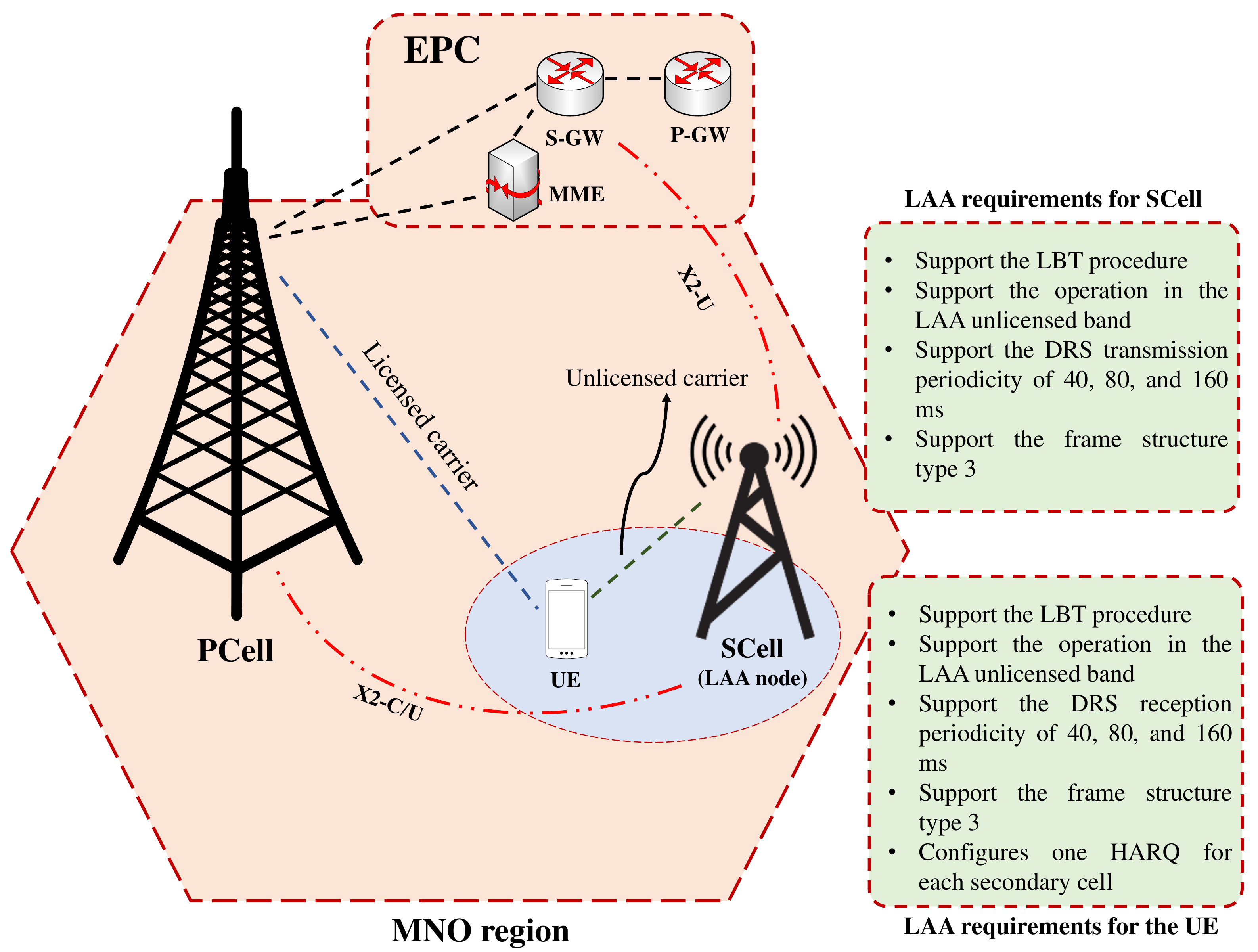}
		\caption{Technical overview of the LAA architecture\cite{Survey_LAA}.} 
		\label{LAA_architecture}
	\end{figure}
	LTE-U and LAA technologies were developed to extend the LTE operation to the unlicensed 5 GHz band. Aggregating the unlicensed bands provided by the secondary cells (SCells) with the legacy licensed bands can provide high-speed and seamless broadband multimedia services. During the transmission, the licensed or primary component carrier (PCC) is combined with several unlicensed or Secondary component carriers (SCCs) (configuration information can be conveyed via PCC to dynamically remove or add SCCs) through CA, which leads to a significant increase in user data rates \cite{zhang2015lte} (see Fig. \ref{LAA_architecture}). 
	
	A key challenge for these technologies is that they must adopt mechanisms to coexist fairly with other technologies in the unlicensed spectrum (e.g., Wi-Fi. In this way forward, LTE-U and LAA apply Carrier Sense Adaptive Transmission (CSAT) and LBT procedures, respectively \cite{lteuforum1, chen2016coexistence}. The principal characteristic of LTE-U is its duty-cycling nature when accessing a channel for transmissions. According to the CSAT procedure, the SCell can adjust the duty-cycle pattern, which is the percentage of time the SCell is in the ON state. When the duty cycle is in ON state, LTE-U will access the channel and act as a legacy LTE network, irrespective of the presence of Wi-Fi transmissions. Therefore, the probability of collisions with co-deployed Wi-Fi networks will increase, and this is the primary concern raised against the LTE-U technology. In other words, LTE-U does not coexist fairly with other technologies Furthermore, LTE-U technology can only be applied in countries such as the United States, Korea, and China, where there is no regulatory limitation for such an approach. Regulations in other countries (Europe, Japan, and India) require the LBT mechanism to ensure fair sharing \cite{bojovic2019evaluating}.
	
\begin{figure*}[!t] 
		\centering
		\includegraphics[width=0.95\textwidth]{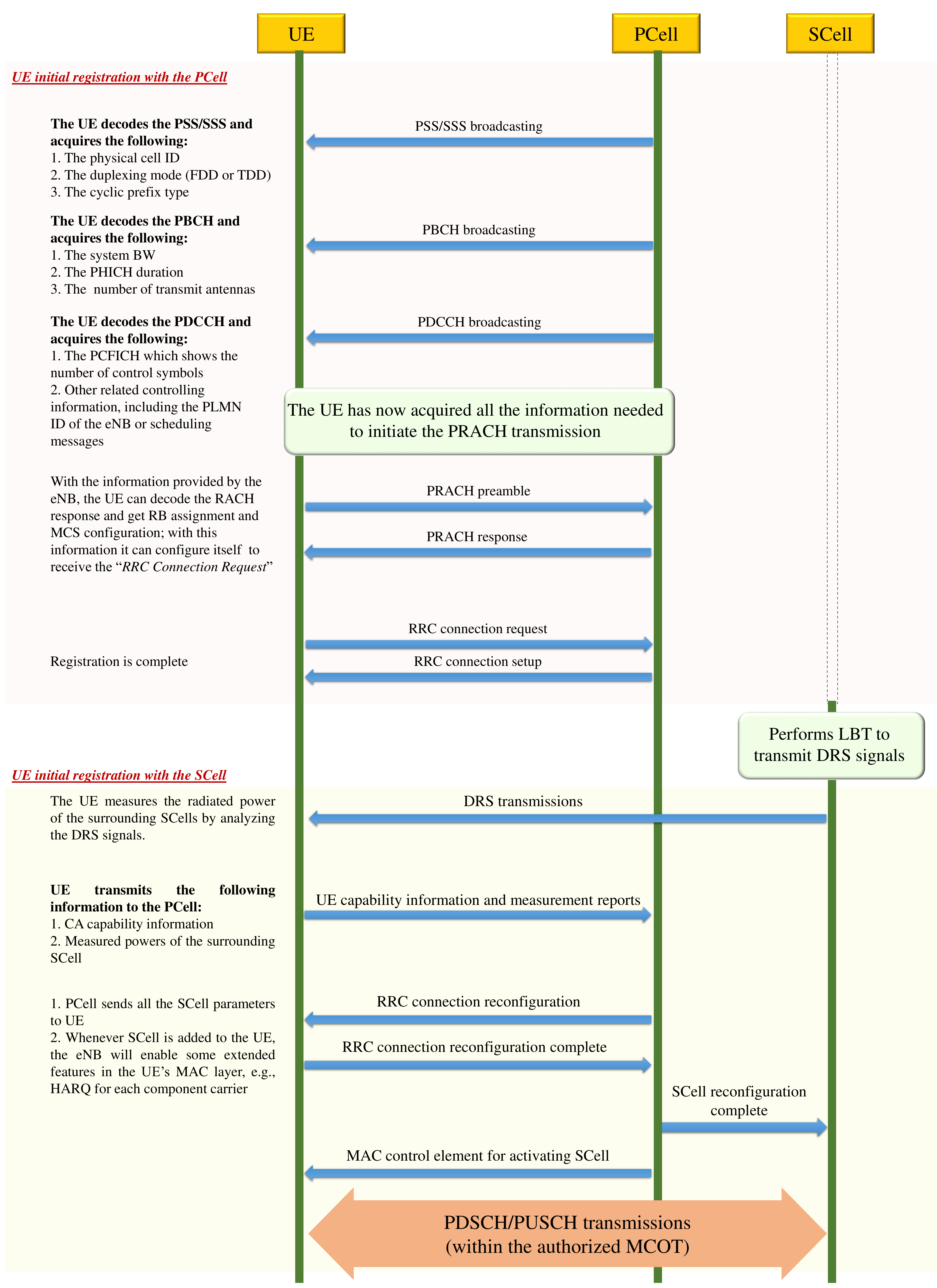}
		\caption{LAA message flow.} 
		\label{LAA_flow}
	\end{figure*}

	LAA is the 3GPP standardized mechanism developed for extending LTE to the unlicensed bands. LAA's first standardization came with Release 13 and later its enhanced version (eLAA) was introduced in 3GPP Release 14 to support uplink operation. LAA combines the licensed and unlicensed frequency bands (5 GHz) to increase the data rate through CA. The advantages of this method can be summarized as follows:
	\begin{itemize}
		\item Increases data rates using a combination of licensed and unlicensed channels.
		\item Releases licensed bandwidth capacity by distributing traffic between licensed and unlicensed bands.
		\item Uses the 5 GHz band more efficiently by facilitating the fair coexistence of Wi-Fi and LTE with an LBT procedure.
		\item Improves internal coverage by exploiting the 5 GHz frequency bands.
	\end{itemize}
	LAA is an LTE booster that aims to improve capacity and efficiency over currently licensed LTE carriers. LAA is not a standalone system; it needs a primary carrier from the LTE licensed spectrum to operate. In essence, the PCCs are used for receiving the measurement reports handling radio resource control (RRC) and non-access stratum procedures. The SCCs are used as supplementary UL or DL carriers for PDSCH or physical uplink shared channel (PUSCH) transmissions.
	Wi-Fi signals also use the 5 GHz frequency band, so algorithms that can facilitate their coexistence are needed. This is possible with LBT \cite{Survey_LAA}. Through this mechanism, the LTE-LAA eNB first senses the environment to detect whether there is an ongoing operation. If the band is idle, the eNB initiates transmissions. Furthermore, frame structure type 3 is introduced to support the partial subframe transmissions, which must be supported by the BS and the user. More details about the standard specifications and implementation features are provided below. For more information regarding the message flow of steps and interactions between different parts of the LAA framework, refer to Fig \ref{LAA_flow}.
	
	\subsection{Standardization}
	
	\begin{table*}[!t]
		\centering
		\caption{3GPP standardization works in the field of LTE-U \& LAA.}
		\begin{tabular}{p{0.75cm} p{2cm} p{14cm}}
			\toprule
			\textbf{Ref.} & \textbf{Standard} & \textbf{Subject}  \\
			\midrule
			\cite{3gpp_36_808}      & 3GPP TR 36.808               &  Evolved Universal Terrestrial Radio Access (E-UTRA); Carrier Aggregation; 
			Base Station (BS) Radio Transmission and Reception\\\midrule
			\cite{3gpp_36_823} 	    & 3GPP TR 36.823               &  Evolved Universal Terrestrial Radio Access (E-UTRA); Carrier aggregation enhancements; User Equipment
			(UE) and Base Station (BS) radio transmission and reception \\\midrule
			\cite{3gpp_36_913}      & 3GPP TR 36.913               &  Requirements for further advancements for Evolved Universal Terrestrial Radio Access (E-UTRA) (LTEAdvanced)\\\midrule
			\cite{3gpp_36_300}      & 3GPP TS 36.300               & Evolved Universal Terrestrial Radio Access (E-UTRA) and Evolved Universal 
			Terrestrial Radio Access Network (E-UTRAN); Overall description; Stage 2\\\midrule
			\cite{3gpp_36_889}      & 3GPP TR 36.889               & Study on Licensed-Assisted Access to Unlicensed Spectrum\\\midrule
			\cite{3gpp_36_213}      & 3GPP TS 36.213               & Evolved Universal Terrestrial Radio Access (E-UTRA);
			Physical layer procedures\\\midrule
			\cite{3gpp_36_211}      & 3GPP TS 36.211 			   & Technical Specification Group Radio Access Network;
			Evolved Universal Terrestrial Radio Access (E-UTRA);
			Physical channels and modulation\\\midrule
			\cite{3gpp_36_789}      & 3GPP TR 36.789               & Evolved Universal Terrestrial Radio Access (E-UTRA);
			Multi-node tests for Licence-Assisted Access (LAA)\\\midrule
			\cite{3gpp_37_213}      & 3GPP TS 37.213			   & Physical Layer Procedures for Shared Spectrum Channel Access\\\midrule
			\cite{3gpp_36_104}      & 3GPP TS 36.104			   & Base Station (BS) Radio Transmission and Reception\\\midrule
			\cite{3gpp_36_141}      & 3GPP TS 36.141			   & Base Station (BS) conformance testing\\\midrule
			\cite{3gpp_36_133}      & 3GPP TS 36.133			   & Requirements for support of radio resource management\\\bottomrule
		\end{tabular}
		\label{LAA_table}
	\end{table*}
	The specifications and structure of LTE-U technology, including the CSAT procedure, were determined by the LTE-U Forum in \cite{lteuforum1,lteuforum2}. The scientific study of LAA was also thoroughly described in 3GPP Release 13 TR 36.889 standard \cite{3gpp_36_889} for Supplemental DownLink (SDL) in the unlicensed band. According to the design, LTE-LAA in the unlicensed spectrum leverages the LTE CA protocols stated by 3GPP in \cite{3gpp_36_808,3gpp_36_823,3gpp_36_913,3gpp_36_300} with one carrier serving as the PCell and others are serving as SCells. Due to the fundamental differences between LTE and Wi-Fi Medium Access Control (MAC) layer protocols, the LBT coexistence algorithm was  investigated in \cite{3gpp_36_889}. This algorithm is a contention-based collision-avoidance mechanism. Four LBT design options were examined in \cite{3gpp_36_889}. The first design was without LBT. The second used LBT without a random backoff, where time duration set for channel sensing was deterministic. The third design used LBT with a random backoff in a contention window of a fixed size. Finally, the fourth design used LBT with a random backoff in a contention window of variable size. The fourth design option was shown to allow for better coexistence with Wi-Fi due to the configuration flexibility of its parameters. According to \cite{3gpp_37_213}, using LBT in 5-GHz unlicensed bands was shown to be necessary;  however, they also showed that the licensed band could optionally be included for signaling in the uplink. 
	Furthermore, to support partial subframe transmissions, a new frame structure (i.e., Type 3) was introduced by 3GPP in Release 13 \cite{3gpp_36_211}, which facilitates  transmissions at  slot boundaries and applies to LAA SCell operation with normal cyclic prefix. Coexistence evaluation plans and multi-node tests involving two Rel-13 LAA BSs or one Rel-13 LAA BS and one other wireless system (e.g., IEEE 802.11) to ensure  acceptable system performance in the unlicensed band was studied in \cite{3gpp_36_789}.
	
	3GPP has further outlined the channel access procedure and LBT mechanism details for both uplink and downlink in TS 36.213 \cite{3gpp_36_213}. In this technical specification, they outlined channel access procedures for transmissions, including discovery signal transmission along with contention window adjustment. They also delineated the channel access process for transmissions on multiple carriers as well as methods for energy detection threshold adaptation.  In \cite{3gpp_36_104} , they specified the channel access parameters for downlink operation in Band 46 (described in \cite{3gpp_36_213}) as well as the CA of component carriers in different operating bands. Chapter 9 of \cite{3gpp_36_141} specified the eNB conformance testing operations for LBT. This procedure is used to evaluate the accuracy of the energy detection threshold, Maximum Channel Occupancy Time (MCOT), which refers to the maximum time a device can utilize the channel, and minimum idle time under normal operating conditions for LAA following the specifications stated in \cite{3gpp_36_213} and \cite{3gpp_36_104}. In \cite{3gpp_36_133}, they specified requirements for support of radio resource management for LAA under frame structure 3.
	
	In Release 14 \cite{3gpp_36_213}, the enhanced version of LAA (eLAA) was introduced that promoted LAA capabilities for uplink transmission while keeping the downlink the same as that of LAA. This specification defined two modes (i.e., Type 1 and Type 2) for for uplink channel access, which are similar to the ones used by the eNodeB. Furthermore, new features and enhancements are currently being researched in the Further Enhanced LAA (feLAA) Study Item \cite{FeLAA}. 
	
	\subsection{Physical Layer Aspects and Implementation}
	\subsubsection{LTE-U}
	Two modes of operation have been examined for LTE-U's: SDL and time-division duplex (TDD). In SDL mode, the unlicensed spectrum is used only for downlink transmission. This mode is suitable for data-hungry downlink operations. The TDD mode incorporates both the downlink and uplink operations, similar to the LTE TDD system in licensed bands. 
	
	As discussed above, based on the CSAT procedure, the eNB can regulate the duty-cycle pattern, which consists of alternating $ T_{ON} $ and $ T_{OFF} $ periods. This flexibility can further improve the LTE-U operation and facilitate a fair ecosystem between different technologies in the unlicensed band. CSAT is an adaptive muting algorithm by which the eNB periodically senses the channel during the OFF period. Depending on the measurement, it can then readjust the duty-cycle pattern. The LTE-U Forum does not specify any algorithm according to which the duty cycle is established, and it is left to vendors to choose their own adaptation algorithm. The CSAT duty cycle can vary over time according to the network traffic and channel usage. However, for proper operation, the following limitations and constraints have been defined \cite{lteuforum1,Extending_LTE}:
	\begin{itemize}
		\item The minimum off-state period in which the eNB stops any type of transmission has to be 1 ms.
		\item The maximum off-state period is determined through the LDS periodicity. The LDS is a similar message to the discovery reference signals (DRSs), which provides the UEs with the SCell tracking and measurements and informs them of the eNB time and frequency adjustment \cite{lteuforum1}. The LDS is comprised of CRS/PSS/SSS and is broadcasted by the LTE-U eNB with periodicities of 40, 80, or 160 ms only at subframe 5 occasions.
		\item When there is data to transmit, the minimum on-state period has to be 4 ms. 
		\item The maximum on-state period should be 20 ms.
		\item The energy detection threshold is -62 dBm.
	\end{itemize}

	\subsubsection{LAA}
	
	\begin{figure}[!t] 
		\centering
		\includegraphics[width=.4\textwidth]{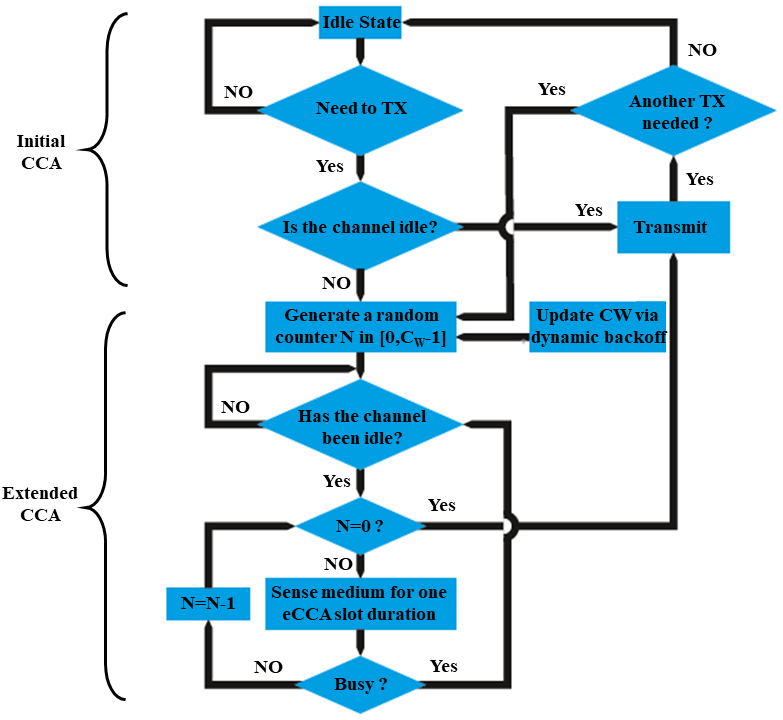}
		\caption{LBT procedure \cite{LTE-U_Systems}.} 
		\label{LBT procedure}
	\end{figure}
	
	\begin{figure}[!t] 
		\centering
		\includegraphics[width=.4\textwidth]{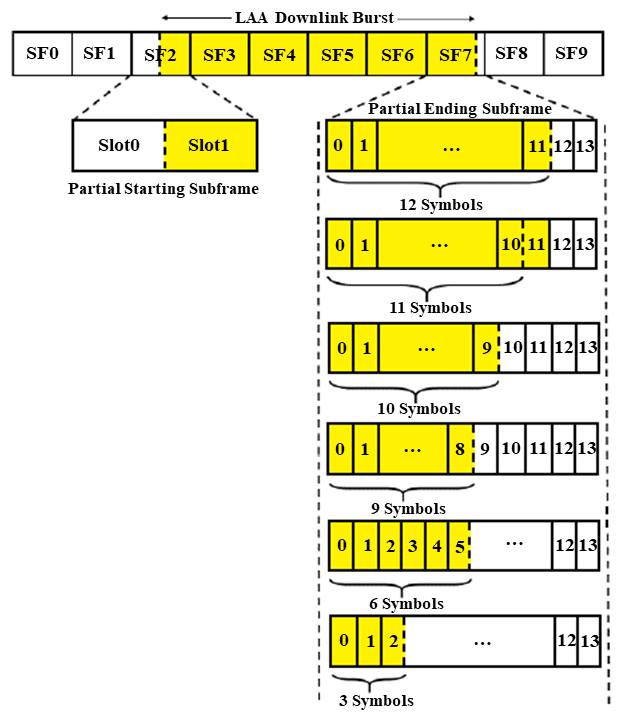}
		\caption{LTE-LAA frame structure type 3  \cite{LTE-LAA}.} 
		\label{LAA_frame}
	\end{figure}
	\begin{figure}[!t] 
		\centering
		\includegraphics[width=.4\textwidth]{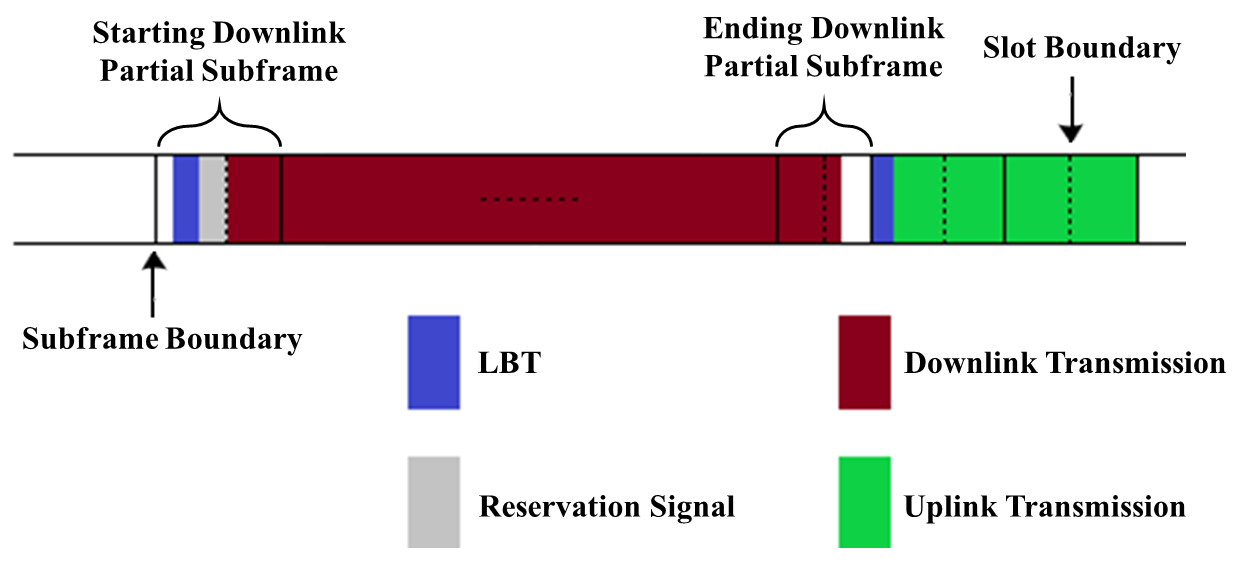}
		\caption{Frame structure type 3  \cite{LTE-LAA}.} 
		\label{Frame_type3}
	\end{figure}

	In this section, the protocol structure for implementing the LAA method is stated. The secondary network, LTE, needs to use remote radio head (RRH) and other equipment to operate in the 5 GHz frequency band for implementing the LAA method. Network kernel, security, authentication, and other service requirements for LAA are similar to conventional LTE technology. Therefore, there is no need to change the core of the network. However, in the access network, the eNB needs to be changed in such a way as to ensure the possibility of exploiting the unlicensed frequency spectrum as well as the conditions for optimal use of the unlicensed frequency. It is also necessary to update the user side devices so that they can use the unlicensed frequency spectrum based on LTE protocols. 
	
%
		
\begin{figure*}[!t] 
		\centering
		\includegraphics[width=.9\textwidth]{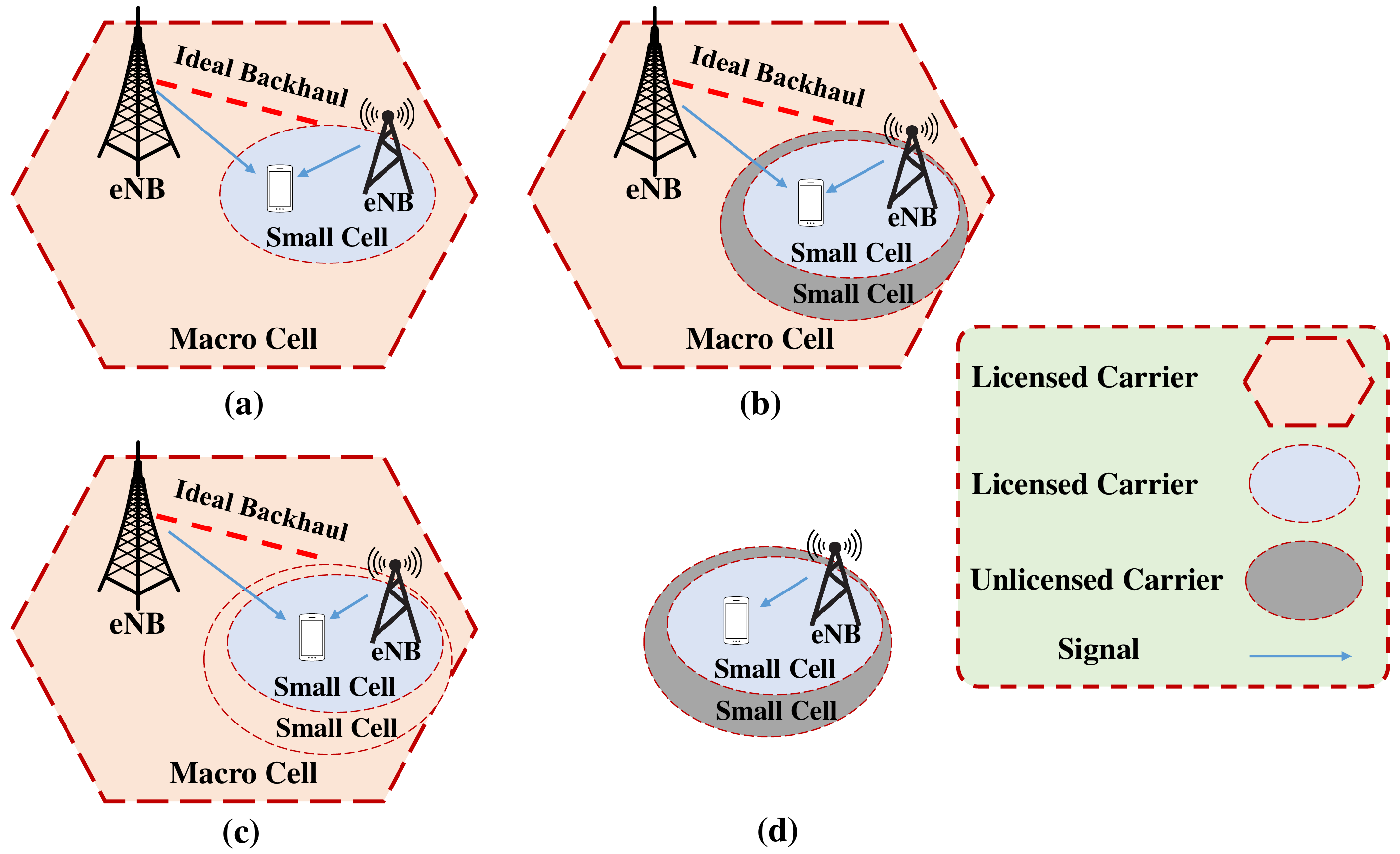}
		\caption{LAA deployment scenarios \cite{3gpp_36889}.} 
		\label{LAA_deployment}
	\end{figure*}

	There are four possible deployment scenarios for LAA \cite{3gpp_36_889}. LAA uses CA, where one or more low power small cells operate in unlicensed spectrum as shown in Fig. \ref{LAA_deployment}. The deployment scenarios include schemes with and without macro coverage, and both outdoor and indoor small cell deployments between licensed and unlicensed carriers.

	An unlicensed small cell only operates in the framework of CA through ideal backhaul with a licensed cell, even though the backhaul between small cells can be non-ideal. By contrast, when CA is performed within a small cell, the backhaul between macro and small cells can be ideal or non-ideal.
	\begin{itemize}
		\item Fig. \ref{LAA_deployment} (a),  In this scenario, the licensed carrier is associated with a macro cell, while the unlicensed carrier is associated with a small cell. The licensed carrier exchanges the RRC signal and data rate. A secondary carrier from an unlicensed band could be employed for transmitting data if a UE is covered by a small cell.
		\item Fig. \ref{LAA_deployment} (b): This scenario illustrates the usage of different licensed carriers for macro and small cells. Licensed macro cell with CA between licensed small cell and unlicensed small cell.
		\item Fig. \ref{LAA_deployment} (c): This scenario shows licensed macro and small cells with CA operating between licensed and unlicensed small cells. The same licensed carrier for both macro and small cells requires synchronization and scheduling.
		\item Fig. \ref{LAA_deployment} (d): In this scenario, we see the CA operating between licensed and unlicensed small cells without macro cell coverage.
	\end{itemize}

	LAA requires modifications to transmitters and receivers. These modifications need to be applied to the physical layer in the physical layer of the BS  and on the user's side. At the base station, it requires category 4, support for the 5 GHz band, frame structure type 3, PSS/SSS in every 5 ms, CRS in every subframe. On the user side, it requires support for the 5 GHz band and frame structure type 3. The chief propellent behind LAA's superior operation is the LBT procedure, which facilitates fair coexistence between LTE and other technologies in the band.  LBT requirements can be categorized as follows \cite{Amdocs}:
	\begin{itemize}
		\item Transmit power control (TPC):
		a regulatory requirement in some regions where transmitting devices should be able to limit their transmit power.
		\item DRS, frequency/time estimation and RRM measurements:
		for RRM measurements and effective reception of information on the unlicensed band, RRM and CSI measurements, including channel and interference measurements, are needed.
		\item Bandwidth support:
		support for a system bandwidth of at least 20 MHz in the 5 GHz band.
		\item Dynamic frequency selection:
		for certain frequency bands, dynamic frequency selection (DFS) is a regulatory necessity; DFS detects and protects radar systems by stopping transmission within a certain period of time and not transmitting again for at least a certain amount of time.
		\item Discontinuous transmission:
		since channel availability cannot always be assured in unlicensed spectrum, duty cycles must be maintained.
	\end{itemize}
	
	\begin{table*}[!t]
		\centering
		\caption{LAA channel access priority classes \cite{3gpp_36_213}.}
		\begin{tabular}{p{4cm} p{1cm} p{1.2cm} p{1.2cm} p{1.6cm} p{3.5cm}}
			\toprule[1.2pt]
			{Channel Access Priority Class (p)} & {$ m_p $} & {$ CW_{\text{min},p} $} & {$ CW_{\text{max},p} $} & {$ T_{\text{mcot},p} $} & {Allowed $ CW_p $} sizes  \\
			\midrule
			1 & 1 & 3 & 7 & $2 \mathrm{~ms}$ & 3,7 \\
			2 & 1 & 7 & 15 & $3 \mathrm{~ms}$ & 7,15 \\
			3 & 3 & 15 & 63 & 8 or $10 \mathrm{~ms}$ & $15,31,63$ \\
			4 & 7 & 15 & 1023 & 8 or $10 \mathrm{~ms}$ & $15,31,63,127,255,511,1023$ \\
			\bottomrule
		\end{tabular}
		\label{laa_priority}
	\end{table*}
	\begin{table*}[!t]
		\centering
		\caption{eLAA channel access priority classes \cite{3gpp_36_213}.}
		\begin{tabular}{p{4cm} p{1cm} p{1.2cm} p{1.2cm} p{1.6cm} p{3.5cm}}
			\toprule[1.2pt]
			{Channel Access Priority Class (p)} & {$ m_p $} & {$ CW_{\text{min},p} $} & {$ CW_{\text{max},p} $} & {$ T_{\text{mcot},p} $} & {Allowed $ CW_p $} sizes  \\
			\midrule
			1 & 2 & 3 & 7 & $2 \mathrm{~ms}$ & 3,7 \\
			2 & 2 & 7 & 15 & $4 \mathrm{~ms}$ & 7,15 \\
			3 & 3 & 15 & 1023 & 6 or $10 \mathrm{~ms}$ & $15,31,63,127,255,511,1023$ \\
			4 & 7 & 15 & 1023 & 6 or $10 \mathrm{~ms}$ & $15,31,63,127,255,511,1023$ \\
			\bottomrule
		\end{tabular}
		\label{elaa_priority}
	\end{table*}
	
	A concise description of the LBT procedure, which is shown in Fig. \ref{LBT procedure}, is as follows: 
	Before initiating any transmissions, the eNB must sense that the channel is idle for an initial defer time period ($ T_{\text{defer}} $). The defer time is composed of a duration $ T_\text{f} = 16 \mathrm{~\mu s} $, immediately followed by $ m_\text{p} $ consecutive slot durations of size $ T_{\text{sl}} = 9 \mathrm{~\mu s} $, as stated in Section 15.1.1 of \cite{3gpp_36_213}. The value of $ m_\text{p} $ is determined through the channel access priority class, as specified in Table \ref{laa_priority}. For example, $ T_{\text{defer}} $ for the priority class 4 is $ 79 \mathrm{~\mu s} $. The channel will be considered idle if the detected power in the band is lower than the energy detection threshold $ E_{\text{th}} $ for at least $ 4 \mathrm{~\mu s} $ (The procedure by which the $ E_{\text{th}} $ quantity is determined is described in Section \ref{laaedt}). 
	If the channel is sensed to be idle during the $ T_{\text{defer}} $ time, then the eNB can start the transmission. Otherwise, the eNB initiates the eCCA during which the eNB chooses a random number $ N $ that is uniformly distributed between 0 and $ CW_p $. $ CW_p $ is the contention window size, that falls within the range of $ CW_{\text{min}}<CW_p<CW_{\text{max}} $, and is determined by the channel access priority class. Before every transmission, $ CW_p $ is set to $ CW_{\text{min}} $. During the eCCA, the eNB will first sense the channel for eCCA defer period, $ T_{\text{defer}} $, and once the channel is free for $ T_{\text{defer}} $, the eNB must register the channel as idle for the duration of an additional period of ($ N + 1 $) times the $ 9 \mathrm{~\mu s} $ slot duration. If the eNB determines that the channel is occupied, the backoff counter will freeze, and the eNB will have to sense the channel for a duration of $ T_{\text{defer}} $. 
	The counter $ N $ is affected by the HARQ process. If more than 80\% of all transmissions in reference subframe $ k $ are detected to be NACK/DTX\footnote{DTX refers to the situation when a UE misses the control message containing scheduling information (i.e., PDCCH).}, $ CW_p $ will be increased to the next possible value (see Table \ref{laa_priority}). 
	If the channel is found to be idle during all the ($ N + 1 $) slot durations and ($ T_{\text{defer}} $), then the eNB will occupy the channel for a maximum duration equivalent to the MCOT. For $ p = 3 $ and $ p = 4 $, $ T_{\text{mcot},p} = 10 \mathrm{~ms} $ if it can be guaranteed that no other technologies are sharing the carrier (otherwise, $ T_{\text{mcot},p} = 8 \mathrm{~ms} $). 
	
	For the case of transmitting a DRS, the eNB will sense the channel for a period of 25 $\mathrm{~\mu s}$, and when the channel is sensed to be idle, the eNodeB can transmit the DRS for a maximum period of 1 ms. DRS consists of the first five subframes of a radio frame and includes PSS, SSS, and CRS signals. Depending on the configuration, it may contain a channel state information reference signal (CSI-RS) and be transmitted with a periodicity of 40, 80, or 160 ms. DRS allows a small cell to quickly transition from the off-state to the on-state and transmit a low-duty cycle signal for RRM purposes. The UE is configured with a DMTC, which is a time window of 6 $ ms $ in duration within which the UE can expect the DRS to be received from the eNB. Since the DRS transmissions are subject to LBT, the fixed periodicity is relaxed for LAA to make it compatible with the LBT requirement. In \cite{3gpp_36_889}, two alternative options were considered for DRS design for LAA:
	\begin{itemize}
		\item Subject to LBT, DRS can be transmitted in a fixed time position within the configured DMTC,
		\item Subject to LBT, DRS is allowed to be transmitted in at least one of the other time positions within the configured DMTC.
	\end{itemize}
	
	Two channel access types were defined for eLAA. The type 1 channel access procedure is identical to the procedure for LAA; however, there are different channel access priority classes defined for the uplink, as specified in Table \ref{elaa_priority}. The type 2  procedure is similar to transmitting DRS in the downlink, where, after  sensing the channel for 25 µs, the device can start its PUSCH transmission \cite{3gpp_36_213}. Whenever the UE wants to initiate a UL transmission, it has to first perform the scheduling request (SR) transmission on the PCell and wait for a grant reception from the eNB. The LBT procedure has to be completed by the eNB prior to its access to the channel for successful UL grant transmission to the UE. The primary concern is that as the eNB schedules data transmission some subframes ahead of time, the eNB does not know whether the channel will be free and accessible for UE transmission. Therefore, it is up to the UE to sense the channel before beginning data transmission in the assigned subframe. If the shared COT between the eNB and UE does not surpass the maximum COT, the UE will transmit UL data after a successful 25 µs LBT. Otherwise, the UE should succeed in category 4 LBT first and if the channel is not free, the UE will avoid transmission. In this situation, the eNB has to reschedule this transmission using either the primary or secondary carrier. Two principal concerns arise here. First, the UE sustains long delays and unnecessary halts before it can attempt to access the channel. Second, two LBTs need to be performed before the UE can transmit. The first LBT is performed by the eNB for the transmission of the UL grant to the UE; the second LBT is performed by the UE associated with the UL grant \cite{kim2020uplink, karaki2017uplink}.
	
	Legacy LTE transmissions start at a subframe boundary. Therefore, other neighboring networks with no such limitation, such as Wi-Fi, can initiate their transmissions anytime. In this case, the LAA-eNB will have to wait until the next subframe boundary to begin transmitting after performing a successful LBT procedure. To cope with this complication, LAA reserves the channel by transmitting reservation signals after a successful LBT. However, reservation signals present unfairness for Wi-Fi and decrease spectral efficiency. Accordingly, a new LTE frame type (i.e., frame structure type 3) was introduced to assist unlicensed spectrum operations with normal cyclic prefixes only. It consists of 10 subframes of 1 ms duration each as shown in Fig. \ref{LAA_frame}. These 10 subframes may be used for uplink or downlink transmission or left blank as illustrated in Fig. \ref{Frame_type3}.
	
	LAA transmission can begin and end at any subframe, and the burst can involve one or more consecutive subframes. LAA downlink transmission can end at any of the downlink pilot time slot (DwPTS) symbols or at the subframe boundary. As a result, the last subframe can be fully filled with 14 OFDM symbols or it can be made up of any DwPTS length symbol. Since one of LAA's aims is to ensure fair coexistence with Wi-Fi \cite{Extending_LTE}, the eNB must first sense the spectrum before transmitting, and it only transmits if the channel is clear. Besides, within frame structure type 3, no PBCH is transmitted \cite{LTE-LAA}.
	
	To report the success or failure of decoding a transmission, LAA uses the HARQ method. The UE sends an acknowledgment (ACK) or negative acknowledgment (NACK) over the licensed band, and the scheduler assigns a retransmission protocol that is vendor-dependent but typically has a higher scheduling priority. The CW size is also doubled if the NACKs for a given reference subframe surpass a certain threshold; otherwise they are reset \cite{Revisiting_LTE}.
	\subsubsection{Energy Detection Procedure in LAA}\label{laaedt}
	Based on Section 15.1.4 of \cite{3gpp_36_213}, the value of $ \mathbb{E}_{\text{th}} $ is determined as follows:
	for the case where the absence of any other technology sharing the carrier can be guaranteed on a long-term basis (as in the case of regulations in certain regions), then this quantity would be
	\begin{equation}
	\mathbb{E}_{\text{th}_{\text{max}}}=\min \left\{\begin{array}{l}
	T_{\text{max}}+10 {~\text{dB}}, \\
	X_{\text{r}},
	\end{array}\right.
	\end{equation}
	where $ X_{\text{r}} $ is the maximum energy detection threshold defined by regulatory requirements in dBm. For the case where there are multiple technologies in the band, $ \mathbb{E}_{\text{th}} $ quantity would be
	
	\begin{equation}
	\begin{aligned}
	& \mathbb{E}_{\text{th}_{\text{max}}}=\max \left\{\begin{array}{l}
	-72 + 10\log_{10}({BW}/20) ~\text{dBm}, \\
	\mathcal{A},
	\end{array}\right.\\
	&  \resizebox{.9\hsize}{!}{$\mathcal{A}=\min \left\{\begin{array}{l}
		T_{\text{max}}, \\
		T_{\text{max}}-T_{\text{A}}+\left(P_\text{H}+10\log_{10}({BW}/20)-P_{\text{T}{x}}\right),
		\end{array}\right.$}
	\end{aligned}
	\end{equation}
	where the bandwidth is in the order of MHz, $ T_\text{A} $ equals $ 10~ \text{dB} $ for transmissions that include PDSCH, and $ 5~ \text{dB} $ for transmissions that include DRS and controlling information. $ P_\text{H} $ is equal to $ 23~ \text{dBm} $, and $ P_{\text{Tx}} $ is the maximum eNB output power in dBm, which the eNB uses over a single carrier irrespective of single or multi-carrier transmission. Finally, $ T_{\text{max}} = 10\log_{10}(3.16228\cdot 10^{-8} \times BW) $.

	\begin{figure}[!t] 
		\centering
		\includegraphics[width=.48\textwidth]{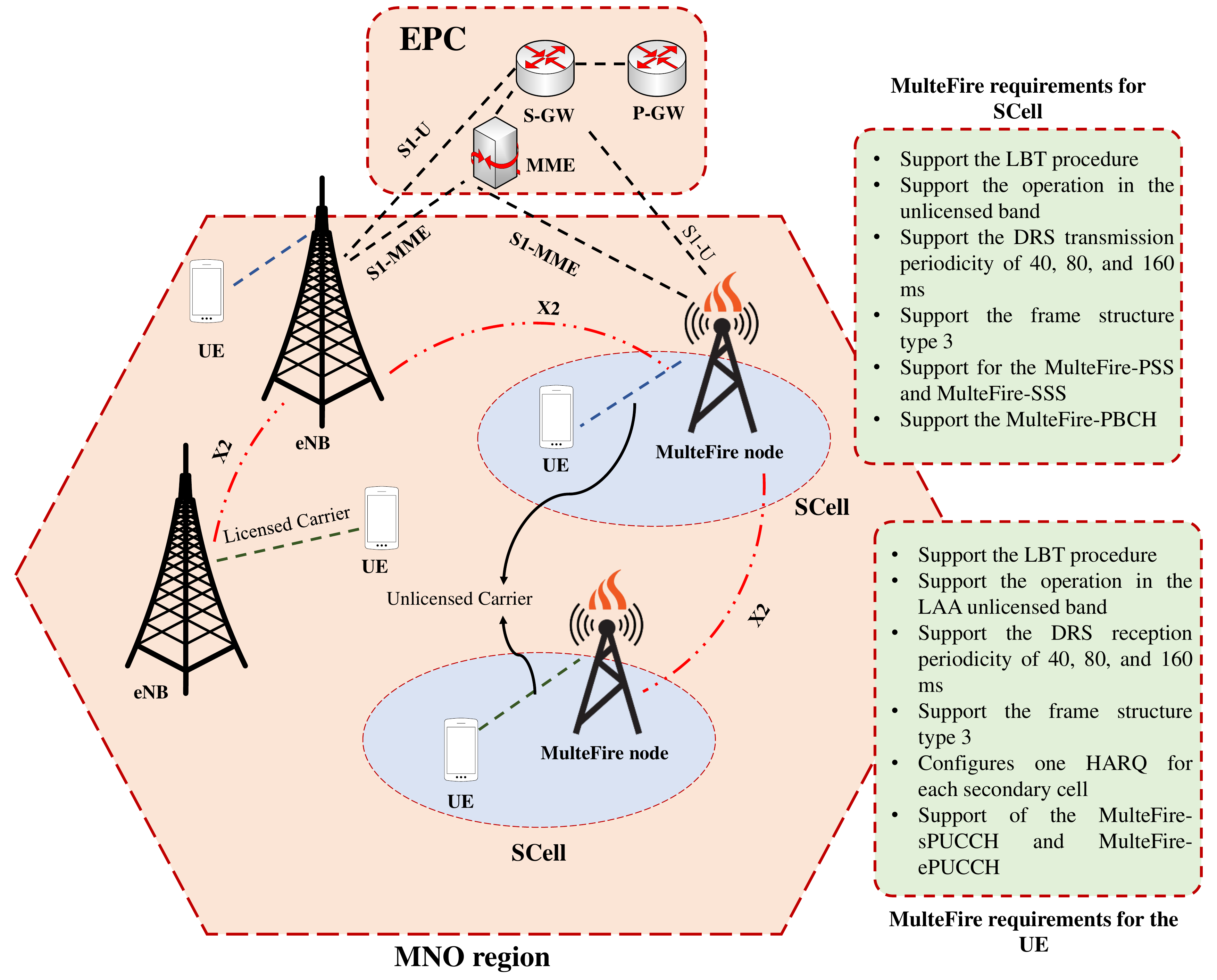}
		\caption{Technical overview of the MulteFire architecture.} 
		\label{multefire_architecture}
	\end{figure}

%

	\subsection{LAA Field Trials}
	Various companies have been combing through the solutions to integrate the LAA sharing into their frameworks. Samsung introduced the Exynos 9 Series (9810) mobile processor’s gigabit LTE modem \cite{samsunglaatrial}. Thanks to the adoption of LAA (6CA) for downlink and eLAA (2CA) for uplink, Exynos 9 can reach up to 1.2 Gbps downlink speeds and 200 Mbps uplink speeds. Also, Huawei and NTT DOCOMO conducted a live demonstration of the co-channel coexistence between LAA and Wi-Fi systems with a  Small Cell \cite{huaweilaatrial}. The results showed that with the assistance of the LBT coexistence mechanism, LAA was able to co-exist fairly with the neighboring Wi-Fi system, while sustaining most of the performance advantages of LTE. Qualcomm in cooperation with SK Telecom also launched an eLAA over-the-air trial \cite{sktelelaatrial}. In this trial, a 2.6 GHz LTE was utilized as the licensed band and a 5 GHz band as the unlicensed band. The results showed that LTE could double the capacity in the unlicensed band while sharing the spectrum fairly with Wi-Fi using the LBT. T-Mobile and Ericsson also announced that they were able to reach speeds of over 1 Gbps on the unlicensed spectrum by using 12-layer LAA \cite{tmobieerricsonlaatrial}. Similarly, AT\&T and Ericsson managed to achieve wireless speeds of more than 750 Mbps by employing LAA \cite{attaalatrial}. Nokia together with T-Mobile reported data rates of up to 1.3 Gbps using the LAA with five-component CA on 14 antenna layers \cite{nokiatraillaa}.

	\section{MulteFire}\label{multefire}
	\subsection{Overview}
	\begin{figure*}[!t]
		\centering     
		\subfigure[PLMN Access Mode Reference Architecture]{\label{plmn_architecture}\includegraphics[width=.49\textwidth]{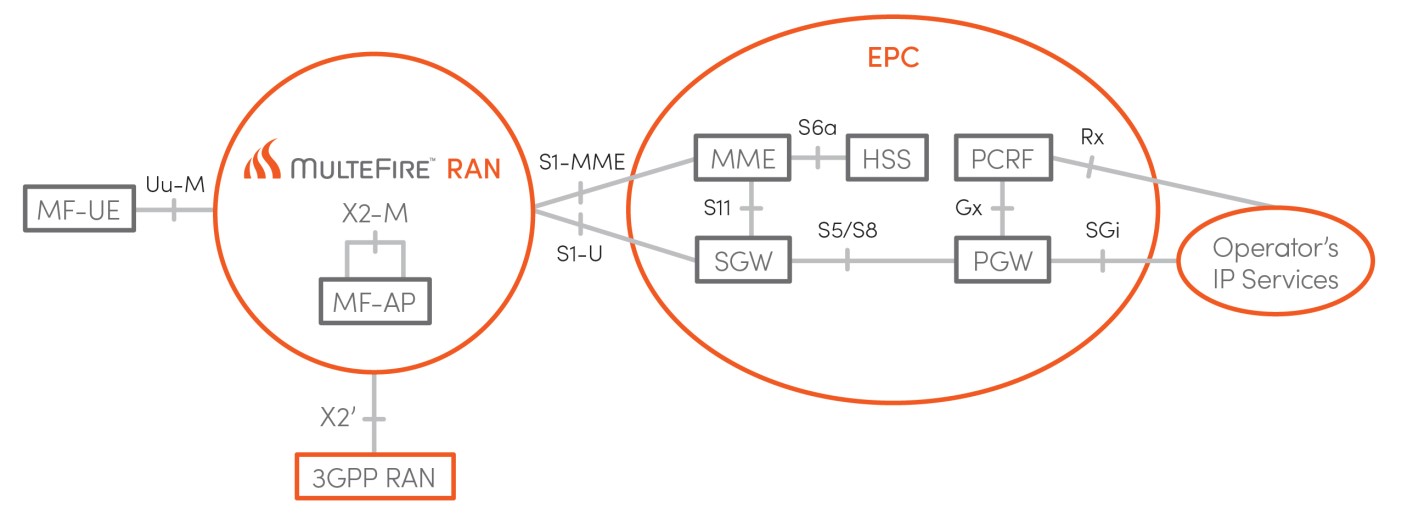}}
		\hspace{3em}
		\subfigure[NHN Access Mode Reference Architecture]{\label{nhn_architecture}\includegraphics[width=.37\textwidth]{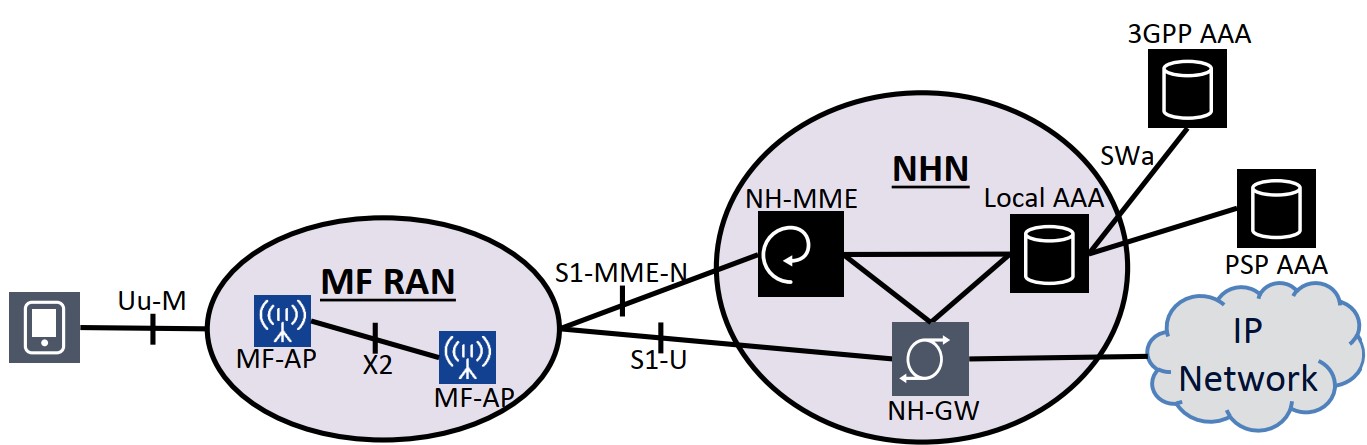}}
		\caption{MulteFire reference architectures \cite{multefire_alliance}.}
		\vspace*{-\baselineskip}
	\end{figure*}
	MulteFire is a new radio access technology built on LTE specifications and developed to facilitate the standalone operation of LTE in the unlicensed band without any dependency on authorized spectrum. The MuLTEfire Alliance is the organization formed to establish and develop the structure of MuLTEfire by defining technology specifications, hardware testing, software design, and evaluating interoperability \cite{multefire_alliance}. MulteFire standardization is completely consistent with the 3GPP standards and follows the specifications of 3GPP Release 13 and 3GPP Release 14 designations for LAA and eLAA. Similar to other spectrum sharing frameworks in the unlicensed band, MulteFire applies the LBT procedure to provide fair coexistence with other spectrum users, such as Wi-Fi or LAA.
	The rationale behind the success of MulteFire can be summed up in the following points:
	\begin{itemize}
		\item MulteFire, unlike the LAA, operates solely in the unlicensed band, without the need for additional regulatory approval, costly spectrum, or specialist expertise, and this can add more flexibility in deployment.
		\item MulteFire allows the private and public vertical venues, businesses, and property owners to deploy, install and operate their own private or neutral host MulteFire network.
		\item MulteFire facilitates the widespread adoption of small cells, especially indoors, and enables neutral host deployments on a much larger scale.
		\item MulteFire promotes a wide range of services (e.g. voice over LTE [VoLTE], IoT) and creates new business opportunities that allow different verticals to profit from LTE technology such as sports and entertainment, public venues (malls, airports), transportation applications, and hospitality.
	\end{itemize}
	More details about the publications on standardization and implementation features are provided in the following sections.
	\subsection{Standardization}
	MuLTEfire Release 1.0 specification \cite{multefire_alliance} stipulates that LTE operates in the unlicensed band while assuring appropriate sharing of spectrum with other users by removing the requirement for an anchor in licensed spectrum. Furthermore, new optimizations for IoT, additional spectrum bands, including MulteFire 1.9 GHz, and further improvements to Release 1.0 were considered in MulteFire 1.1 \cite{multefire_alliance_1}. A summary of MulteFire was presented in \cite{rosa2018standalone}, which elaborated on the principal challenges due to LBT and the standalone operation of MuLTEfire, while proposing solutions to overcome such impediments.
	\subsection{Physical Layer Aspects and Implementation}
	Based on MulteFire 1.0 specification \cite{multefire_alliance}, to enable various deployments and cover multiple services, two reference architectures have been developed for MulteFire,
	\begin{itemize}
		\item \textbf{Public Land Mobile Network (PLMN) Access Mode}: 
		The MulteFire cell is connected to 3GPP EPC as an additional radio access network (RAN) for the PLMN in a similar manner as an evolved universal terrestrial radio access network (E-UTRAN a.k.a. LTE RAN) is connected to an EPC, Fig. \ref{plmn_architecture}.
		\item \textbf{Neutral Host Network (NHN) Access Mode}: The MulteFire cell is connected to an NHN, which is a self-contained network deployment providing IP services, Fig. \ref{nhn_architecture}.
	\end{itemize}
	A MulteFire cell may also support PLMN access mode for specific PLMNs and NHN access mode for a specific NHN, simultaneously.


	\subsubsection{PLMN Access Mode	Architecture}
	Fig. \ref{plmn_architecture} shows the reference architecture model for MulteFire PLMN access mode. Accordingly, the MulteFire RAN is connected to a 3GPP EPC via an S1 interface (S1-U and S1-MME), and its functionality is quite similar to the functionality of E-UTRAN in LTE. From the EPC’s point of view, the MulteFire access point (AP) plays the role of the LTE eNB. Seamless mobility is established between MulteFire and 3GPP RAN. Intra-MulteFire RAN mobility and mobility from a MulteFire RAN to a 3GPP RAN are supported in the RRC connected mode and idle mode. Mobility can also be supported from a 3GPP RAN to a MulteFire RAN; however, this requires MulteFire support in 3GPP RANs. Based on the standardization, the PLMN access mode has two use cases. The MulteFire can be used as a supplementary RAN connected to an existing core network to increase the MNO's network efficiency and coverage. This could be used in situations where the licensed spectrum is not available or to improve the overall capacity. In addition, an operator can deploy an EPC with a MulteFire RAN without using any licensed band.

	\subsubsection{NHN Access Mode Architecture} 
	Fig. \ref{nhn_architecture} depicts the reference architecture for the NHN access mode. A MulteFire cell is connected to an NHN. The NHN communicates with external participating service providers (PSPs) to facilitate services for their subscribers. PSPs are logically separate from the administrator of the NHN and provide subscription, billing, and the associated authentication, authorization, and accounting (AAA) services. PSPs can be either 3GPP PLMNs (e.g., MNOs) that employ 3GPP specified AAA servers that interact with the NHN using the Non-3GPP SWa or STa access reference points, or they can also be other entities using generic AAA servers that interact with the NHN using generic diameter or radius signaling. The functions of the newly introduced logical entities are as follows: 
	
	NH-MME and NH-GW provide similar functionalities for NHN as the MME and combined SGW/PGW provides in LTE. 
	
	The local AAA proxy is a single contact point for external AAA servers.
	
	\subsubsection{MulteFire Radio Air Interface}
	In order to provide a robust anchor carrier, MulteFire has introduced enhancements to DRS, which was first presented in LTE Release 12. Due to the non-guaranteed channel availability caused by LBT, the synchronization process between the UE and the MulteFire AP to acquire the system timing and other system information has to be fast and effective. In response, compared to LTE, the synchronization signals for MulteFire are slightly modified.
	
	\begin{figure*}[!t] 
		\centering
		\includegraphics[width=.95\textwidth]{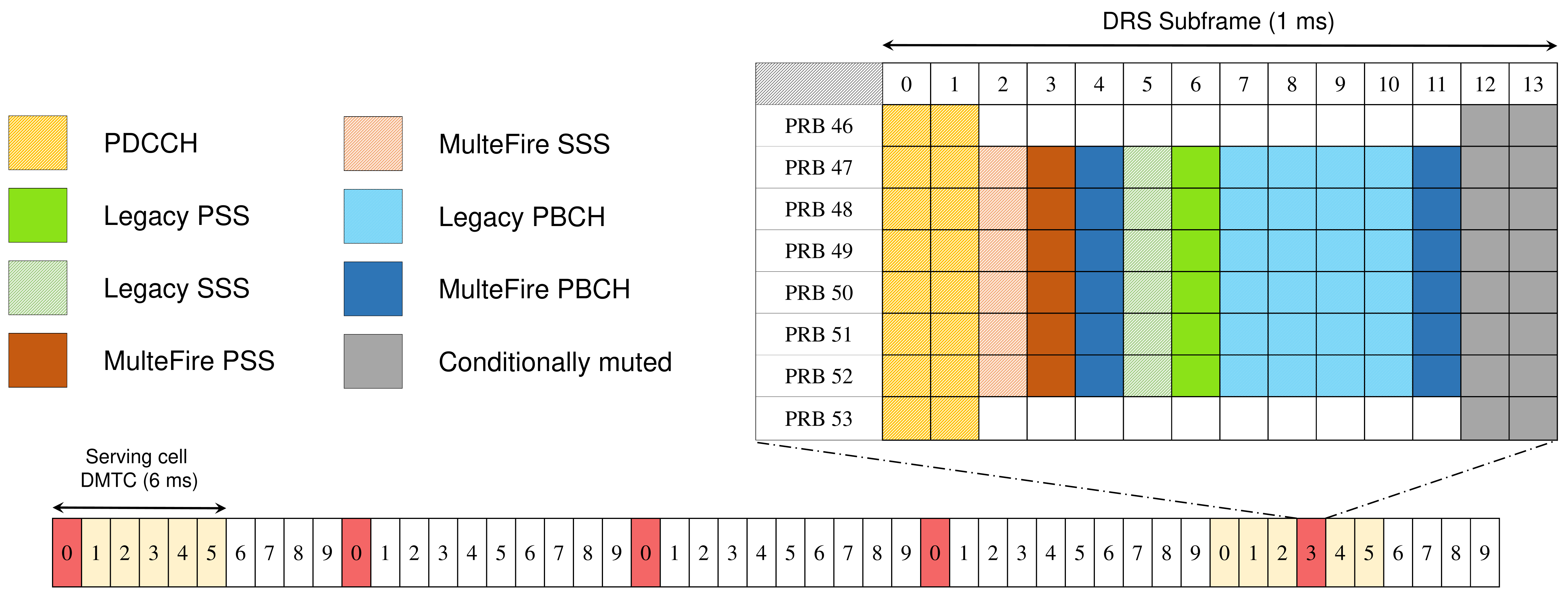}
		\caption{MulteFire physical layer design: serving cell DMTC, synchronization signals, and broadcast channel\cite{multefire_alliance}.} 
		\label{multefire_frame}
	\end{figure*}
	Unlike the frame structure employed by Rel-13 LAA, where the PSS and the SSS were transmitted on the seventh and sixth OFDM symbols, MulteFire also includes a MulteFire primary synchronization signal (MulteFire-PSS) and a MulteFire secondary synchronization signal (MulteFireSSS), which are transmitted on the fourth and third OFDM symbols of a DRS subframe, as illustrated in Fig. \ref{multefire_frame}. The MulteFire-PSS and MulteFire-SSS facilitate fast timing and frequency synchronization and cell search. The MulteFire-PSS sequence also leads to a supplementary processing gain and robustness during the initial acquisition, which allows the UE to identify MulteFire eNB from an LAA eNB, and can enhance the reliability of RRM measurements \cite{multefire_alliance}. 
	The MulteFire physical broadcast channel (MF-PBCH), which carries the master information block (MIB), is also transmitted during the DRS subframe \cite{rosa2018standalone}.  In Rel-13 LAA, this information is conveyed over the primary licensed carrier. However, in MulteFire, the transmission of PBCH may not be guaranteed as it performs on an unlicensed band. In order to address this issue, MulteFire enhanced PBCH (MulteFire-PBCH) is used to transmit the enhanced master information block (MIB-MulteFire) on six OFDM symbols, namely symbols 4, 7, 8, 9, 10 and 11, in comparison with Rel-8 PBCH, which uses four OFDM symbols.
	DRS transmissions within a DRS measurement timing configuration (DMTC) are conditional on successful LBT of 25 $\mu$sec. To overcome this, the MulteFire AP can transmit the DRS subframe during a network configurable time window, which is repeated periodically every 40, 80, or 160 ms, and it may be configured up to 10 ms in length, during which UEs expect to receive DRS transmissions. 
	Furthermore, to cope with the potential DRS transmission-blocking due to LBT, the network can transmit the DRS signal in any subframe within the DMTC occasion. This means that the DRS transmission is not strictly periodic; however, it leads to a better probability of successful DRS transmission. 
	In addition to DMTC transmissions, the opportunistic transmission of DRS outside the DMTC is only allowed on SF0 if the eNB successfully clears eCCA on those subframes.

	As mentioned earlier, MulteFire follows the LBT policy for channel access procedures. Typically, DL and UL data transmissions are based on Category 4 LBT with an energy detection threshold similar to LAA and eLAA. After successfully decoding the system information, the UE executes the random access procedure, which mimics the same four-step procedure as in LTE. Still, the LBT has to be performed before every transmission. MulteFire also supports the UL transmission without LBT within 16 μs of the preceding successful DL transmission. The UE and the network can proceed with the exchange of information in the unlicensed band, after which the RRC connection is authenticated successfully.

	In the physical layer, MulteFire adopts a very flexible TDD based frame structure to dynamically conform to any DL and UL traffic fluctuations. In essence, unlike the conventional LTE-TDD, any subframe can be either UL or DL without a predefined designation of DL or UL direction to any subframe. The eNB determines the subframe's identity through the common physical downlink control channel (C-PDCCH). Similar to LTE-TDD, the DL-UL switching is based on a special subframe that includes the following: i) a DL partial ending subframe; ii) a guard period for DL-UL switching; iii) an uplink LBT; and iv) a shortened UL subframe consisting of four OFDM symbols. In order to conform to the ETSI regulatory requirements in the 5 GHz band, block-interleaved frequency-division multiple access (B-IFDMA) has been adopted as MulteFire's baseline uplink transmission scheme, where one carrier is divided into N interlaces (N = 5 for 10 MHz, and N = 10 for 20 MHz carrier), and each interlace incorporates M equally spaced physical RBs (M = 10 for both 10 MHz and 20 MHz carriers) \cite{multefire_alliance}.
	
	\begin{figure}[!b] 
		\centering
		\includegraphics[width=0.48\textwidth]{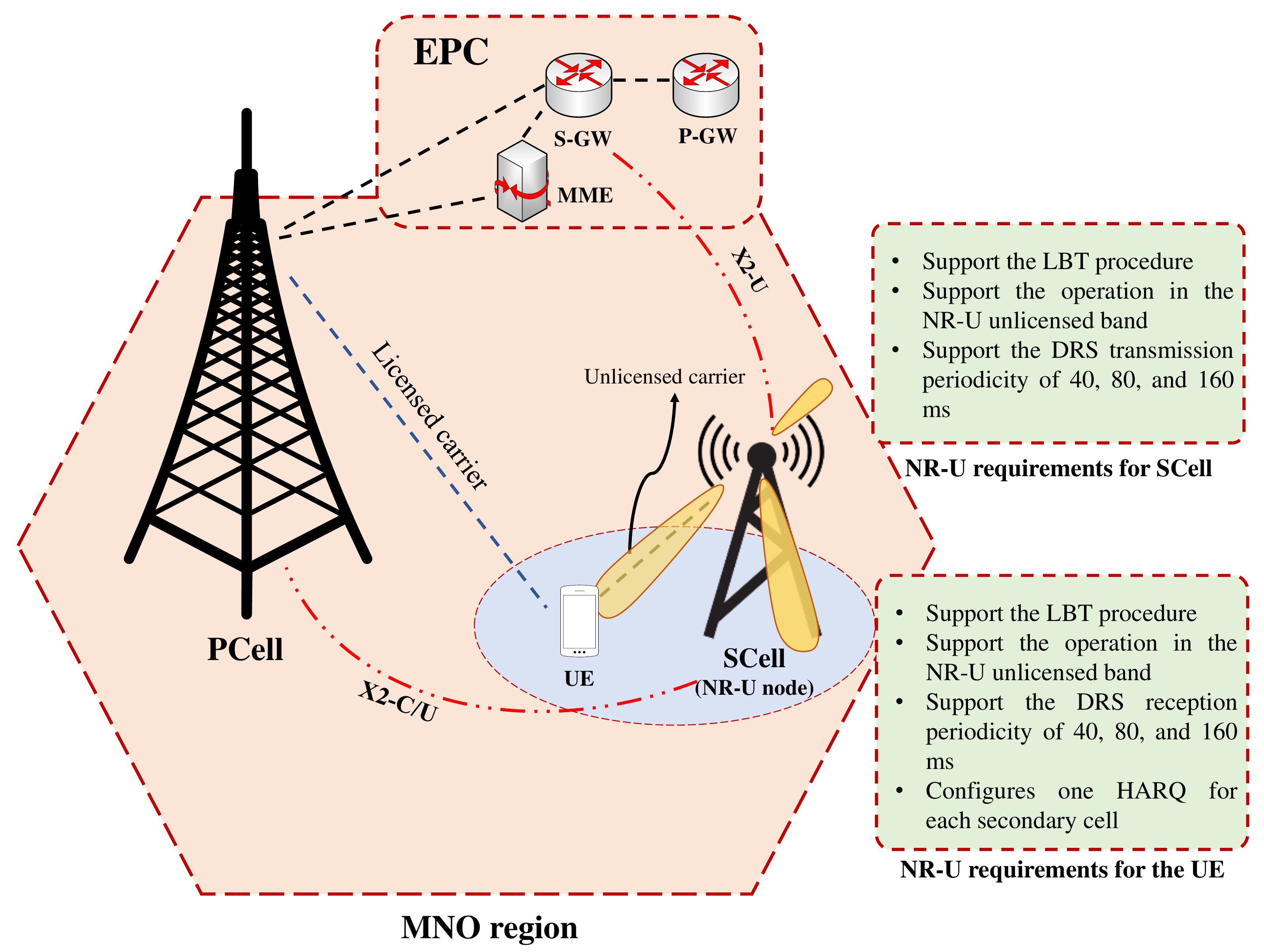}
		\caption{Technical overview of the NR-U architecture.} 
		\label{NRUecosystem}
	\end{figure}
	\begin{figure*}[!t] 
		\centering
		\includegraphics[width=1\textwidth]{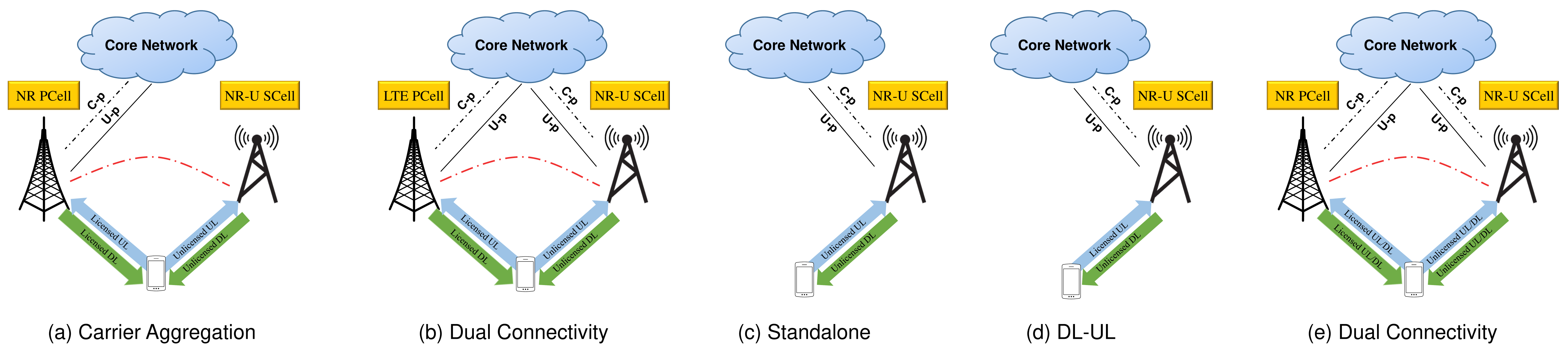}
		\caption{NR-U deployment scenarios.} 
		\label{NRUscenarios}
	\end{figure*}
	Since MulteFire was designed to operate in the unlicensed band, all transmissions are subject to LBT, and this necessitated a partial redesign of the legacy LTE PUSCH for the transmission of UL data. One of these modifications is the introduction of the MulteFire short PUCCH (MF-sPUCCH) which is transmitted over four OFDM symbols in the UL subframe where the other 10 symbols of the subframe are not used for uplink operation. The UE transmits MF-sPUCCH during the last four OFDM symbols of a DL subframe in the gap between DL and UL transmission without performing LBT. According to the ETSI regulations in the unlicensed band, the UE is allowed to operate within 16 $ \mathrm{~\mu s} $ after the DL transmission to carry small control information such as reception acknowledgments (ACK and NACK). However, compared to the LTE PUCCH, the MulteFire-sPUCCH format maintains a limited payload for the transmission of uplink control information. As a result, MulteFire Extended PUCCH (MulteFire ePUCCH) has also been introduced, which can span an entire subframe (1 ms) and can be multiplexed with PUSCH to transmit large control information such as the CSI \cite{labib2017extending}.
	\section{New-Radio Unlicensed}\label{nr_u}
	\subsection{Overview}
	Similar to what was planned for LTE in unlicensed bands, the idea of allowing NR to operate in unlicensed bands was introduced by the 3GPP in a study item of NR Rel-16 \cite{nrubefore,3gpp_38_889}. The principal goal of NR-U is to extend the applicability of NR to unlicensed bands and employ a pattern that provides fair coexistence across different radio access technologies (RATs). Unlike the CA-based deployment of LAA and LTE-U and the standalone operation of MuleFire in the 5 GHz band, NR-U supports various deployment modes and technologies, including CA, DC, and standalone operation in unlicensed spectrum bands. Furthermore, NR-U can operate in various bands: 2.4 GHz (unlicensed worldwide), 3.5 GHz (shared in the USA, also designated to be accessible by SAS), 5 GHz (unlicensed worldwide, accessible through LAA, LTE-U, MulteFire), 6 GHz (unlicensed in the USA and Europe), 37 GHz (shared in the USA), and 60 GHz (unlicensed worldwide). The 3GPP has further classified these frequency bands into low-frequency bands below 7 GHz and high-frequency mmWave bands at 60 GHz, incorporating the 37 and 60 GHz bands.

	NR-U leverages beam-based transmissions, which enhances spatial reuse but complicates interference management. The directional nature of beams in NR exacerbates the hidden node and exposed node problems in the unlicensed bands \cite{subramanian2010addressing}, \cite{lagen2018lbt}, and therefore, the coexistence framework with other RATs in the unlicensed bands becomes more challenging. NR-U, like LAA and MulteFire, adheres to LBT requirements to access a channel. However, in directional transmissions, LBT performance may diminish. In essence, when omnidirectional LBT (omniLBT) is used to sense activity in a channel at the same time as a beam-based transmission is employed it can lead to exposed node problems. One solution is to apply directional LBT (dirLBT) in which the direction of the intended communication is known. However, this solution may lead to hidden node problems \cite{huaweihisiliconnru}. This phenomenon, known as the omniLBT/dirLBT trade-off, has been detailed in \cite{lagen2018listen}. The authors there showed that for low network densities, dirLBT functions better than omniLBT, while for high network densities, omniLBT is a better solution. By considering the impact of narrow beam transmissions in NR-U, we will outline the NR-U scenarios and LBT procedures designed to cope with the aforementioned challenges in the following sections.
		
	\subsection{Standardization}
	The standardization of NR-U began in a study item of NR Rel-16 \cite{nrubefore,3gpp_38_889}, which targeted the unlicensed national information infrastructure (UNII) bands at 5 GHz and 6 GHz \cite{rpran20193gpp}. Future specifications will extend NR-U functionality to unlicensed mmWave spectrum bands at 60 GHz in later releases (i.e., NR Rel-17 and beyond). Other enhancements, such as multichannel operation, frequency reuse, and improvements to the initial access procedure, have been included in \cite{3gpp_38_889}. 
	
	\begin{figure*}[!t] 
		\centering
		\includegraphics[width=.95\textwidth]{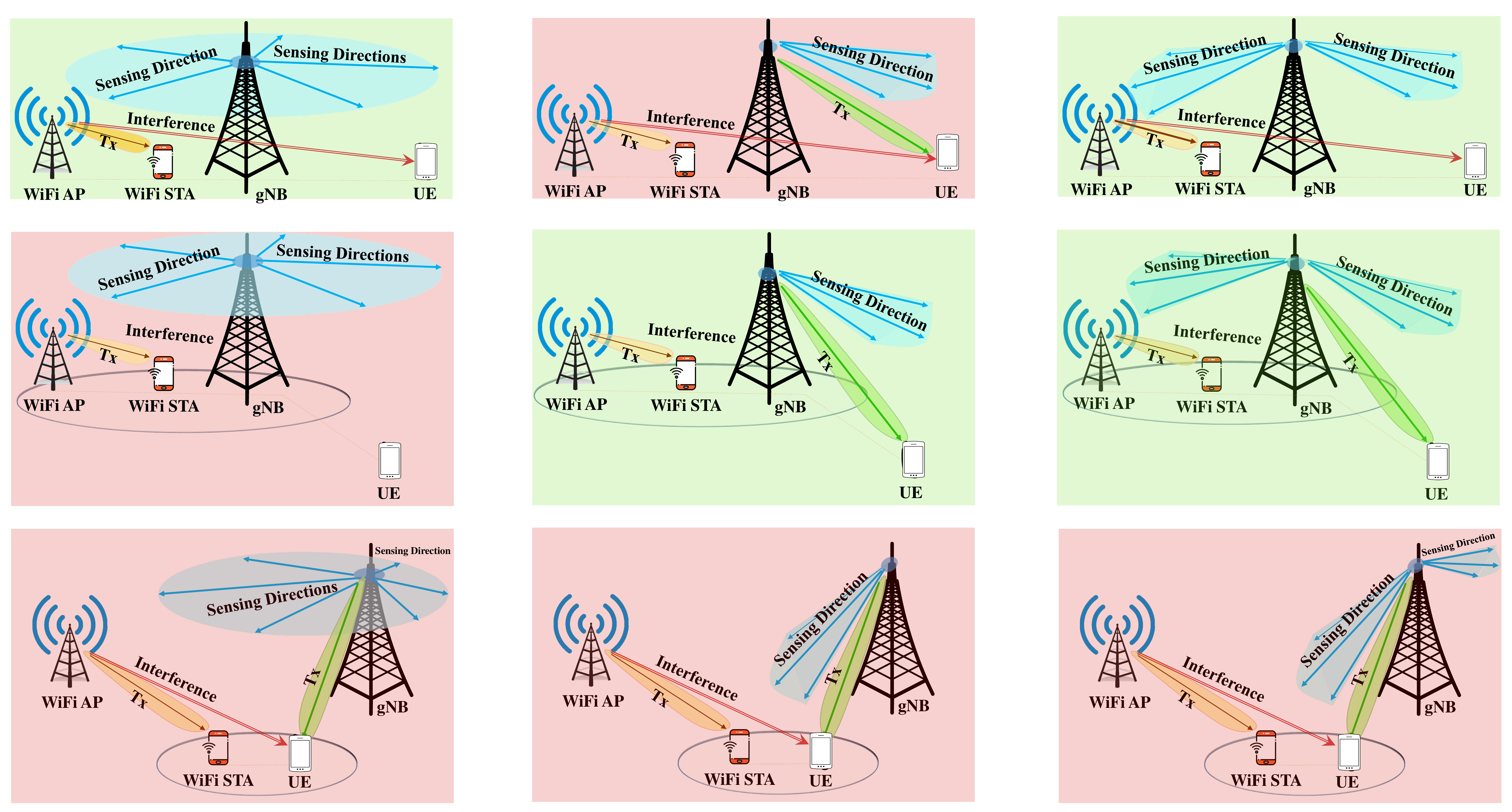}
		\caption{Behavior of (a) omniLBT, (b) dirLBT, and (c) pairLBT techniques, assuming beam-based transmissions, for fully-aligned (top), aligned transmitters (middle), and aligned receivers (bottom) deployment configurations \cite{lagen2019new}.} 
		\label{NRUlbtscenarios}
	\end{figure*}
	
	ETSI regulation has harmonized the specifications for the 5 GHz and 60 GHz bands in \cite{ETSI_EN_301_893} and \cite{ETSI_EN_302_567}, respectively, and any technology, such as NR-U or LAA, that attempts to access these unlicensed bands has to fulfill these regulatory requirements. One of these conditions is LBT. 
	The specifications for this LBT procedure were described in \cite{ETSI_EN_301_893} and \cite{ETSI_EN_302_567}. In essence, the CCA slot duration is 5 µs in the 60 GHz band and 9 µs in the 5 GHz band, respectively. The energy detection threshold that determines the activity of other signals on the 5 GHz and 60 GHz bands are set to -72 dBm and -47 dBm, respectively. The other prerequisite is related to the MCOT, which refers to the maximum time a device can utilize the channel. The MCOT in the 60 GHz band is restricted to 9 ms, and in the 5 GHz band is set to 2 ms, 4 ms, or 6 ms, although it may be extended up to 8--10 ms in some circumstances. Moreover, 5 GHz and 60 GHz bands can share the COT with connected UEs, which means that the UEs can skip the CCA examination and directly transmit in response to the received signals from the gNB (i.e., Cat1 LBT) if there is a gap in between DL and UL transmissions of less than 16 µs. 
	When the gap is between 16 µs and 25 µs, only a short sensing (i.e., Cat 2 LBT) is required at the UEs. Otherwise, when the gap is greater than 25 µs, regular LBT (i.e., Cat 4 LBT for data) must be applied at responding devices. Besides, NR-U supports multiple DL/UL switching points within the COT.
	The third condition pertains to occupied channel bandwidth (OCB), which is bandwidth that contains 99\% of the signal power and mandates the unlicensed technologies to utilize a significant portion of the channel bandwidth when they access the channel. According to ETSI specifications, for the 5 GHz band, the OCB must be between 70\% and 100\% \cite{ETSI_EN_301_893}, and in the 60 GHz band, it must be between 80\% and 100\% of the NCB \cite{ETSI_EN_302_567}. Other requirements include dynamic frequency selection (DFS) and frequency reuse (FR), which allows the same frequency channel to be reused by different devices simultaneously (see \cite{ETSI_EN_302_567,ETSI_EN_301_893,lagen2019new} for more details).
	\subsection{Physical Layer Aspects and Implementation}

	In this section, we elaborate more on the implementation scenarios and solutions to the aforementioned challenges related to NR-U described in \cite{3gpp_38_889}.	
	Based on the deployment and propagation environment conditions, four layout scenarios have been proposed for NR-U:
	\begin{itemize}
		\item indoor sub 7 GHz; 
		\item indoor mmWave;
		\item outdoor sub 7 GHz; 
		\item outdoor mmWave.
	\end{itemize} 
	Furthermore, the following deployment scenarios have also been defined for NR-U:
	\begin{itemize}
		\item CA between licensed band NR and unlicensed band NR-U (similar to LAA);
		\item DC between LTE in the licensed band and NR-U in the unlicensed band;
		\item standalone NR-U (similar to MulteFire);
		\item NR with DL in the unlicensed band and UL in the licensed band;
		\item DC between licensed and unlicensed band NR-U.
	\end{itemize}
	These deployment scenarios, which are shown in Fig. \ref{NRUscenarios}, can be applied to each of the NR-U layout scenarios described earlier. 
	
	\begin{figure*}[!t] 
		\centering
		\includegraphics[width=1\textwidth]{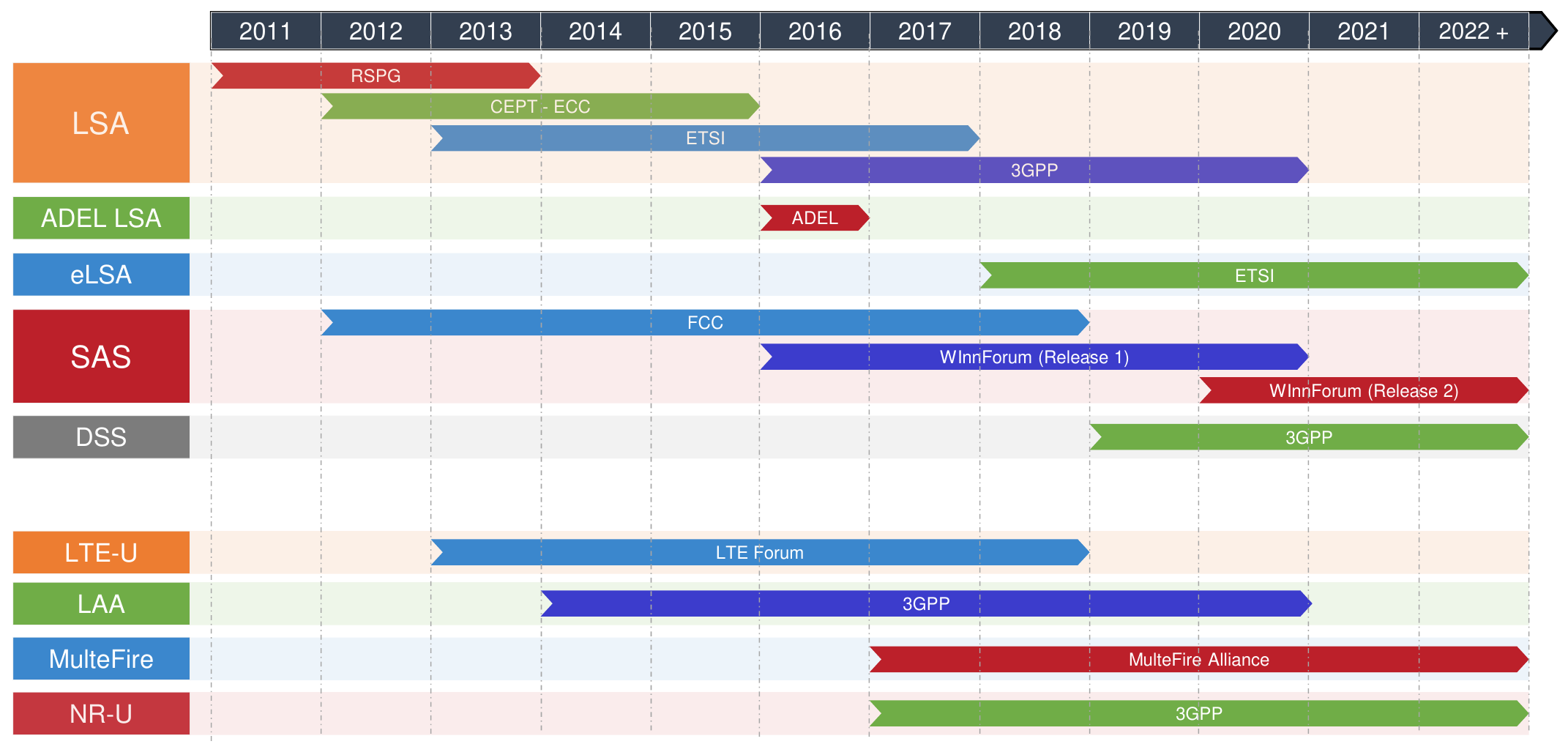}
		\caption{Standardization timeline of spectrum sharing methods.} 
		\label{standardtimeline}
	\end{figure*}
	
	Despite the lower interference and latency due to the beam-based transmissions provided by NR, further flexibility and improvements are required to facilitate the worldwide deployment of NR-U, as it is anticipated to operate in more frequency bands compared to LAA and MulteFire technologies. 
	Below, we examine three design principles that have been addressed in technical standardization publications or proposed in various scientific articles.
	\subsubsection{LBT in Beam-Based Transmissions}
	LBT is a critical steps before accessing the channel; however, as mentioned earlier, under beam-based transmissions, LBT's performance may degrade and can lead to hidden node and exposed node problems. Two solutions have been proposed to address this impediment and guarantee fair coexistence, namely the omniLBT and dirLBT \cite{huaweihisiliconnru}.

	A comprehensive analysis of omniLBT/dirLBT performance is shown in Fig. \ref{NRUlbtscenarios} for various scenarios. In the fully-aligned scenario, the AP, gNB, STA, and UE are arranged along the same spatial line; in aligned transmitters, the gNB is in the coverage area of the AP, and in aligned receivers, the UE is in the coverage area of the AP.
	As we can see in Fig. \ref{NRUlbtscenarios}, (a), top, under directional transmissions where the receivers and transmitters are fully aligned, the omniLBT procedure is advantageous and prevents interference on the UE side, which lies within the antenna boresight. However, in situations like Fig. \ref{NRUlbtscenarios}, (a), middle, where only the transmitters are aligned, the omniLBT leads to an unnecessary halt in the transmission process for gNB-UE, which could have reused the spectrum (known as an exposed node problem).
	By contrast, since dirLBT only senses the spatial direction in which the transmission will be performed, it will permit the activity of both the transmitters (see Fig. \ref{NRUlbtscenarios} [b], middle). Nevertheless, the Achilles heel of this approach is that it cannot recognize the ongoing activities which take place in the vicinity. Therefore, unlike omniLBT, it will allow transmission for the fully aligned scenario, which will subsequently lead to a hidden node problem (see Fig. \ref{NRUlbtscenarios} [b], top), which is not favorable.

	\cite{lagenpairlabt} addressed the omniLBT/dirLBT trade-off by introducing a distributed solution called pairLBT. In pairLBT, which is shown in Fig. \ref{NRUlbtscenarios} (c), the sensing is performed in paired directions which is similar to performing dirLBT in the transmitting and opposite direction. Fig. \ref{NRUlbtscenarios} (c) shows how applying this procedure can address both the hidden node and exposed node problems. However, the pairLBT procedure is only suitable for indoor scenarios; for outdoor scenarios, following \cite{lagenpairlabt} a new dimension (height) should be included in the definition and optimization. Based on the results in \cite{lagenpairlabt}, dirLBT performs better than omniLBT when the network density is low, while with higher densities omniLBT performs more efficiently.
	Another technique was also proposed in \cite{lagen2018lbt} to deal with the omniLBT/dirLBT trade-off, called the LBTswitch, which is a dynamic switching method between omniLBT and dirLBT.
	
	Both the pairLBT and LBTswitch methods can handle hidden node and exposed node problems (for both the fully aligned and aligned transmitters case). However, for the case of aligned receivers, since the gNB cannot discern activities near the receivers through any LBT-based schemes, hidden node problems are unavoidable (see Fig. \ref{NRUlbtscenarios}, bottom). Accordingly, employing techniques that can utilize information from the UEs (also known as receiver-assisted LBT procedures) might be a promising solution. One such method was proposed in \cite{mmMAGIC4_1} (Sec. 8.2.2), called listen after-talk (LAT). This method requires the UE to sense and detect any collision in the channel and give feedback to the transmitter about whether to stop or proceed with the transmission. However, LAT is not compliant with the legacy LBT procedure, since it does not impose any constraints on the gNB transmissions unless the gNB receives a collision report from the UE. Nevertheless, it is applicable to regions where there is no obligation for LBT, like the USA and China.
	
	Another possible solution is the listen-before-receive (LBR) \cite{lagen2018listen}, which is based on the physical carrier sense of RTS/CTS, in which the gNB triggers the UE to sense the channel, and only if the UE responds, the gNB can begin the transmission. The aforementioned mechanism referred to as receiver-assisted LBT can use any of the directional, omnidirectional, paired, or switching sensing strategies.  Nonetheless, receiver-assisted LBT imposes extra signaling overhead due to the additional message exchange between gNB and UE.

	\begin{table*}[!t]
		\setlength\lightrulewidth{0.051pt}
		\scriptsize
		\caption{Comparison of spectrum sharing methods.}
		\label{sharing_comparison}
		{\renewcommand{\arraystretch}{1.2}
			{}\begin{tabular}{p{1.5cm} p{1.5cm} p{1cm} p{2.3cm} p{2.2cm} p{2cm} p{1.5cm} p{1.2cm} c}
				\toprule[1.1pt]
				&	\textbf{Standardization body} & \textbf{Ecosystem} & \textbf{Deployment scenario} & \textbf{Operational bands} & \textbf{coexistence scheme} & \textbf{Deployment \newline possibility} & \textbf{architecture} & \textbf{sensing} \\
				\midrule
				\multirow{2}{*}{\textbf{LSA}} & RSPG, CEPT,\newline ETSI, 3GPP & Fig. \ref{LSA_archtecture} & standalone,\newline licensed bands & 2.3-2.4 GHz & database assisted & EU region & centralized & \xmark \\\midrule[0.5pt]
				\multirow{2}{*}{\textbf{ADEL LSA}} & ADEL & Fig. \ref{ADEL_LSA_fram} & standalone,\newline licensed bands & 2.3-2.4 GHz & database and \newline sensing assisted & EU region & centralized & \cmark \\\midrule[0.5pt]
				\textbf{eLSA} & ETSI & Fig. \ref{LSA_archtecture} & standalone, \newline licensed bands & not specified & database assisted & EU region & centralized & \xmark\\\midrule[0.5pt]
				\multirow{2}{*}{\textbf{SAS}} & FCC, \newline WInnForum & Fig. \ref{SAS_architecture} & standalone,\newline licensed bands & 3.55-3.7 GHz & database and\newline sensing assisted & USA & centralized & \cmark\\\midrule[0.5pt]
				\textbf{DSS} & 3GPP & Fig. \ref{DSS_architecture} & standalone,\newline licensed bands & Existing LTE bands & 4G and 5G\newline interworking &worldwide& decentralized & \xmark\\\midrule[0.5pt]
				\multirow{2}{*}{\textbf{LTE-U}} & LTE-U Forum & Fig. \ref{LAA_architecture}\footnotemark[1] & CA between licensed and unlicensed bands & 5 GHz & CSAT &China, Korea,\newline India, USA& decentralized & \cmark\\\midrule[0.5pt]
				\multirow{2}{*}{\textbf{LAA}} & 3GPP & Fig. \ref{LAA_architecture} & CA between licensed and unlicensed bands & 5 GHz & omni LBT &worldwide& decentralized & \cmark\\\midrule[0.5pt]
				\textbf{MulteFire} & MulteFire \newline Alliance & Fig. \ref{multefire_architecture} & standalone, unlicensed bands & 1.9, 3.5, 5 GHz & omni LBT &worldwide& decentralized & \cmark\\\midrule[0.5pt]
				\multirow{4}{*}{\textbf{NR-U}} & \multirow{4}{*}{3GPP} & \multirow{4}{*}{Fig. \ref{NRUecosystem}} & standalone,\newline unlicensed bands, CA\newline DC between licensed and unlicensed bands & \multirow{4}{=}{2.4, 3.5, 5, 6, 37, 60 GHz} & \multirow{3}{*}{dir/omni LBT}&\multirow{4}{*}{worldwide}& \multirow{4}{*}{decentralized} & \multirow{4}{*}{\cmark} \\
				\bottomrule[1.1pt]
		\end{tabular}}
		
		\footnotemark[1]{LTE-U and LAA ecosystems are similar; however, in LTE-U unlike LAA, the SCell employs CSAT procedure instead of LBT.}
	\end{table*}
	
	\subsubsection{COT Structure for NR-U}
	After a successful LBT, the UE can access the frequency spectrum band for a duration of the MCOT, which is 9 ms in the 60 GHz band. Owing to the flexible slot structure, NR-U can utilize the channel more efficiently. As an example, with an SCS of 120 kHz (slot duration = 0.125 ms), 72 slots fit within an MCOT of 9 ms, each of which is comprised of 14 OFDM symbols. 
	Therefore, 3GPP has defined two different COT structures, namely a COT with a single DL/UL switch and a COT with multiple DL/UL switches (\cite{3gpp_38_889}, Sec. 7.2.1.1).
	The signaling overhead induced by the COT with a single DL/UL switch is low as it requires only one guard band. However, it leads to more delayed HARQ feedback. Accordingly, this configuration is suitable for high-throughput conditions, like eMBB traffic. The COT with multiple DL/UL switches simplifies the HARQ timing; however, it sustains high overheads as it involves multiple guard bands, which leads to lower
	spectral efficiency. This configuration is suitable for delay-sensitive services, like URLLC.
	\subsubsection{Initial Access Procedures for NR-U}
	Similar to LTE, the initial access procedure for NR-U include primary and secondary synchronization signals (PSS and SSS) used by the UE to synchronize and recognize a network. In addition, there is a PBCH that carries the MIB alongside the PSS/SSS. In NR-U, the combination of PSS, SSS, and PBCH is called a synchronization signal (SS) block; it can also be referred to as the NR-U DRS. This block is primarily used for synchronization, initial access, and resource management \cite{3gpp_38_211}(\cite{3gpp_38_300}, Sec. 5.2.4). Due to the LBT requirement and channel occupation, the transmission of SS blocks can be interrupted. One solution to this problem is to transmit the SS blocks inside the periodically occurring time window. This has also been employed in LAA DRS transmissions. According to regulatory requirements in the 60 GHz band, the OCB has to contain 80\% to 100\% of the NCB \cite{ETSI_EN_302_567}. However, the SS blocks only occupy a single portion of the channel, which violates this regulation. The multiplexing of the SS blocks and data can mitigate this issue. Other alternatives include replicating the SS blocks in multiple frequency locations or redesigning the SS blocks to meet the regulatory demands. Interested readers can also refer to \cite{lagen2019new} for a comprehensive survey of NR-U, which includes a detailed description of existing challenges and potential solutions.
	
	A high-level standardization timeline of the spectrum sharing methods has been demonstrated in Fig. \ref{standardtimeline}. Furthermore, a comprehensive comparison of these methods is shown in Table \ref{sharing_comparison}.
	
	\begin{table*}[!t]
		\centering
		\caption{A summary of approaches using AI in spectrum sensing.}
		\begin{tabular}{p{0.75cm} p{0.5cm} p{12.5cm} p{2.5cm}}
			\toprule
			\textbf{Ref.} & \textbf{Year} & \textbf{Summary} & \textbf{Algorithm}  \\
			\midrule
			\cite{s1} & 2020 &  Use transfer learning strategies to improve the performance for real-world signals.
			Proposed method is better than traditional methods, i.e., maximum-minimum eigenvalue ratio and frequency domain entropy. & Deep learning classification
			\\\midrule
			\cite{s2} & 2020 &  Cooperative Spectrum Sensing for NOMA in cognitive radio & DAG-SVM \\\midrule
			\cite{s3} & 2020 &  Analyzed spectrum access and spectrum sensing in IoT network.
			Centralized and distributed AI-enabled. & DRL
			\\\midrule
			\cite{s4} & 2020 &  Combine RL with spectrum sensing based on three optimization objectives: Energy Optimization, Accuracy and Energy Joint Optimization and Throughput Optimization. & RL \\\midrule
			\cite{s5} & 2018 &  Classify input signal as primary user or secondary user. & DAENN and SVM
			\\\midrule
			\cite{s6} & 2020 &  Considered Spectrum sensing as classification problem.
			Used cooperative spectrum sensing scenario to classify the sensed signal from primary user as available or busy. & CNN
			\\\midrule
			\cite{s7} & 2020 &  Compute the PU activity based on spectrum sensing learning the implicit features from the spectrum data. & LSTM \\\midrule
			\cite{s8} & 2019 &  Formulate the spectrum sensing problem into multi-class classification in an OFDM system. & NBC \\
			\bottomrule
		\end{tabular}
		\label{sensing}
	\end{table*}

	\section{Application of AI and ML in Spectrum Sharing}\label{app_AI}
	6G communication networks will transform societies by providing wireless connectivity everywhere. Recent advances in AI have facilitated a wide range of new technologies. Such innovations and developments are possible with advanced AI models, large datasets, and high computing power. Furthermore, the proliferation of IoT devices has increased the demand for spectrum, leading to complicated radio resource management scenarios. Conventional optimization methods are incapable of solving these complex sharing scenarios, and AI can play a crucial role in resolving such issues \cite{white_paper}.
	
\begin{figure}[!b] 
		\centering
		\includegraphics[width=.45\textwidth]{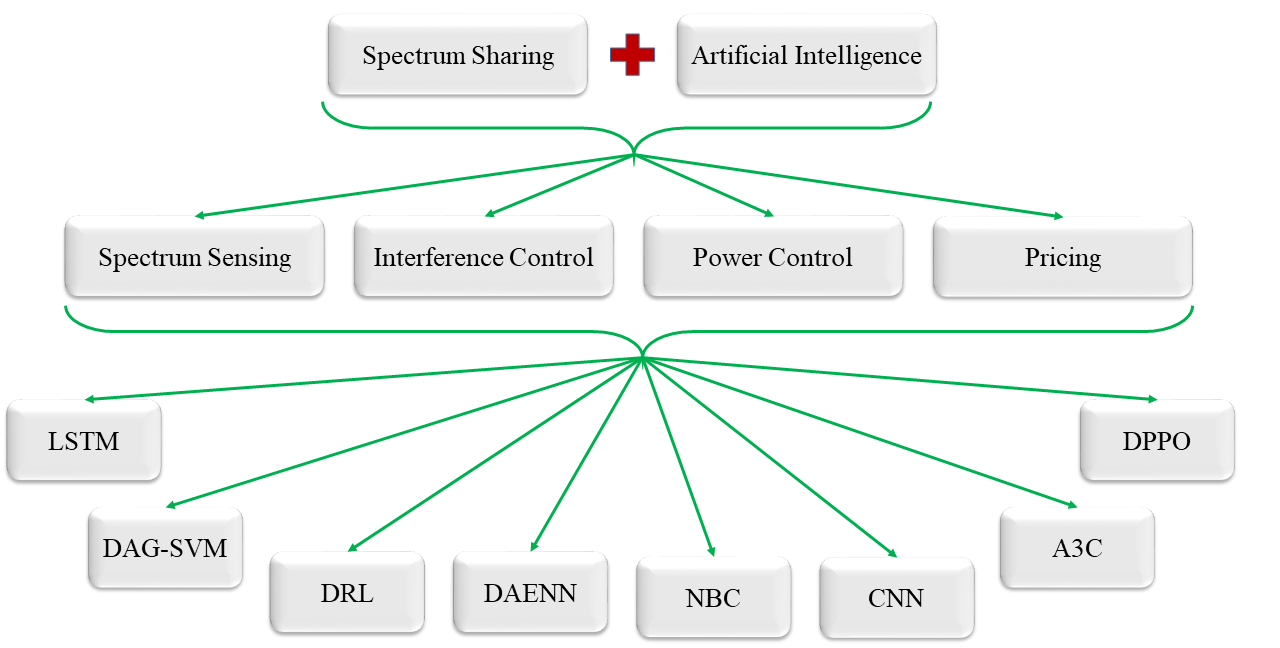}
		\caption{Taxonomy of spectrum sharing with AI.} 
		\label{AI_Fig}
	\end{figure}

	AI represents an essential capability for ensuring the efficiency of future wireless communication networks. Current wireless networks rely heavily on mathematical models that define the structure of a communication system. Such mathematical models often do not represent the exact structure of the systems accurately. As the mathematical modeling of such systems become accurate, it becomes increasingly challenging to analyze the system they represent. Spectrum sharing covers a range of services and requires synchronization and consistency between numerous entities on a large scale. Network optimization processes, for their part, require elaborate mathematical solutions that are often inefficient in terms of computation time and complexity. Also, they require substantial computational resources \cite{white_paper}. Therefore, ML algorithms must be deployed and trained at different levels of the network (i.e., management layer, kernel, and radio BSs) to address a variety of functions, including radio interface optimization and network management. 
	The critical point is that when using any spectrum sharing methods (i.e., LSA, LAA, SAS, etc.), AI can be adopted in different parts of the network. For instance, when a secondary user wants to utilize a primary user's spectrum, spectrum sensing is required, which can be done with the help of AI to detect ongoing transmissions in the medium more efficiently. Also, when a user's session ends, AI can be used in the spectrum sharing system for billing, channel estimation, synchronization, and user positioning. This last will be necessary for a spectrum sharing system, especially with mobility in the system. We illustrate taxonomy of spectrum sharing with AI in Fig \ref{AI_Fig}.
	
	\subsection{Spectrum Sensing}
	One of the critical actions to be taken in spectrum sharing and cognitive radio networks is the spectrum sensing process. However, one of the main challenges in this method is the data needed to train ML-based models. The first step would be to either synthesize the data based on some models for Tx/Rx and the propagation channel or to capture the data over the air. In either case, however, the data used for training might not be representative of the actual real-world scenario. It should be noted that this problem can only be handled by properly collecting or synthesizing training data for all possible instances. In \cite{RF_Dataset}, the authors developed a dataset consisting of synthetically generated waveforms of the incumbent for CBSD band. The goal was to detect whether the signal was present or absent regardless of its type. In \cite{MAC_protocol}, a novel database-assisted MAC protocol was presented, which let an incumbent operator share spectrum with a licensee operator under agreed terms and conditions. Hence, the licensee operator was able to use a portion of the available spectrum while adhering to the incumbent operator's minimum throughput standards.
	
	Assuming proper ML-based spectrum sensing training, in order to access the spectrum, secondary users must first be aware of any primary users in the frequency band so as not interfere with them. In recent years, various methods have been proposed to measure spectrum usage, each with its own set of advantages and disadvantages. However, with the development of AI algorithms, conventional methods have been replaced by ML and deep learning solutions, and recently these solutions have been applied in many fields. The reason can be summed up as follows:
	\begin{itemize}
		\item Better performance with AI, especially when dealing with complex systems.
		\item Faster inferences after the trained model is deployed.
		\item No need to design complex algorithms from scratch; just need to prepare appropriate training data.
	\end{itemize}

	In \cite{s1}, the authors modeled the spectrum sensing process as a deep learning classification problem and demonstrated that their method outperformed the other two conventional spectrum sensing methods (i.e., maximum-minimum eigenvalue ratio and frequency domain entropy). They further evaluated their results with colored noise and confirmed that their method performed better. In \cite{s2}, the authors proposed a cooperative spectrum sensing framework for NOMA in cognitive radio and pointed to a trade-off between sensing accuracy, training data volume, power allocation, and the number of secondary users.
	
	The frameworks of centralized and distributed network architectures based on AI as well as random access and spectrum sensing for IoT networks with deep reinforcement learning (DRL) were discussed in \cite{s3}. In \cite{s4}, the combination of spectrum sensing with RL was analyzed in three categories: energy optimization, joint accuracy and energy optimization, and throughput optimization. The authors also showed that throughput optimization generally obtains better average sensing accuracy. Spectrum sensing based on a deep auto-encoder neural network (DAENN) and support vector machine (SVM) was analyzed in \cite{s5}. First, the auto-encoder neural network was used for automatic feature learning (i.e., feature extraction), and then SVM was used for classification. With this method, they achieved up to 95\% spectrum sensing accuracy. In \cite{s6}, the authors applied a spatio-temporal system model of cooperative spectrum sensing to classify sensed signal and detect primary users based on a convolutional neural network (CNN). The authors examined different scenarios depending on how the primary users were placed and showed that the proposed model performed well with different level of noise. 
	A long short-term memory (LSTM)-based network capable of learning features from the spectrum data drawn from a correlation between a present and past timestamp to analyze PU activity statistics was proposed in \cite{s7}. The results showed that the proposed model improved the detection performance and accuracy at low SNR.
	The authors in \cite{s8} formulated a spectrum sensing problem in an orthogonal frequency-division multiplexing (OFDM) system and solved it with a naive Bayes classifier (NBC). They showed that spectrum sensing accuracy is higher than conventional methods in low SNRs by translating the Bayes error rate into a spectrum sensing error rate.
	In \cite{DeepRadar}, the authors presented a novel deep-learning-based environmental sensing system for detecting radar signals and determining their spectrum occupancy in the CBRS band. The method adapted its computations on the basis of available computing resources to make real-time decisions.
	
	\subsection{Pricing}
	The methods introduced so far require a contract between an incumbent and MNO, where all the features including frequency band, channel characteristics, transmission rules, and pricing, are stated. Here, we discuss pricing, which is another aspect of spectrum sharing, and we review the application of AI in it. In various spectrum sharing methods, the secondary user leases spectrum from the primary user based on a contract that includes the duration of use, the required parameters, and the specified price. But managing the cost of spectrum becomes more complex as the number of users increases. Also, due to managing the varying demand levels and the number of networks and applications, static methods for pricing are not useful. Therefore, it is possible to use dynamic pricing using artificial intelligence. Because it will be more efficient and more compatible with new generation networks when they have a high number of users. AI also facilitates continuous prediction and can adjust price quickly. Moreover, it is automatic, flexible, and completely scalable. \cite{Dynamic_AI}.
	
	In \cite{V2V_Price}, the authors studied the allocation of resources for a computation task among vehicles and designed a dynamic pricing scheme that motivated vehicles to contribute computing resources according to the price they received. The scheme was formulated as a sequential decision making problem, which could be solved with deep reinforcement learning. \cite{IIoT_MEC} proposed a dynamic pricing model to maximize MNO revenue and industrial IoT (IIoT) operators’ economic cost. A discrete finite Markov decision process (MDP) was used for the model and a Q-learning algorithm was used to solve the problem. The optimal strategy for leasing spectrum by a virtual network operator was considered in \cite{Dynamic_Price}. Spectrum sensing and leasing were used by multiple cognitive virtual network operators (CVNOs) to obtain spectrum resources from a mobile network operator. The authors aimed to find an efficient spectrum sensing and leasing strategy for a CVNO that would maximize its long-term utility. The problem was formulated as a sequential decision process that considered the dynamics of user activities and spectrum prices. The authors solved the problem with a deep reinforcement learning algorithm. In \cite{Smart_Price}, the authors studied a revenue maximization problem for a mobile edge computing (MEC) system, where an access point was equipped with an MEC server, which provided a job offloading service for multiple users while charging them a service fee. The authors proposed a policy gradient (PG)-based reinforcement learning algorithm which enabled continuous pricing.
	
	\subsection{Spectrum Sharing with AI and ML}
	Several articles have investigated spectrum sharing between primary and secondary users based on AI in cellular and IoT networks. Issues investigated include power control, resource allocation and fair sharing without causing interference for the primary user. AI is an appropriate solution for this problem due to the complexity of emerging coexistence scenarios involving many potential users with diverse access technologies and capabilities.
	
	In \cite{ss1}, the authors proposed spectrum sharing between LTE-U and Wi-Fi on the basis of RL to maximize LTE-U utilization. They compared the Almost Blank Subframe allocation (ABS), Markov modulated batch Poisson process (MMBPP), and RL. They showed that their proposed algorithm performed better than ABS and similarity to MMBPP without requiring extensive computations, although parameter estimation was needed in MMBPP. A spectrum sharing network for an IoT environment was considered in \cite{ss2} on the basis of deep Q-network. The proposed system was compared with random allocation and allocation based on request. The authors showed that their proposed system performed better, especially in an unknown environment. In \cite{ss3}, the authors proposed a coexisting LTE-LAA system based on a deep Q-Network. The neural network learned the traffic pattern of the LTE-LAA system to optimize its transmission time. This showed that LTE could operate on the unlicensed band without causing a disturbance for Wi-Fi and achieving maximum spectrum usage. \cite{ss4} considered a spectrum sharing system consisting of UAVs and licensed terrestrial networks, where the UAVs were categorized into two groups: relaying and sensing. Some UAVs acted as sensors and others acted as relays, with the UAVs deciding locally which task to perform. The problem was solved by a distributed reinforcement learning algorithm. In \cite{ss5}, the authors proposed a reinforcement learning-based subchannel selection technique where multiple LTE-LAA and Wi-Fi links competed for spectrum access in 5 GHz frequency band intelligently. The access points and eNBs selected the best subchannel based on MAC protocols and physical layer parameters. A combination of Q-learning and HARQ for controlling interference between LTE-LAA and Wi-Fi was considered in \cite{ss6}. The authors showed that there was a trade-off in priority between the LTE-LAA and Wi-Fi accessing the channel based on ACKs and NACKs received. If the number of ACKs was used for the reward function, LTE-LAA took precedence over Wi-Fi, and if the number of NACKs was used for the reward function, Wi-Fi took precedence over LTE-LAA. In both \cite{ss7-1} and \cite{ss7-2}, the authors considered a cognitive radio system consisting of a PU, a SU, and wireless sensors, where the SU could not sense the spectrum and needed to use the information of the wireless sensors to adjust its power. The power control problem was solved by distributed proximal policy optimization (DPPO), which is an asynchronous actor-critic (A3C) with adaptive moment (Adam) optimization. In \cite{ss8}, the authors performed the Category 4 algorithm introduced by 3GPP with RL techniques in two cooperative and non-cooperative groups. The proposed learning algorithms maintained transmission fairness between LTE-LAA and Wi-Fi and improveed the total throughput performance compared to the LBT procedure. In the DARPA spectrum collaboration challenge, the GatorWings team developed a collaborative intelligent radio network (CIRN) that could independently sense the entire communication environment, including RF channel characteristics, to obtain an understanding of the environment. It was also shown that the CIRN could make decisions and activate a spectrum sharing strategy which could optimize a set of given communication targets \cite{DARPA_AI}.
	
	\section{Pricing models and Economic issues}\label{price}
	Spectrum sharing between heterogeneous wireless networks largely depends technological innovation. In addition, the implementation of spectrum sharing methods can increase network efficiency and agility. Spectrum sharing markets are expected to change wireless economics and give rise to both predictable and as yet unanticipated implications for the valuation of spectrum \cite{lehr2020economics}. Here, we consider how the emergence of spectrum sharing markets and related economic issues will influence the economics of wireless networks and technology. For a better understanding of the value of the spectrum, first we discuss the economic value of the spectrum. Second, we discuss the growing demands for spectrum and the importance of spectrum sharing with economic considerations. Third, we discuss spectrum sharing markets. And finally we describe a number of spectrum pricing models in terms of how they are used in industrial projects and academic works.
	\begin{figure}
		\centering
		\includegraphics[width=1\linewidth]{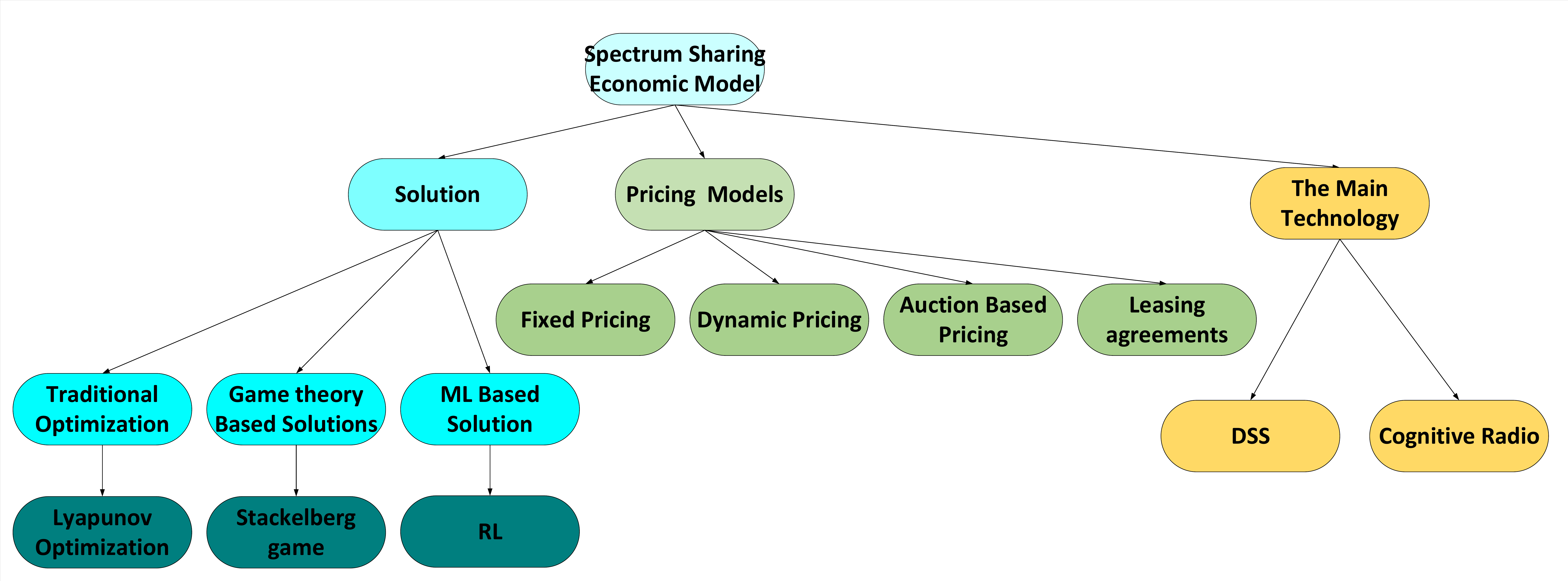}
		\caption{The main spectrum sharing economic model in the literature.}
		\label{fig:spectrum-sharing}
	\end{figure}
	
	\subsection {Estimating Spectrum Value}
	Spectrum plays an important role in the success of wireless technologies. Due to rapid growth and ubiquity of wireless technologies, emerging new concepts and innovations, like IoT, V2X communication, and mobile edge communication, there has been significant growth in spectrum value. In addition,  ITU in \cite{alden2012exploring} introduced and studied some of the main factors that have a significant influence on the valuation of frequency spectrum.
	\begin{table}[ht]
		\centering
		\caption{The main factors for spectrum  pricing \cite{alden2012exploring}.}
	\scalebox{.8}{
			\begin{tabular}{p{2cm} p{1.5cm}  p{1.5cm} p{4cm}}
				\toprule[1.1pt]
				&\textbf{Natural \newline features of \newline spectrum} & 	\textbf{Extrinsic features of \newline spectrum}  & \multirow{3}{*}{\textbf{Factors}}    \\
				\toprule
				\multirow{5}{*}{\textbf{Physical factors}}
				&\cmark & \xmark & Propagation
				characteristics
				\\
				&\cmark & \xmark & Sharing capacity
				\\
				&\cmark & \xmark & Harmonization of global and regional policies
				\\
				&\xmark&\cmark  & Geography  and climate
				\\\midrule		
			\multirow{6}{=}{\textbf{Socioeconomic factors}}
			&\xmark	& \cmark & Population density\\
				&\xmark		&\cmark & Income distribution
				\\
				&\xmark		& \cmark& Economic growth rate
				\\
				&\xmark		&\cmark &Political stability
				\\
				&\xmark			&\cmark & Absence of corruption
				\\	
				&\xmark			&\cmark & Rule of law
				\\\midrule	
			\multirow{14}{=}{\textbf{Policy and regulation}} &\xmark		 &\cmark & Independent regulatory
				agency
				\\
				&\xmark	&\cmark &Competition policy
				\\
				&\xmark	&\cmark& Infrastructure sharing
				\\
				&\xmark	&\cmark&Rules of protection of the 	public against electromagnetic waves
				\\	
				&\xmark	&\cmark& Open access rules
				\\
				&\xmark	&\cmark& Technology neutrality	\\
				&\xmark	&\cmark& Limitation of and protection \newline against interference	\\
				&\xmark	&\cmark&Licensing framework
				\\
				&\cmark	& \cmark&	Dispute-resolution mechanisms
				\\
				\bottomrule
			\end{tabular}
		}
		\label{price_factor}
	\end{table}
	\subsection{Growing Demand for Spectrum}
	As Cisco reported in \cite{cisco2020cisco}, by 2023, worldwide 5G devices and connections will account for more than 10 percent of mobile devices and connections. In addition, the global mobile device market is expected to grow from 8.8 billion devices in 2018 to 13.1 billion devices by 2023--about 1.4 billion of which will be 5G enabled. As a result, efficient use of the frequency spectrum and spectrum sharing methods are vital for the future of telecommunication networks.
	
	\begin{figure}[!b]
		\centering
		\includegraphics[width=0.9\linewidth]{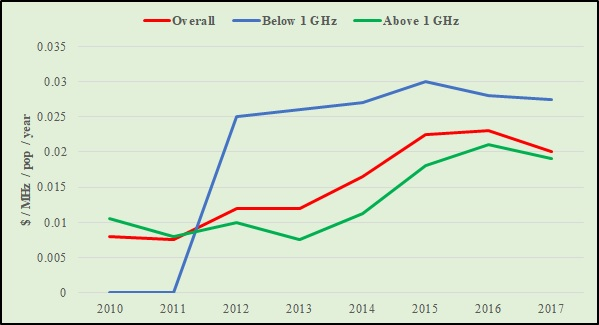}
		\caption{4G final spectrum price on the rise in development  country \cite{ITU_Spectrum_Management}.}
		\label{fig:spectrum-price}
	\end{figure}
	\subsection {Secondary Markets and Sharing}
	An important enabler for spectrum sharing in the future is the emergence of more robust secondary markets for trading bundles of spectrum rights. Secondary market trading allow spectrum resources to be shared across heterogeneous networks, business models, and use cases.
	Today, secondary markets have not yet seriously entered into economic activities related to spectral sharing markets.
	 If efficient secondary markets did exist for spectrum, such markets would serve a number of important functions. First, such markets would provide a continuously available mechanism	for balancing aggregate spectrum supply and demand over time. In so doing, they would help	ensure that spectrum is continuously assigned to its most valuable use. The transaction data	provided by such markets would be useful for estimating the value of spectrum and a signal to all market participants of mismatches between supply and demand. The availability of such market price signals would help promote competition and enable market participants to better target their wireless investments and business planning. Spectrum users would be forced to confront the opportunity cost of using the spectrum, and the price signals would impose stronger incentives to use spectrum efficiently. Accordingly, each spectrum sharing method has its own particular characteristics and requirements. 
		\begin{table}[!t]
		\setlength\lightrulewidth{0.051pt}
		\caption{Leasing agreement on 24 GHz frequency ban for 5G in EU \cite{Design_of_the_European}.}
		{\renewcommand{\arraystretch}{1.2}
			\begin{tabular}{p{0.8cm} p{1.2cm} p{0.9cm} p{1cm} p{1cm} p{1.4cm}}
				\toprule[1.1pt]
				\textbf{Country} & \textbf{Date of reward} & \textbf{Total paid} & \textbf{Price MHz/pop} & \textbf{Duration}&\textbf{Spectrum cap}\\\midrule
				Finland& Nov. 2016& \texteuro66.3m&\texteuro0.2&17 years&2$ \times $10 MHz\\\midrule
				France&Nov. 2015&\texteuro2.8bn&\texteuro0.69&20 years&2$ \times $15 MHz (2$ \times $10 MHz for some)\\\midrule
				Germany&June 2015&\texteuro1.0bn&\texteuro0.2&18 years&None\\\midrule
				Italy&Oct. 2018&\texteuro2.0bn&\texteuro0.5&20 years&2$ \times $15 MHz\\\midrule
				Sweden&Dec. 2018&\texteuro0.27bn&\texteuro0.677&22 years&40 MHz\\
				\bottomrule[1.1pt]
		\end{tabular}}
		\label{fig:screenshot001}
	\end{table}

	\subsection{Spectrum pricing model}
	Pricing models depend on whether they are for industry or academic applications. Although the basis of these methods is the same, their implementation and application can be different. Especially in commercial and industrial applications, organizations or legislative institutions can implement either method with a unique approach. In this subsection, we evaluate industrial and academic approaches to spectrum pricing. As we can see in Fig. \ref{fig:spectrum-sharing}, the economic model of spectrum sharing can be categorized into four classes, namely fixed pricing, dynamic pricing, auction-based pricing and leasing agreement. Different pricing methods have been studied in the literature drawing on various solutions, such as ML-based algorithms, game theory, and Lyapunov optimization methods. In \cite{niyato2008competitive,hu2021distributed,qian2020multi}, the authors studied a spectrum sharing framework that could include PUs and SUs or MNOs and infrastructure providers (InPs) by deploying technology like DSS. In contrast to academic applications, pricing models for industrial applications are more focused on fixed pricing auctions, and leasing agreements. In the following, we explain the difference between pricing models in industrial and academic applications in more detail:
	\begin{table}[!t]
		\centering
		\raggedleft
		\tiny
		\setlength\lightrulewidth{0.051pt}
		\caption{Leasing agreement on 3.5 GHz frequency ban for 5G in EU \cite{Design_of_the_European}.}
		{\renewcommand{\arraystretch}{1.2}
			\begin{tabular}{p{0.6cm} p{1.2cm} p{0.8cm} p{0.8cm} p{0.6cm} p{2.2cm}}
				\toprule[1.1pt]
				\textbf{country} & \textbf{date of reward} & \textbf{total paid} & \textbf{price MHz/pop} & \textbf{duration}&\textbf{spectrum cap}\\\midrule
				
				\multirow{3}{*}{Czech}&          &                &              &         &Incumbent Operators: 40 MHz\\
				&July 2017 &\texteuro38.4m  &\texteuro0.02 &15 years & New entrants: 80 MHz \\
				&          &                &              &         & No new entrants: 80 MHz \\ \midrule
				Finland& Oct. 2018 & \texteuro77.6m &\texteuro0.04 &15 years &130 MHz \\\midrule
				Hungary& April 2016& \texteuro2.9m&\texteuro0.004 &18 years &100 MHz, with sub-caps \\\midrule
				Ireland& May 2017&\texteuro60.5m &\texteuro0.04 &15 years &150 MHz per region \\\midrule
				Italy& Oct. 2018& \texteuro6.5bn &\texteuro0.35 &18 years & 100 MHz\\\midrule
				Norway& Feb. 2016&\texteuro2m & \texteuro0.002&6 years & No\\\midrule
				Romania& Oct. 2015 & \texteuro10m& \texteuro0.002& 10 years&No \\\midrule
				\multirow{2}{*}{Slovakia}&March 2015 &\texteuro1.17m&\texteuro0.002&10 years&No\\
				&Aug. 2015&\texteuro2.43m&\texteuro0.004 &10 years&No \\\midrule
				\multirow{2}{*}{Spain}&May 2016 & \texteuro20m&\texteuro0.01&14 years&No\\
				&July 2018&\texteuro437.64m&\texteuro0.05 &20 years&120 MHz (3.4-3.8 GHz band) \\\midrule
				UK& April 2018 & \texteuro1.33bn &\texteuro0.134 &Indefinite & 340 MHz on overall spectrum held\\\midrule
				Austria& March 2019 & \texteuro188m &\texteuro0.06 &20 years &National+regional \\
				\bottomrule[1.1pt]
		\end{tabular}}
		\label{fig:screenshot002}
	\end{table}
\begin{table*}[!t]
	\centering
	\caption{Summary of works on spectrum pricing for academic applications.}
	\begin{tabular}{p{1cm} p{1cm}  p{13cm}}
		\toprule[1.1pt]
		\textbf{Ref.} & \textbf{Year} 
		& \textbf{Method / Scenario}   \\
		\midrule
		\cite{zhao2018price}  & 2018 & Stackelberg game model / Spectrum sharing and power allocation 
		\\\midrule
		\cite{sengupta2009economic}  & 2009 &  Knapsack problem /  Pricing and  dynamic spectrum for wireless service providers
		\\\midrule
		\cite{matinmikko2020spectrum}& 2020&   Spectrum sharing and related enabling technologies are discussed in relation to the upcoming 6G era.
		\\\midrule
		\cite{joshi2017dynamic}   &2017& Lyapunov Optimization / Dynamic Inter-Operator Spectrum Sharing	\\\midrule
		\cite{li2017joint}		& 2018& Stackelberg
		game model /  Pricing for	multi beam satellite systems  	\\\midrule
		\cite{qian2020leveraging} &2020& Spectrum Sharing in 5G-VANET / Stackelberg Pricing Game\\
		\bottomrule
	\end{tabular}
	\label{pricing_methods}
\end{table*}
	\subsubsection {Dynamic Pricing}
To the best of our knowledge, there is no practical project under dynamic pricing. In this part, we evaluate some scientific papers. The reason for this can be the value of significant investments made in telecommunication networks according to their rate of return on investment. Implementing a variable price-based framework can overshadow the profit margins of MNOs. 
In addition, considering the role of other network actors such as ISPs or InPs can complicate economic conditions.
According to these assumption, in this section, we  review some works in the field of dynamic pricing.
	In \cite{vslapak2021cost}, the authors proposed an approach for resource allocation on edge nodes in a  multi-access edge computing (MEC) system model with multitier nodes. Using Bayesian optimization, MEC was shown to achieve 20--40$\%$ higher performance than pseudo-random optimization, even under budget constraints.
	In addition, some works have studied resource pricing jointly with resource allocation. In this context, the unit price of each of the resources is assumed as an optimization variable. Moreover, to provide the partners' profits in the network, such as MNOs, InP, and ISPs, several works have studied game-based algorithms for scenarios such as cooperation and competition. In \cite{zhao2018price}, the authors studied a price-based power allocation problem in a cellular network  with spectrum sharing. They propose a new problem based on pricing to manage the interference and optimize resource utilization in macrocell and femtocell coexistence. By deploying the Stackelberg game scenario and considering the macrocell and femtocell as the game players, they solved the proposed problem. Their results showed that spectrum sharing between macro and femtocells can increase the resource utility and profit under power constraints. 
	\subsubsection {Auction-Based Pricing}
\begin{figure}
		\centering
		\includegraphics[width=0.7\linewidth]{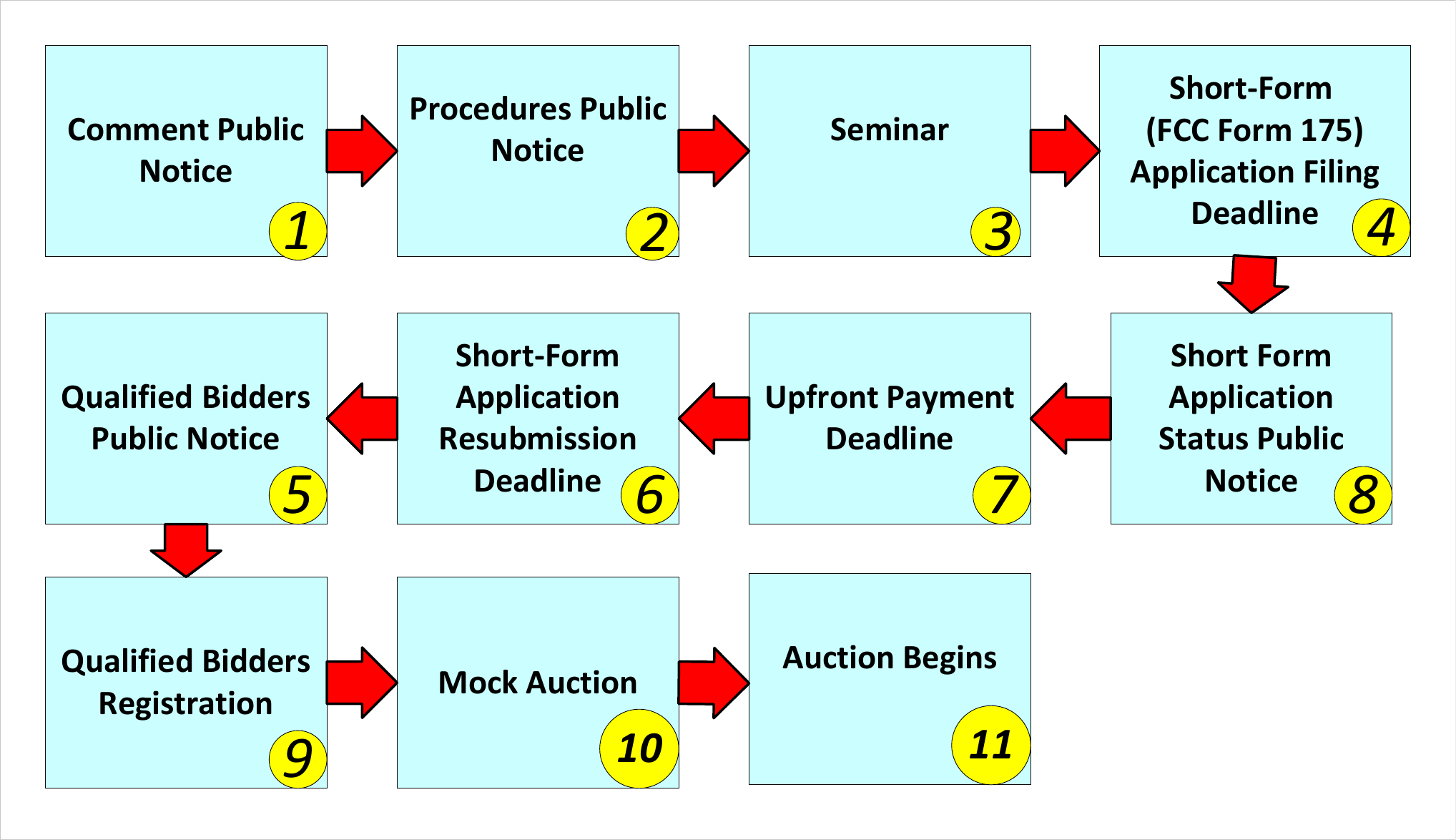}
		\caption{FCC pre-auction process \cite{FCCPRE}.}
		\label{fig:fcc2}
	\end{figure}

	This is one of the most common models of pricing for spectrum sharing where it is significant in real economic and industrial projects.
	\paragraph{Industrial Projects} Regulatory organizations hold  auctions  at different intervals, but the way these auctions are held is different in different organizations. Despite the existence of a wide variety of regulatory organizations that are very old in this field, here, we review some important organizations and their regulations for auctions.
	\begin{itemize}
		\item Ofcom: Buyers must first apply to participate in the auction for obtaining a license to use the spectrum bands. Next, Ofcom examines each application and determines whether the application should be qualified to participate in the award. The identities of qualified bidders are then made public. After the bidding opens, a number of rounds follow. Ofcom sets the prices for each round, and each bidder bids for the amount of spectrum they are willing to purchase at the price for that round. Spectrum prices generally increase until the amount of spectrum available matches demand \cite{ofcom_auc}. 
	\end{itemize}
	\begin{figure}[!b]
		\centering
		\includegraphics[width=0.7\linewidth]{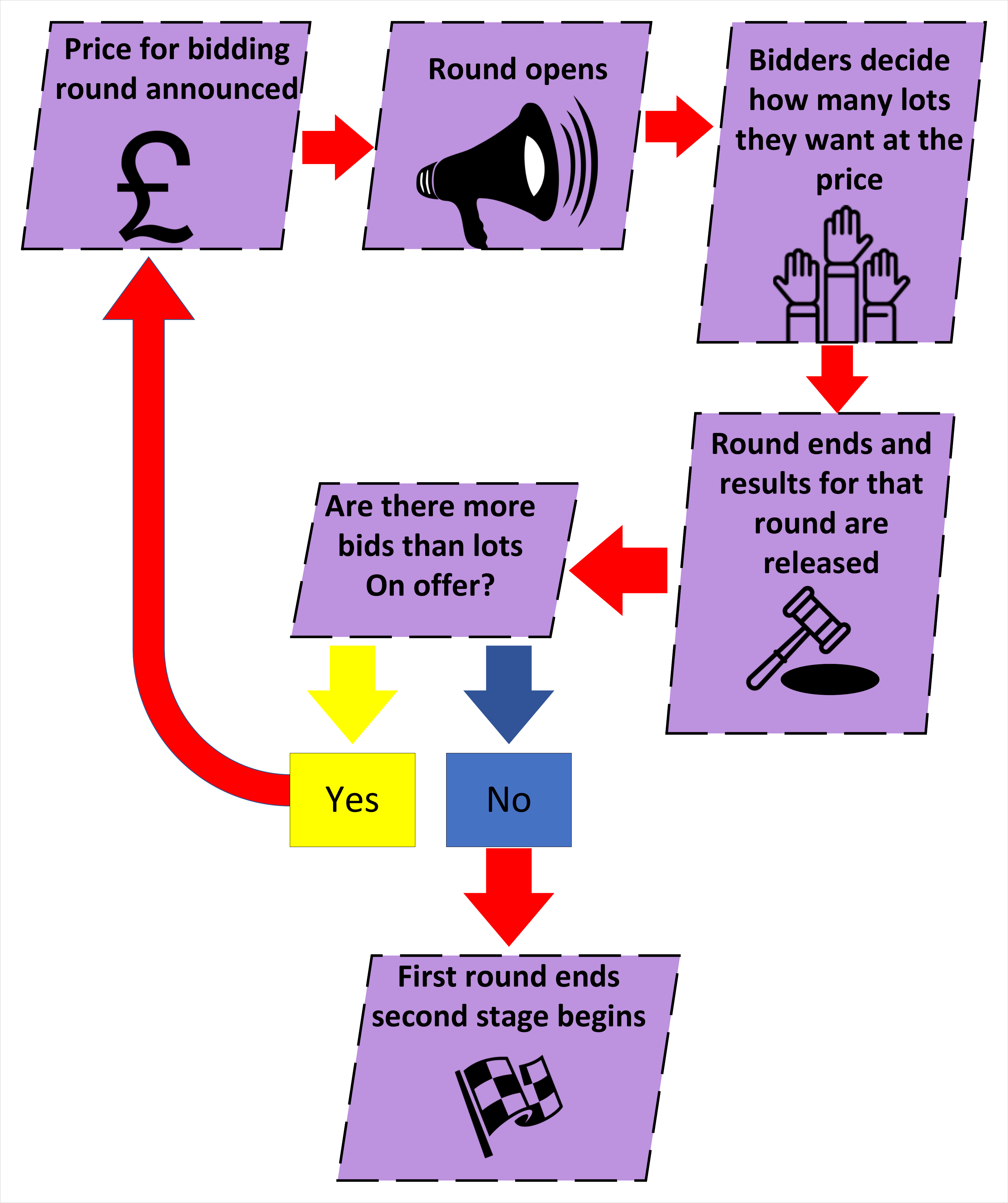}
		\caption{Spectrum auction process for ofcom \cite{OFCOMPERIOD}.
	}
		\label{fig:ofcom}
	\end{figure}
	\begin{itemize}
		\item FCC: The FCC is one of the most experienced and reputable regulatory organizations for auctions. There are two stages for auctions held by the FCC: pre-auction and auction. The pre-auction stage can be broken down into ten smaller phases \cite{fcc_auc}:
	\begin{enumerate}
		\item Public notice for comments: phase is held 4–6 months prior to the auction. In this phase,  according to the Budget Act of 1997, a public notice is released to seek comments on auction procedures, such as activity rules, upfront payment amounts, and minimum opening bids.
		\item Public notice for procedures (3–5 months prior to auction): Procedures public notice: Potential participants are informed about the procedures, terms, and conditions of the auction event in a public notice. This phase is held approximately 3--5 months prior to auction.
		\item Seminar: To introduce potential participants to the auction rules and process, the FCC offer a free pre-auction seminar. During the presentation, participants are given a chance to ask questions and demonstrate the bidding software.
		\item Short-form  application filing deadline: In this stage, the information of the bidders is collected. This stage occurs 45 to 60 days before the auction.
		\item Public notice for short-form application status: After the deadline for filing, the FCC reviews all short-form applications and deem them accepted, incomplete, or rejected. This public notice announces the status of the applications. Each applicant receives a package containing the public notice from the FCC. If an application is incomplete, a letter explaining the reason for the "incomplete" designation is included. There is also a contact person at the FCC, in case the applicant needs more explanation. This stage is held 30--40 days prior to the auction.
		\item Upfront payment deadline: In order to participate in the auction, bidders must submit a refundable deposit, which is used to purchase eligibility (bidding units). Participants pay the prepayment amount 3 to 4 weeks before the auction.
		\item Short-form application resubmission deadline: This deadline, which often coincides with the upfront payment deadline, must be accompanied by form 175 applications that have been deemed incomplete.
		\item Qualified bidders public notice: The notice includes the list of qualified bidders, their FCC registration numbers, and claims for bidding credit. This public notice may also list the bidding unit eligibility of each bidder, along with the items that they selected on their FCC form 175 in the case of some auctions. Public notices also include information about the mock auction, the auction schedule for the first day, and other auction-specific details.
		\item Qualified bidders registration: In this phase, seven days prior to the auction, the participants are sent mail that includes information for logging into the Integrated Spectrum Auction System (ISAS).
		\item Mock auction: Five days before the auction, the FCC conducts mock auctions for qualified bidders to ensure they know how auctions work. During the mock auction, bidders become familiar with the process of bidding.
	\end{enumerate}
		After completing the pre-auction, the FCC holds the auction, which is structured as a simultaneous multiple-round (SMR) auction. Bidding in SMR auctions is discrete and successive, unlike most other auctions where the length of each round is announced in advance. The results of each round are released after the round closes. When bidders learn about the other bids placed by other bidders, they can make new bids. As a result, all bidders will have information about the value of the licenses, which increases the possibility of the licenses being awarded to the highest bidders. Bidders can adjust their bidding strategies between auction rounds. In addition, the FCC allows package bidding, where introduced another  method for auctions known as package bidding, where participants can bid on either individual licenses or groups of licenses. As a result, bidders can benefit from combining complementary items that may exist between licenses and avoid losing out on partial licensing. The use of package bidding is generally appropriate when some bidders have complimentary licenses, and the complementarities vary among bidders. A package bidding process produces a good outcome in these circumstances, since licenses are secured by those bidders who are most able to pay. The FCC's pre-auction process is outlined in Fig. \ref{fig:fcc2}.
	\end{itemize}

	\begin{figure}[!b]
		\centering
		\includegraphics[width=0.7\linewidth]{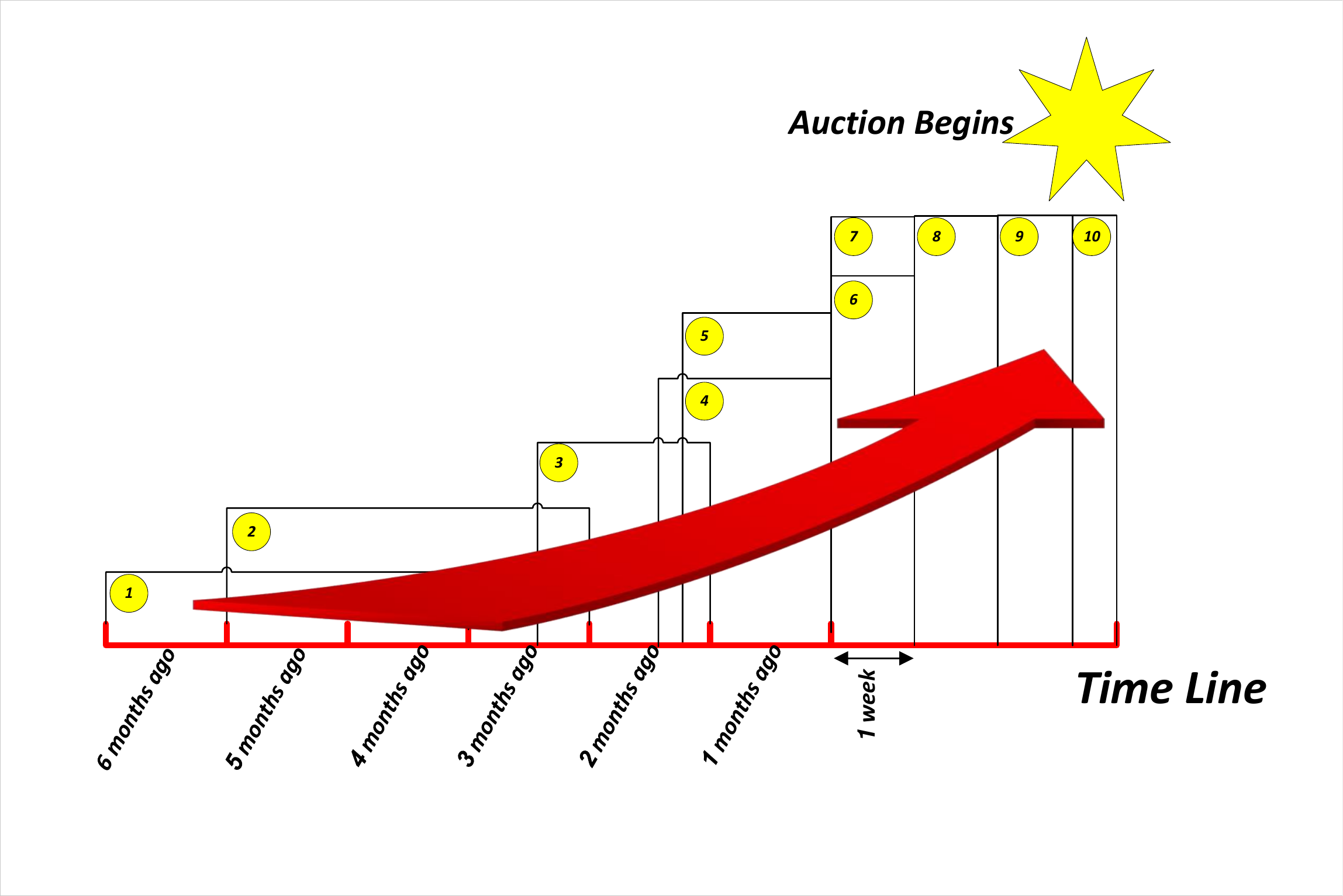}
		\caption{Pre-auction timeline for FCC \cite{FCCPRE}.}
		\label{fig:fcc3}
	\end{figure}
	
	\begin{itemize}
		\item GSMA: In recent years, the GSMA has published valuable reports in the field of pricing and spectrum management. This reports have included reviews of the situations in different countries. In \cite{GSMA_auc1} the GSMA described different types of auctions and the processes involved. In this report, the GSMA highlighted the following points:
		\begin{enumerate}
			\item Mobile services that are affordable and of high quality should be the number one goal for spectrum auctions.
			\item Government policies that maximize revenue at the expense of consumers may cause serious harm.
			\item Choosing an auction design should not increase risk and uncertainty for bidders.
			\item Inefficient outcomes can result from wrongly selected lot sizes or inflexible packages of spectrum lots.
		\end{enumerate}
		To study and address the spectrum pricing challenges in the Latin America countries, the GSAM in \cite{GSMA_auc2} provided some suggestions that could help efficient spectrum pricing  and the telecommunication development.
According to this report,
despite the fact that auctions are viewed as a way to increase state revenue, this report disproves the claim.
		  Because of high spectrum prices can create disincentives for investment in networks in developing countries. Therefore effective pricing has a two-way effect on government revenue and network development.
	\end{itemize}

	\paragraph{Academical Works} As discussed above, the process of creating and holding a frequency spectrum auction is a long process. Yet most academic works assume that the access time to the spectrum is very short. To support the coexistence of rate guaranteed cellular users and massive IoT devices,  the authors in \cite{qian2020multi} examined the dynamic spectrum sharing problem between multiple operators. With the Stackelberg pricing game, they introduced a wireless spectrum provider (WSP) for spectrum trading among mobile network operators (MNOs). They showed that coexisting IoT devices and cellular device users under a single MNO could be ensured with coexisting access rules that honor QoS for users and prioritize their needs. Stackelberg equilibrium (SE) solution and its ability to maximize MNO and WSP payoffs simultaneously are of particular interest in this study. IEEE 802.22-based WRANs were described in \cite{niyato2009dynamic} with regard to the major issues around dynamic spectrum sharing. The authors showed that WRAN services provided by IEEE 802.22-based solutions can operate on the exclusive-use model of dynamic spectrum access where TV broadcasters grant the WRAN service providers the exclusive right to use spectrum. By modeling spectrum sharing as a market with multiple buyers and sellers, the authors  applied a sealed-bid double auction model to determine the number of IEEE 802.22 service bands procured by vendors as well as their costs. As WRAN service providers procure TV bands, they compete with each other to charge WRAN users a fair price for their services. In addition, to obtain the Nash equilibrium for competitive spectrum bidding and spectrum pricing strategies, the authors developed non-cooperative games.

	\subsubsection {Leasing Agreement}
\begin{figure}[!t]
		\centering
		\includegraphics[width=0.7\linewidth]{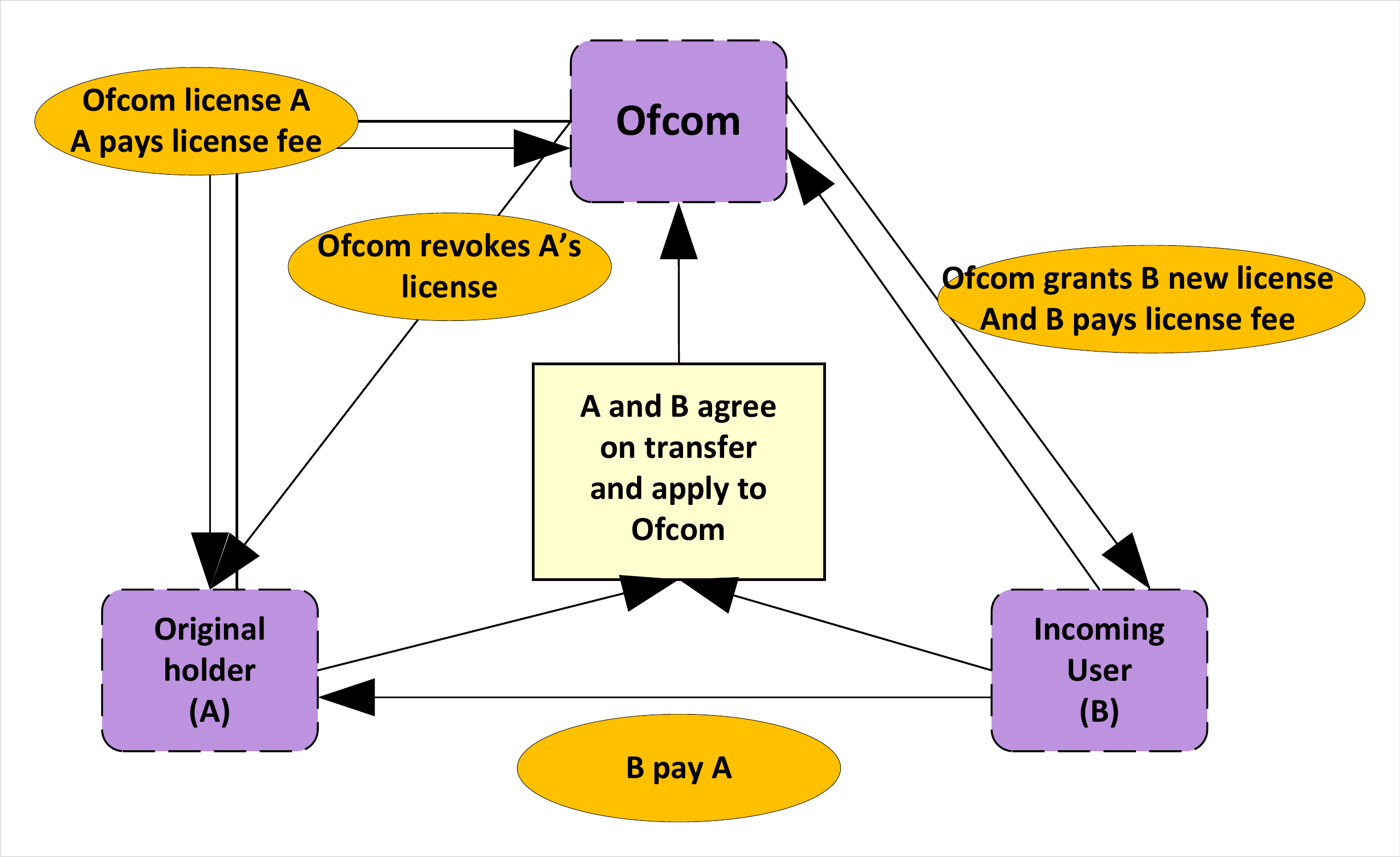}
		\caption{Spectrum transform process for Ofcom \cite{OFCOMLEAS}.}
		\label{fig:ofcom-2}
	\end{figure}

	One of the most important pricing methods and ways of accessing the frequency spectrum is the leasing method. During a process, a spectrum holder gives permission to use its spectrum under certain conditions.
	\paragraph{Industrial Projects}
	here, we review some documents for leasing agreement.
	\begin{figure}[!t]
		\centering
		\includegraphics[width=0.7\linewidth]{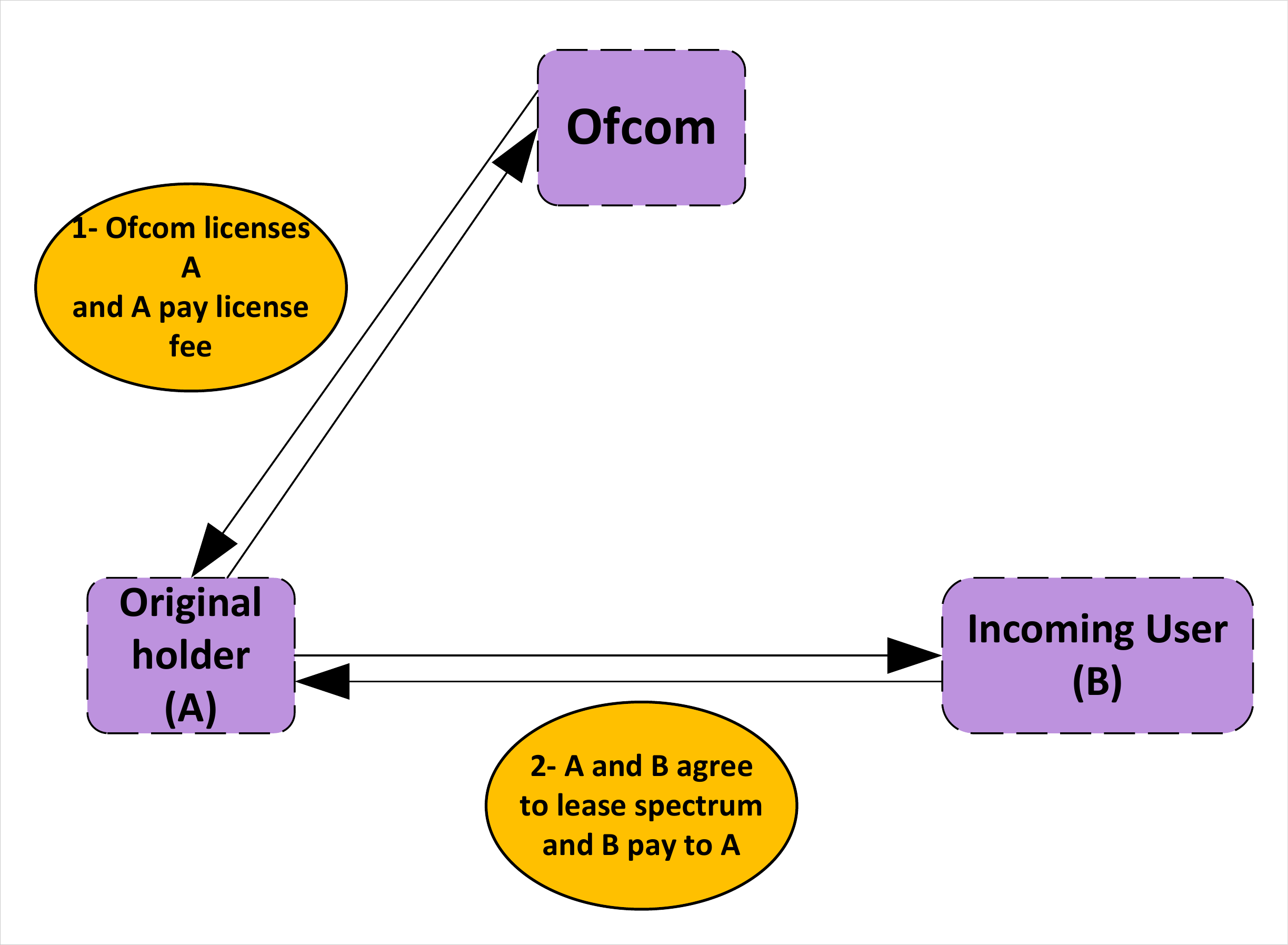}
		\caption{Spectrum leasing process for Ofcom \cite{OFCOMLEAS}.}
		\label{fig:ofcom-3}
	\end{figure}
	\begin{itemize}
		\item Ofcom: To facilitate access to the frequency spectrum, Ofcom regulates spectrum leasing \cite{ofcom_leas1}. They proposed a framework called ``spectrum transfer'' that is illustrated in Fig. \ref{fig:ofcom-2}. There are two primary difference between spectrum transfer and leasing:
		\begin{enumerate}
			\item Spectrum leasing is an agreement between the parties that do not require us to issue a new license. Leaseholders (incoming users) are not given licenses by Ofcom; rather, they are authorized to use the spectrum based on a lease agreement with a licensee (lessor). When the leaseholder does not comply with license terms and conditions, the lessor may, depending on the circumstances, act directly against the leaseholder. Leases will initially be permitted and regulated by license terms and conditions. Leases can only be granted if a license contains the necessary terms and conditions.
			\item Spectrum transfer proceeds by the transfer of license rights and obligations and necessarily involves the grant of a new license to the transfer. The transfer may be for all or part of the license duration. Ofcom refers to the latter as limited time transfer spectrum. 
		\end{enumerate}
	\end{itemize}

	\begin{itemize}
		\item FCC: FCC spectrum leasing is part of its secondary market initiatives aimed at removing regulatory barriers and opening up spectrum access \cite{fcc_leas}. In this context, companies that have an exclusive license for certain spectrum bands can lease spectrum to third parties using two different arrangements:
		\begin{enumerate}
			\item spectrum manager leasing: spectrum manager leases may
			be entered into by licensees and lessees provided that (1) the
			licensee retains de jure control over its license according to
			the spectrum leasing standard or a  de facto control over the leased
			spectrum exists. The licensee or lessee must also meet the other requirements for spectrum manager leases (if necessary) according to the FCC rules.
			\item De facto transfer leasing: Under a de facto transfer lease, the licensee retains de jure control of the license and has de facto control over the use of the leased spectrum, provided the licensee and lessee also adhere to all other applicable requirements set forth in the commission's rules.
		\end{enumerate}
		In addition, as part of its new rules on spectrum leasing, the FCC made an amendment to allow certain spectrum leasing applications to be processed in an immediate manner (as well as license assignments and transfers). For certain spectrum leasing arrangements, these changes removed the public comment period completely.
	\end{itemize}
	\paragraph{Academical Works} 
	In \cite{wang2010cooperative}, the authors assumed an SU selects a set of secondary users to serve as cooperative relays for a PU, with the PU selecting the secondary users to serve as the primary users' cooperative partners. As part of this agreement, the PU leases a certain amount of SU access time for their transmissions. In the case of cooperative relays, the PU decides how much channel access time it will leave to the SUs (i.e., the selected relays), and the relays decide their power level. The PU transmits to achieve proportional access time to a channel. Based on their assumptions, the PU and SU achieve maximum utility by following rational and selfish rationality. Since an SU’s utility is a function of its own transmission rate and the power cost for PU’s transmission, it will choose a proper power level to meet the trade off between transmission rate and power cost. To encourage cooperative relays to employ higher power levels, the PU will select a proportion of channel access time for them. The authors showed that the proposed game converges to a Stackelberg equilibrium when the PU and SUs play it non-cooperatively. The unique equilibrium point can be reached by using an iterative updating algorithm.
	
	\subsubsection {Fixed Pricing Model}
		Regardless of how the spectrum is accessed, in this method the company or organization operating the frequency spectrum pays a fixed amount periodically, such as annually or monthly.
	\paragraph{Academical Works} 
		In academic works, authors assume that the duration time for the payment is much less than in industrial projects. To meet the dynamic on-demand needs of RAN-slicing under a spectrum sharing scenario, the authors in \cite{li2020service} studied a spectrum-aware service-oriented RAN-slicing trading scheme (SSRT). SSRT slices multi-dimensional resources, including heterogeneous spectrum (licensed and unlicensed), time, and network infrastructure (nodes, radios, powers). Based on the cases discussed in this section, access to the frequency spectrum through methods other than spectrum sharing can be exceedingly time-consuming or costly. Hence, the importance of using spectrum sharing methods is very important.
	\subsubsection{Spectrum Discussing for 6G Future Network}
		A new era of 6G technology in the 2030s will bring new challenges because the current  spectrum band allocations and spectrum management approaches are fragmented in a way that creates drastically different propagation characteristics \cite{matinmikko2020spectrum}. 6G network spectrum discussions have just begun. Obtaining new spectrum bands to support mobile communication networks is a global process conducted at the ITU level through the WRC. In the past decades, accessing new spectrum bands created rivalry among the holders of spectrum access rights to the new bands envisioned for mobile communications. Spectrum sharing, simultaneous wireless information, and power transfer (SWIPT) are efficient techniques to increase spectrum and energy efficiency. These techniques were studied \cite{lu2020swipt} for 6G-enabled cognitive IoT networks. To address a SWIPT-based problem, the authors proposed a two-phase scenario. In the first phase, the IoT device transmitter (DT) in a cognitive IoT network decodes the received RF signal and harvests energy from it. To avoid interference in phase two, the signals of the DT and the primary system are transmitted with orthogonal subcarriers using the harvested energy.
	
	\section{Open Problems and Issues } \label{problem}
	In the previous sections, we thoroughly surveyed existing spectrum sharing methods both in the licensed and unlicensed bands. In this section, we will discuss the challenges and open problems that remain for each sharing framework. These challenges have to be considered before attempting to apply these sharing schemes.
	
	As discussed above, the LSA framework targets the 2.3 GHz band. 2.3 GHz deployment leads to a decrease in investment in extra base
	stations. Most of the expenditures that LSA deployments incur for current cellular systems are due to LR and LC deployments and the administrative costs associated with defining new rights and issuing licenses. Furthermore, interference management in this method is static and based on the predefined information in the LR. There is also the issue of the highly dependent nature of the LSA framework on the incumbents' partnership. Participating incumbents must agree to report their spectrum usage to the LR, and there must be a concerted effort to influence these incumbents to take part in the sharing process. Besides, there are still some concerns that have not been addressed in standardization works, issues such as how much time is needed for proper evacuation of the LSA licensees from the band, or what happens if there are other incumbents in the band who are unwilling to share. 
	
	These predicaments also apply to the SAS framework. Aside from SAS's complexity and intricate implementation, it has not been specified how mobility management is conducted or whether there will be any interface between the CBSDs and the MNOs' already deployed BSs. In addition, the protocol for the interaction of SAS and ESC is still under standardization.
	
	By extending the spectrum sharing methods into the unlicensed spectrum, MNOs face more restrictions. These bands are heavily crowded with Wi-Fi users, and MNOs irrespective of any technology that they are employing in the unlicensed spectrum (LTE-U, LAA, NR-U), should not expect a significant revolution in their QoS levels and services. All BSs or devices aimed at operating in these bands have to employ the LBT procedure. However, in scenarios related to the NR-U implementations, as mentioned earlier, hidden node and exposed node problems emerge due to the beam-based transmissions. Nonetheless, one must evaluate whether it is worth investing and updating the devices to be compatible to work in these bands and analyze the balance between the number of users and the operational expenditure (OPEX) for the small cell deployments required.
	
	Finally, due to the increasing demand from 5G cellular users, one method that can mitigate the problem of frequency spectrum scarcity is dynamic spectrum sharing. This can allow MNOs to use DSS as a way of gradually evolving their networks towards 5G. Meanwhile, according to what vendors and MNOs have reported, the main way to implement DSS is to upgrade related equipment software. Accordingly, we can infer that CAPEX is one the main challenges for MNOs in updating their equipment. In addition, MNOs need to assess the rate of return on investment to implement this method given the increasing number of 5G users over time. Technically, it can be concluded that vendor lock can hamper the ability of start-up networks that need DSS. Moreover, DSS implementation needs two of the main components of the cellular networks (eNB and gNB) to be synchronized and coordinated. Therefore, the signaling data resulting from this coordination can affect the performance of the entire cellular network.
	\section{Conclusion}\label{Conclusion}
	In this article, we provided a comprehensive survey of existing spectrum sharing methods both in the licensed and unlicensed regime for cellular systems with a focus on two principal aspects: standardization and implementation. We took a critical look at the standardization activities as well as the steps required for the efficient design and deployment of each sharing scheme. It is hoped that this bilateral view will help to narrow the gap between theory and practice, since when it comes to real-world implementation, the vendors and telecom companies can adopt different approaches, while referring to the same standard. Although progress in spectrum sharing shows promise, it is clear that it can exert severe pressure on incumbents and MNOs alike. f course, spectrum sharing involves interference in system operations when MNOs access shared spectrum. Furthermore, both incumbent and MNOs need to have a decent estimation of the potential economic profits, which means that defining a proper pricing model is of great importance. In this regard, incentives are needed to encourage both sides to participate in the sharing process. Even though there has been a concerted effort in academia and industry to mitigate these problems, as the number of requests from MNOs to cooperate in sharing processes increases, we can anticipate an inexorable rise in the complexity of sharing frameworks. In this regard, AI and ML algorithms can bring a lot of benefits to the table. In this survey, we outlined the recent progress in the application of AI methods at various levels of a sharing process, from intelligent interference management systems to effective power control scenarios. We saw that the full potential of AI has yet to be exploited and that further studies are needed to evaluate its capacity to facilitate more complex sharing scenarios.
	
	\bibliographystyle{IEEEtran}
	\bibliography{References}
\end{document}